\documentclass[11pt,a4paper]{article}
\usepackage{jheppub}
\usepackage{amsmath}
\allowdisplaybreaks[4]
\usepackage{amssymb}
\usepackage{graphicx}
\usepackage{booktabs}
\usepackage{enumerate}
\usepackage{bm}
\usepackage{multirow}
\usepackage{psfrag}
\usepackage[normalem]{ulem}
\usepackage{mathtools}
\usepackage{overpic}
\usepackage{slashed}
\usepackage[utf8x]{inputenc}
\usepackage{printlen}
\usepackage{color}
\usepackage{subcaption}

\makeatletter
\def\@fpheader{~}
\makeatother
\definecolor{purple}{rgb}{0.3,0,1}
\newcommand{\spac}{{\hspace{0.3mm}}}
\definecolor{RoyalBlue}{cmyk}{1, 0.50, 0, 0}

\usepackage[dvipsnames]{xcolor}

\title{Low-energy theory of jet processes and PDF factorization}

\author[a]{Thomas Becher,}
\author[b]{Patrick Hager,}
\author[a]{Sebastian Jaskiewicz,}
\author[b,c]{Matthias Neubert,}
\author[a]{and Dominik Schwienbacher}
\affiliation[a]{Albert Einstein Center for Fundamental Physics, Institut f\"ur Theoretische Physik,\\ 
Universit\"at Bern, Sidlerstrasse 5, CH-3012 Bern, Switzerland}
\affiliation[b]{PRISMA$^+$ Cluster of Excellence {\em \&} Mainz Institute for Theoretical Physics,\\ 
Johannes Gutenberg University, 55099 Mainz, Germany}
\affiliation[c]{Department of Physics, LEPP, Cornell University, Ithaca, NY 14853, U.S.A.}

\emailAdd{becher@itp.unibe.ch}
\emailAdd{pahager@uni-mainz.de}
\emailAdd{sebastian.jaskiewicz@unibe.ch}
\emailAdd{matthias.neubert@uni-mainz.de}
\emailAdd{schwienbacher@itp.unibe.ch}

\date{\today}

\preprint{\begin{flushright}
MITP-25-054\\ 
September 8, 2025
\end{flushright}}

\abstract{The consistency of collinear factorization violation with PDF factorization has been an outstanding challenge and subject of considerable debate. In this work we demonstrate their compatibility using a factorization theorem for non-global jet observables. Our analysis relies on consistency relations derived from renormalization conditions in effective field theory. We verify these relations through an explicit computation at three-loop order and show that the double-logarithmic evolution sourcing the super-leading logarithms reduces to single-logarithmic DGLAP running below the lowest perturbative scale. The crucial ingredient reconciling the two evolutions is a perturbative Glauber contribution to the low-energy matrix elements which breaks soft-collinear factorization at the cross section level but restores PDF factorization.}

\usepackage[Q=yes,pverb-linebreak=no]{examplep}

\def\as{\alpha_s}

\def\nb{\bar{n}}

\makeatletter
\DeclareFontFamily{OMX}{MnSymbolE}{}
\DeclareSymbolFont{MnLargeSymbols}{OMX}{MnSymbolE}{m}{n}
\SetSymbolFont{MnLargeSymbols}{bold}{OMX}{MnSymbolE}{b}{n}
\DeclareFontShape{OMX}{MnSymbolE}{m}{n}{
    <-6>  MnSymbolE5
   <6-7>  MnSymbolE6
   <7-8>  MnSymbolE7
   <8-9>  MnSymbolE8
   <9-10> MnSymbolE9
  <10-12> MnSymbolE10
  <12->   MnSymbolE12
}{}
\DeclareFontShape{OMX}{MnSymbolE}{b}{n}{
    <-6>  MnSymbolE-Bold5
   <6-7>  MnSymbolE-Bold6
   <7-8>  MnSymbolE-Bold7
   <8-9>  MnSymbolE-Bold8
   <9-10> MnSymbolE-Bold9
  <10-12> MnSymbolE-Bold10
  <12->   MnSymbolE-Bold12
}{}

\let\llangle\@undefined
\let\rrangle\@undefined
\DeclareMathDelimiter{\llangle}{\mathopen}%
                     {MnLargeSymbols}{'164}{MnLargeSymbols}{'164}
\DeclareMathDelimiter{\rrangle}{\mathclose}%
                     {MnLargeSymbols}{'171}{MnLargeSymbols}{'171}
\makeatother

\usepackage{scalerel,stackengine,amsmath}
\newcommand\equalhat{\mathrel{\stackon[2pt]{=}{\stretchto{%
    \scalerel*[\widthof{=}]{\wedge}{\rule{0.9ex}{2.9ex}}}{0.6ex}}}}

\DeclareMathOperator*{\SumInt}{%
\mathchoice%
  {\ooalign{$\displaystyle\sum$\cr\hidewidth$\displaystyle\int$\hidewidth\cr}}
  {\ooalign{\raisebox{.14\height}{\scalebox{.7}{$\textstyle\sum$}}\cr\hidewidth$\textstyle\int$\hidewidth\cr}}
  {\ooalign{\raisebox{.2\height}{\scalebox{.6}{$\scriptstyle\sum$}}\cr$\scriptstyle\int$\cr}}
  {\ooalign{\raisebox{.2\height}{\scalebox{.6}{$\scriptstyle\sum$}}\cr$\scriptstyle\int$\cr}}
}

\def\epsilon{\varepsilon}

\newcommand{\Hm}{\bm{\mathcal{H}}_{m}}
\newcommand{\Wm}{\bm{\mathcal{W}}_{m}}

\newcommand{\Gammabar}{\overline{\bm{\Gamma}}}
\newcommand{\GammaC}{\bm{\Gamma}^C}
\newcommand{\GammaHO}{\bm{\Gamma}^H_0}
\newcommand{\VG}{\bm{V}^G}

\newcommand{\Hij}{\bm{\mathcal{H}}_{ij\to m}}
\newcommand{\Wij}{\bm{\mathcal{W}}_{ij\to m}^{h_1h_2}}

\newcommand{\Tr}{\mathrm{Tr}}

\DeclareFontFamily{U}{wncy}{}
\DeclareFontShape{U}{wncy}{m}{n}{<->wncyr10}{}
\DeclareSymbolFont{mcy}{U}{wncy}{m}{n}
\DeclareMathSymbol{\Sh}{\mathord}{mcy}{"58} 

\DeclareSymbolFont{dutchcal}{U}{dutchcal}{m}{n}
\DeclareSymbolFontAlphabet{\mathdcal}{dutchcal}
\usepackage[mathscr]{euscript}

\begin{document}
\maketitle
\newpage 

\section{Introduction}
This year marks the 40th anniversary of a seminal paper  \cite{Collins:1985ue} by Collins, Soper and Sterman (CSS) which established the factorization of the inclusive Drell-Yan cross section into partonic cross sections $\hat{\sigma}$ convoluted with parton distribution functions (PDFs). This factorization theorem takes the form
\begin{equation}\label{eq:PDFfactorization}
    \sigma = \sum_{k,l=q,\bar{q},g}  \hat{\sigma}_{kl}(\mu) \ast f_k(\mu)f_l(\mu)\, ,
\end{equation}
where $f_i$ is the distribution of parton $i$ inside the colliding hadron and the asterisk symbol indicates the integrals over the momentum fractions. This theorem is the bedrock on which precision predictions for hadron collider physics rest. It implies that low-energy interactions among the incoming hadrons are suppressed at large scattering energies, and allows to isolate the non-perturbative physics into PDFs, simple single-hadron matrix elements, and the partonic cross sections can be computed order by order in perturbation theory. The factorization theorem involves a factorization scale $\mu$ which separates the perturbative and non-perturbative part of the cross section and the associated scale evolution is driven by the single-logarithmic DGLAP equation~\cite{Gribov:1972ri,Dokshitzer:1977sg,Altarelli:1977zs}. For Deep Inelastic Scattering (DIS) this can be rigorously derived using the operator product expansion~\cite{Gross:1973zrg}. The derivation for the Drell-Yan process is much more involved. It was achieved using diagrammatic methods in~\cite{Collins:1985ue} and the most difficult part of the proof is to show that the interactions of low-energy Glauber gluons between the incoming hadrons cancel out in the cross section~\cite{Bodwin:1981fv}. The necessary arguments are subtle and were further refined in~\cite{Collins:1988ig}. 

The theorem~\eqref{eq:PDFfactorization} was established only for inclusive Drell-Yan production, but is used to compute arbitrary infrared safe cross sections. One may argue that any hard cross section becomes inclusive below the lowest perturbative scale and that the arguments in~\cite{Collins:1985ue,Collins:1988ig} should hold also beyond the Drell-Yan case. However, this intuitive argument has never been formalized, and, over the years, some authors have expressed doubt whether PDF factorization applies to more general observables such as jet cross sections~\cite{Collins:2007nk,Gaunt:2014ska,Zeng:2015iba}. In addition, the discovery of collinear factorization violation for space-like collinear splittings~\cite{Catani:2011st,Forshaw:2012bi,Schwartz:2017nmr,Cieri:2024ytf,Henn:2024qjq,Guan:2024hlf,Duhr:2025lyg} has brought the problem of PDF factorization back into focus. While the breaking of collinear factorization in the amplitudes does not immediately imply invalidation of PDF factorization, it calls for an explanation how the two can be reconciled. 

\begin{figure}
    \centering
    \includegraphics{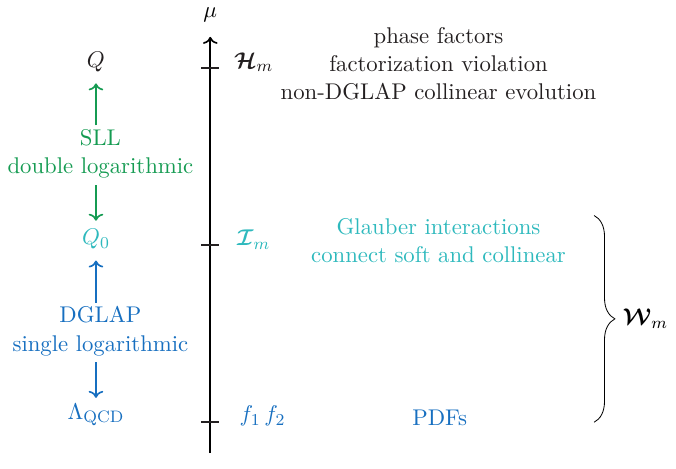}
    \caption{Structure of the factorization formula for gap-between-jets observables.}
    \label{fig:evolution}
\end{figure}

Exclusive jet cross sections are well suited to study this issue in detail since they suffer from Super-Leading Logarithms (SLLs)~\cite{Forshaw:2008cq} which are directly related to factorization breaking effects due to complex Glauber phases.
Over the past years, the all-order structure of the leading SLLs was obtained using Renormalization Group (RG) evolution~\cite{Becher:2021zkk,Becher:2023mtx,Boer:2024hzh}. The evolution is derived from a factorization theorem~\cite{Becher:2016mmh,Balsiger:2018ezi,Becher:2023mtx}
\begin{equation}\label{eq:factorization-theorem-lit}
    \sigma = \sum_{m} \int\!\! d\Pi_{m} \langle\bm{\mathcal{H}}_m(Q,\mu)\,\bm{\mathcal{W}}_m(Q_0,\mu)\rangle
\end{equation}
for the production of jets at energy $Q$ with a veto on radiation outside of the jets at the scale $Q_0$. The hard functions $\bm{\mathcal{H}}_m$ are squared amplitudes with $m$ partons in the final state and $\bm{\mathcal{W}}_m$ are the corresponding low-energy matrix elements. These functions have open color indices which are traced by $\langle \dots \rangle$ after integrating over the phase space of the hard partons.  While for lepton colliders Glauber modes cancel \cite{Forshaw:2012bi,Becher:2021zkk,Becher:2023mtx}, this is no longer true for hadron colliders since the initial-state partons carry color charge. Therefore, beyond leading-order, $\bm{\mathcal{W}}_m$ must in general  be understood  as truly soft-collinear matrix elements, due to possible soft-collinear interactions. The SLLs were resummed by solving the RG equations for the hard functions order-by-order in perturbation theory and reducing the resulting color structures to a basis of operators in color space. Super-leading refers to the fact that, starting at four-loop order, double logarithms arise in the cross section. These emerge from cusp terms in the anomalous dimensions which first contribute at this order. This double logarithmic evolution is in sharp contrast to DGLAP evolution, which is single logarithmic. In order for PDF factorization to hold at small $\mu$, the low-energy matrix elements must factorize as
\begin{equation}
 \bm{\mathcal{W}}_m (Q_0,\mu)= \bm{\mathcal{I}}_{m}(Q_0,\mu) \ast f_1(\mu) \,f_2(\mu)
\end{equation}
and the perturbative physics at scale $Q_0$ must turn the double logarithmic running above $Q_0$ into single logarithmic running below this scale. The form of the factorization formula and the running of the different ingredients is illustrated in Figure~\ref{fig:evolution}.

The RG equations for $\bm{\mathcal{H}}_m$ and $f_i$ imply a set of order-by-order constraints on the renormalization of the perturbative low-energy matrix elements $\bm{\mathcal{I}}_{m}$, which we have verified recently up to three loops in \cite{Becher:2024kmk}. The main finding of this work was the identification of a genuine Glauber contribution to $\bm{\mathcal{I}}_{m}$, well-defined without additional regulators and  not contained within other regions, at the three-loop level. These Glauber terms generate phase terms not present in the soft or collinear matrix elements, i.e.\ the associated loop integration cannot be contour deformed out of the Glauber region. 
The Glauber gluons connect the soft and collinear partons. Without soft-collinear interactions, the collinear part of $\bm{\mathcal{I}}_{m}$ would be scaleless since the only perturbative scale is $Q_0$. This scale does not enter the collinear matrix elements as they are not sensitive to the veto at large angles. However, the collinear contribution is needed to generate a rapidity logarithm via the collinear anomaly. Evaluating the Glauber contribution, we showed in our earlier paper~\cite{Becher:2024kmk} that it indeed cancels the terms from the double logarithmic running.

In the present paper, we provide more details on the relevant computation and extend the analysis in several ways. First of all, we verify that the evolution below $Q_0$ is not just single logarithmic, but indeed given by DGLAP, both for the diagonal and off-diagonal channels. This is quite non-trivial, since the single-logarithmic part of the collinear evolution above $Q_0$ contains non-DGLAP terms, sensitive to the color structure of all partons in the scattering process. The Glauber contribution has the form of a commutator term, which moves the collinear evolution after the Glauber phase where it becomes color-diagonal.
Our computations reveal a concrete mechanism which reconciles collinear factorization violation and PDF factorization. Our results also show that while PDF factorization is restored for $\mu < Q_0$, there are interesting physics effects arising from the factorization violation above~$Q_0$. Indeed, through the SLLs, the collinear factorization violation has an important,  numerically significant effect on the physics of jet observables~\cite{Becher:2024nqc}. 

We also perform a detailed analysis of the subleading poles in the Glauber contribution to $\bm{\mathcal{I}}_{m}$, which reveals the presence of interesting new Lorentz structures for hard-scattering processes involving gluons in the initial state. Due to the interactions with the soft sector, beyond two loops the collinear matrix elements become sensitive to the directions of the soft partons. To properly take this into account, the factorization formula in \cite{Becher:2023mtx} needs to be formulated in terms of objects with open Lorentz indices for the initial-state partons. For single logarithms arising at three-loop order, the projectors inserted into the matrix elements in \cite{Becher:2023mtx} do not capture the full structure present in the low-energy theory. Therefore, in Section~\ref{sec:factorization} we go over a detailed rederivation of the factorization formula and amend it so that it captures the full structure present at higher logarithmic order.

Once the factorization theorem is in place, we perform a series of consistency checks, which we present in detail in Section~\ref{sec:outline}. The first one is to verify that the ultraviolet~(UV) structure of the low-energy matrix elements match the hard anomalous dimensions. To obtain the UV poles of $\bm{\mathcal{W}}_m$ at one-loop order, we evaluate the matrix elements with parton masses as infrared~(IR) regulators. The relevant computations are performed in Section~\ref{sec:1_loop_calc} and fully reproduce the hard anomalous dimensions. This shows that the UV behavior of the low-energy matrix elements is consistent with the hard functions, but cannot address the question whether it factorizes into a perturbative part convoluted with PDFs. To verify this, we then perform a series of computations of the leading divergences in the matrix elements up to three loop order. We first derive all purely soft terms in Section~\ref{sec:soft_matrix_element_three_loop} and then analyze the Glauber contributions in Section~\ref{sec:soft_collinear_matrix_element}. 
We find full consistency with PDF factorization and our formalism based on  Soft-Collinear Effective Theory (SCET)~\cite{Bauer:2001yt,Bauer:2002nz,Beneke:2002ph,Beneke:2002ni} provides a field-theoretical basis for the argument that for $\mu< Q_0$ the cross section becomes inclusive and the arguments of \cite{Collins:1985ue} should apply. In the effective theory, we can integrate out $\bm{\mathcal{I}}_{m}$ and match onto an effective theory which only contains fields with virtualities of order $\Lambda_{\mathrm{QCD}}$. This low-energy theory is identical to the one for the inclusive Drell-Yan process, so the traditional arguments for PDF factorization can then be applied. Further implications of our results are discussed in Section~\ref{sec:discussion}, before we conclude in Section~\ref{sec:conclusion}.

\section{Factorization for jet cross sections}
\label{sec:factorization}

\begin{figure}
\centering
\includegraphics[trim={0 0 0 .35cm}, clip]{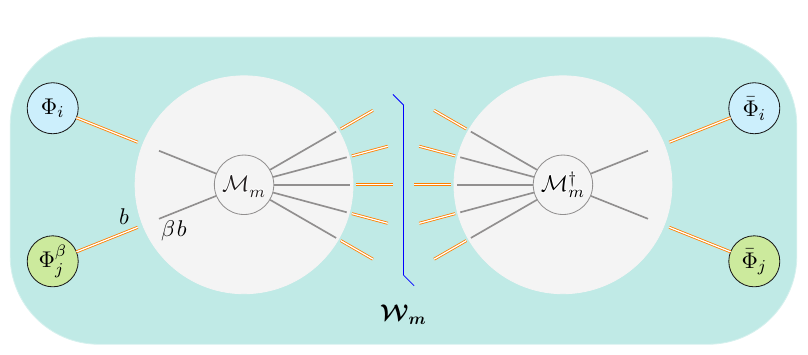}
\caption{Schematic form of the factorization theorem before PDF factorization: the hard part is depicted in gray, while the soft-collinear matrix elements are highlighted in blue. The soft Wilson lines are denoted by orange double lines. The indices of the collinear building blocks and their associated Wilson lines are directly connected to the hard functions. In order to avoid clutter in the figure, we give only one set of indices explicitly.}
\label{fig:factorization_1}
\end{figure}

For gap-between-jet cross sections, the generalization of the previous factorization theorem~\cite{Becher:2016mmh,Balsiger:2018ezi,Becher:2023mtx} is given by 
\begin{align}\label{eq:factorizationTheoremNew}
    \sigma_{2\to M} &= \sum_{m=M}^\infty\int\!\! d\Pi_{m}\int dx_1\, dx_2 \, \llangle[\Hij]\substack{ab|\bar{a}\bar{b}\\\alpha\beta|\bar{\alpha}\bar{\beta}}(\{\underline{p}\},Q,\mu) \nonumber \\
    &\quad[\Wij]\substack{ab|\bar{a}\bar{b}\\\alpha\beta|\bar{\alpha}\bar{\beta}}(\{\underline{n}\},Q_0,x_1,x_2,\mu)\rrangle\,,
\end{align}
consisting of hard functions $\Hij$ and soft-collinear matrix elements $\Wij$, where
the sum over $m$ allows for additional emissions of hard partons into the wide-angle jet~\cite{Becher:2016mmh,Becher:2015hka}.
The two energy scales arising in these processes are denoted by $Q$ for the hard and $Q_0$ for the soft scale, and, despite factorization, the low-energy matrix elements are sensitive to the high energy scale $Q$ via the collinear anomaly~\cite{Becher:2024kmk}. 
As above, $\mu$ denotes the factorization scale, and a sum over the initial states $i,j \in \{q,\bar{q},g\}$ is to be understood. We define $\{\underline{p}\}=\{p_1,p_2,\dots,p_{m+2}\}$ as the partonic momenta and in the factorization theorem \eqref{eq:factorizationTheoremNew} the incoming ones are $p_1 = x_1 P_1 $ and $p_2 = x_2 P_2 $, where $P_1$ and $P_2$ are the momenta of the hadrons. 
Analogously $\{\underline{n}\}=\{n_1,n_2,\dots,n_{m+2}\}$ contains the respective beam directions $n_1, n_2$ and the final state directions $n_3,\dots,n_{m+2}$. The double angle brackets $\llangle\dots\rrangle$ denote sums over the color and polarization indices of all final-state partons, but importantly they {\em do not\/} imply color and polarization averages for the initial-state particles. The phase-space integral over the final-particle states of the hard function is defined as
\begin{align}\label{eq:phase_space}
   \int\!\!d\Pi_{m}\equiv\int\prod_{k=3}^{m+2} \frac{d^d\spac p_k}{(2\pi)^d}2\pi\spac \delta(p_k^2)\theta(p_k^0)=\int\prod_{k=3}^{m+2} 
   \frac{ dE_k}{\tilde{c}^\varepsilon(2\pi)^2}\spac E_k^{d-3}\,\tilde{c}^\varepsilon\frac{d^{d-2}\spac\Omega_k}{2(2\pi)^{d-3}}\,,
\end{align}
where for convenience we inserted a factor of $\tilde{c}=e^{\gamma_E}/\pi$ into the angular integrations \cite{Becher:2021urs}. 
The hard functions $\Hij$ can be thought of as squared partonic amplitudes with open indices, where the final-state particles are defined in color-helicity space~\cite{Catani:1996vz} while the color and Lorentz indices for the initial-state partons $i,j$ are stated explicitly
\begin{align}\label{eq:hard_function_definition}   [\Hij]\substack{ab|\bar{a}\bar{b}\\\alpha\beta|\bar{\alpha}\bar{\beta}}(\{\underline{p}\},Q,\mu)
    &=\Bigl[|\mathcal{M}_{ij\to m}(\{\underline{p}\})\rangle 
    \langle\mathcal{M}_{ij \to m}(\{\underline{p}\})|\Bigr]\substack{ab|\bar{a}\bar{b}\\\alpha\beta|\bar{\alpha}\bar{\beta}} \times {\Delta}(\{\underline{p}\},Q)\,. 
\end{align}
In earlier works and when appearing in figures, $\mathcal{M}_{ij\to m}$ is abbreviated to $\mathcal{M}_{m}$. 
In the above definition, $\Delta$ contains the momentum conservation and hard phase-space constraints 
\begin{equation}\label{eq:ps-constraints}
 {\Delta}(\{\underline{p}\},Q)=    (2\pi)^d\spac \spac \delta(\bar{n}_1\cdot p_\mathrm{tot} - \bar{n}_1 \cdot p_1)\spac\delta(\bar{n}_2\cdot p_\mathrm{tot} - \bar{n}_2\cdot p_2)\spac\delta^{(d-2)}(p_\mathrm{tot}^\perp)\,\mathcal{F}(Q,\{\underline{p}\})\Theta_\mathrm{hard}(\{\underline{n}\})\,.
\end{equation}
Here, $p_\mathrm{tot}$ is the total final-state momentum, with $p_\mathrm{tot}^\perp$ denoting its $(d-2)$ transverse components. The function $\mathcal{F}$ defines the hard scale $Q$ in terms of the partonic momenta, for example through the transverse momenta of jets. 
The angular constraint $\Theta_{\mathrm{hard}}\equiv1-\Theta_{\mathrm{veto}}$ defines the jet region, imposing that the hard partons cannot enter the veto region. A pictorial representation of the factorization theorem \eqref{eq:factorizationTheoremNew} is shown in Figure~\ref{fig:factorization_1}; the inner gray area in the figure corresponds to the hard functions, and the outer part, shaded in blue, indicates the low-energy matrix elements. 

The low-energy matrix elements $\Wij$ describe the soft-collinear interactions between the initial-state particles including both perturbative and non-perturbative low-energy dynamics. They are obtained from the matrix element
\begin{align} \label{eq:W_function_definition}
   &[\widetilde{\bm{\mathcal{W}}}_{ij\to m}^{h_1h_2}]{\substack{ab|\bar{a}\bar{b}\\[-0.68mm]\alpha\beta|\bar{\alpha}\bar{\beta}}} (\{\underline{n}\},Q_0, t_1, t_2,\mu)
   ={} \hspace{0.15cm}\int\limits_{X}\hspace{-0.54cm}\sum\,\, \theta( Q_0 - E^\perp_{\mathrm{\, out}}) \\
   &\times
   \langle  h_1(P_1) h_2(P_2)  |\, \bar{\Phi}_{i}^{\bar{\alpha}\bar{c}}(t_1 \bar{n}_1) \, \bar{\Phi}_{j}^{\bar{\beta}\bar{d}}(t_2 \bar{n}_2) \, [\bm{S}_1^\dagger]^{\bar{c}\spac\bar{a}} \,[\bm{S}_2^\dagger]^{\bar{d}\spac\bar{b}}\, \bm{S}_3^\dagger\, \dots\,  \bm{S}_{m+2}^\dagger\,  |X \rangle \nonumber \\
   &\times \langle X | \,[\bm{S}_1]^{ac}\,[\bm{S}_2]^{bd} \,\bm{S}_3\,  \dots\,  \bm{S}_{m+2}\, \Phi_{i}^{\alpha c}(0) \, \Phi_{j}^{\beta d}(0) \, |h_1(P_1) h_2(P_2) \rangle \,,\nonumber
\end{align}
where $\Phi_i \in \{\chi,\bar{\chi},\mathcal{A}_\perp \} $ 
denote the usual collinear gauge-invariant building blocks for quarks and gluons~\cite{Hill:2002vw}, and $\bm{S}_i$ are the soft Wilson lines along the directions $n_i$ of the hard partons~\cite{Bauer:2001yt}. 
The use of color space formalism ensures that appropriate representations appear for the external partons and the differing $i0$-prescription for incoming (legs $i=1,2$) and outgoing Wilson lines is left implicit~\cite{Becher:2023mtx}. 
The external hadronic states are denoted by $h_i(P_i)$ and the theta function in the top line implements the constraint on soft radiation which is imposed on $E^\perp_{\mathrm{\, out}}$, the total transverse momentum permitted in the gap region. In experimental measurements, the veto is typically imposed on the transverse momentum of the leading jet inside the gap region, rather than the total transverse momentum, to reduce sensitivity to pileup and underlying event \cite{ATLAS:2011yyh}. The logarithmic terms considered in our work do not depend on the precise definition of the gap energy scale. Note that the open color index is carried by the respective soft Wilson line of the partons $i,j$, while the open Lorentz index is the one of the collinear building block $\Phi_i^{\alpha c}$, whose color index is contracted with the respective Wilson line, see  Figure~\ref{fig:factorization_1}.
The final-state Wilson lines are in color space with indices left implicit. The function entering the factorization theorem is the Fourier transform of the above matrix element, defined as
\begin{equation}
  \label{eq:W_function_definition_FT}
    \Wm(\{\underline{n}\},Q_0,x_1,x_2,\mu) = \int_{-\infty}^{\infty} \frac{dt_1}{2\pi}\,e^{- i x_1 t_1 \bar{n}_1\cdot P_1}  \int_{-\infty}^{\infty} \frac{dt_2}{2\pi}\,e^{-i x_2 t_2 \bar{n}_2\cdot P_2} \,\widetilde{\bm{\mathcal{W}}}_m(\{\underline{n}\},Q_0, t_1, t_2,\mu)\,,
\end{equation}
and we abbreviated $\Wij \equiv \Wm$.

\subsection{Low-energy matrix elements at leading order}
\label{sec:factorization-thm-connection}

At lowest order in the perturbative expansion, i.e.\ setting all soft Wilson lines $\bm{S}_i\to\bm{1}$, the matrix elements $\Wm$ reproduce the collinear PDFs with their usual Lorentz and color structure. Explicitly, we have 
\begin{align}
\label{eq:PDFSpinSums}
  \hspace{0.15cm}\int\limits_{X}\hspace{-0.54cm}\sum \langle h(P)|\overline{\chi}^{\bar{a}}_{\bar{\alpha}}(t\bar{n}) |X \rangle \langle X| \chi^{a}_{\alpha}(0)|h(P)\rangle &= \frac{\delta^{a\bar{a}}}{N_c}\int_0^1 \frac{d\xi}{\xi}\:
    \Bigl[\frac{\slashed{n}}{4}\spac
    \xi\spac \bar{n}\cdot P\Bigr]_{\alpha\bar{\alpha}}\,e^{i\xi t\spac \bar{n}\cdot P} f_{q/h}(\xi,\mu)\,,\nonumber\\
  \hspace{0.15cm}\int\limits_{X}\hspace{-0.54cm}\sum  \langle h(P)|\mathcal{A}^{\bar{\alpha},\bar{a}}_{\perp}(t\bar{n}) |X \rangle \langle X| \mathcal{A}^{\alpha,a}_{\perp}(0)|h(P)\rangle &= \frac{\delta^{a\bar{a}}}{N_c^2-1}\int_0^1 \frac{d\xi}{\xi}\: 
    \Bigl[-\frac{g_\perp^{\alpha\bar{\alpha}}}{d-2}\Bigr]\spac
    e^{i\xi t\spac \bar{n}\cdot P} f_{g/h}(\xi,\mu)\,,
\end{align}
where $\alpha,\bar\alpha$ are spinor indices in the first relation and Lorentz indices in the second. In these matrix elements and throughout the paper, we only consider unpolarized scattering, i.e.\ we average over the hadron spins.
These relations can be used in the low-energy theory also in higher orders, as long as there are no factorization-breaking interactions between soft and collinear particles. Such interactions arise from the exchange of Glauber gluons and contribute first at three-loop order. We will come back to them in Section~\ref{sec:soft_collinear_matrix_element}.
The above identities can be conveniently written in terms of generic collinear building blocks as 
\begin{equation}\label{eq:PDFSpinSumGeneric}
\begin{aligned}
   &\int\limits_{X}\hspace{-0.54cm}\sum 
   \langle h_i(P_i)|\bar{\Phi}^{\bar{a}}_{j,\bar{\alpha}}(t\bar{n}_i)|X \rangle \langle X| \Phi^{a}_{j,\alpha}(0)|h_i(P_i)\rangle \\
   &= \frac{\delta^{a\bar{a}}}{\mathcal{N}_j}\int_0^1 \frac{d\xi_i}{\xi_i}\:
    \Bigl[\sum_{s}\bm{u}^{(j)}_{s,\alpha} \bar{\bm{u}}^{(j)}_{s,\bar{\alpha}}\Bigr](\xi_i P_i)\,e^{i\xi_i t\spac \bar{n}_i\cdot P_i} f_{j/h_i}(\xi_i,\mu)\,,
\end{aligned}
\end{equation}
where $i=1,2$ labels the initial-state hadron, and $\bm{u}^{(j)}_{s,\alpha}$ are the polarization vectors/spinors for parton $j$. The factor $\mathcal{N}_j$ accounts for the spin and color average in the explicit results in~\eqref{eq:PDFSpinSums}. It equals $2N_c$ for quarks and anti-quarks, and $(d-2)(N_c^2-1)$ for gluons. Using this result, and neglecting factorization-breaking Glauber interactions, we obtain 
\begin{equation}\label{eq:simpleRelation}
\begin{aligned}
    [\Wij \{\underline{n}\},Q_0,x_1,x_2,\mu)]{\substack{ab|\bar{a}\bar{b}\\[-0.68mm]\alpha\beta|\bar{\alpha}\bar{\beta}}} 
   &= \frac{f_{i/h_1}(x_1,\mu)\,f_{j/h_2}(x_2,\mu)}{x_1 x_2\spac s}\,
    [{\bm{\mathcal{S}}}_{ij\to m}]{\substack{ab|\bar{a}\bar{b}\\ \phantom{X}}}(\{\underline{n}\},Q_0,\mu) \\
   &\quad\times
    \frac{1}{\mathcal{N}_1}\Bigl[ \sum_{s}\bm{u}^{(1)}_{s,\alpha} \bar{\bm{u}}^{(1)}_{s,\bar{\alpha}} \Bigr](x_1 P_1)\,
     \frac{1}{\mathcal{N}_2}\Bigl[\sum_{s}\bm{u}^{(2)}_{s,\beta} \bar{\bm{u}}^{(2)}_{s,\bar{\beta}} \Bigr](x_2 P_2) \,,
\end{aligned}
\end{equation}
where
\begin{align} 
   [{\bm{\mathcal{S}}}_{ij\to m}]{\substack{ab|\bar{a}\bar{b}\\ \phantom{X}}} &(\{\underline{n}\},Q_0,\mu)
   ={} \hspace{0.15cm}\int\limits_{X_s}\hspace{-0.54cm}\sum\,\theta( Q_0 - E^\perp_{\mathrm{\, out}}) \\
   &\times
   \langle 0 |\,[\bm{S}_1^\dagger]^{c\spac\bar{a}} \,[\bm{S}_2^\dagger]^{d\spac\bar{b}}\, \bm{S}_3^\dagger\, \dots\,  \bm{S}_{m+2}^\dagger\,  |X_s \rangle  \langle X_s | \,[\bm{S}_1]^{ac}\,[\bm{S}_2]^{bd} \,\bm{S}_3\,  \dots\,  \bm{S}_{m+2}\, | 0 \rangle \,,\nonumber
\end{align}
and we have used that $\bar n_1\cdot P_1\,\bar n_2\cdot P_2=s$.

Pulling these factors out from the low-energy matrix elements, contracting their indices with the indices of the hard functions, moving the flux factor $1/(x_1 x_2 s)$ into the hard functions, and integrating over the energies over the final-state particles (keeping their directions fixed), one obtains from \eqref{eq:factorizationTheoremNew}
\begin{align}
    \sigma_{2\to M} &= \sum_{m=M}^\infty \int dx_1\, dx_2 \,
     f_{i/h_1}(x_1,\mu)\,f_{j/h_2}(x_2,\mu) \notag\\
    &\quad\times \langle\spac\overline{\bm{\mathcal{H}}}_{ij\to m}(\{\underline{p}\},Q,\mu)\otimes
    {\bm{\mathcal{S}}}_{ij\to m}(\{\underline{n}\},Q_0,\mu)\rangle \,, 
\end{align}
where the single brackets imply a sum (average) over final-state (initial-state) color and polarization indices, and the symbol $\otimes$ indicates the angular integrals over the directions of the final-state particles of the hard function. The open color indices on the hard and soft functions live in color space and are not written out explicitly in this result. The new hard functions are defined as in \cite{Becher:2023mtx},\footnote{These hard functions were denoted as $\bm{\mathcal{H}}_m$ in~\cite{Becher:2023mtx}. We use the overline to distinguish them from the un-integrated ones defined in~\eqref{eq:hard_function_definition}.} i.e.\
\begin{align} \label{eq:hard_function_spin_averaged}
     \overline{\bm{\mathcal{H}}}_{ij\to m}
    = \frac{1}{2x_1 x_2\spac s} \prod_{i=3}^m \int\!\frac{dE_i\,E_i^{d-3}}{{\tilde{c}}^\epsilon \,(2\pi)^{2}}\,
    |\mathcal{M}_m(\{\underline{p}\})\rangle \langle\mathcal{M}_m(\{\underline{p}\})| \times 2\spac {\Delta}(\{\underline{p}\},Q) \,,
\end{align} 
and include the flux factor for the partonic cross section. 
The integration over the energies can always be carried out for the fixed angular constraints considered below. However, the general form with open integrations is useful for analyzing clustering effects~\cite{Becher:2023znt} or massive quarks~\cite{Balsiger:2020ogy,inprep}. 

This successfully recovers the factorization theorem of~\cite{Becher:2016mmh,Balsiger:2018ezi,Becher:2023mtx}, valid up to two-loop order (three-loop for pure QCD processes). 
Beyond this order, non-trivial Lorentz structures arise from Glauber-gluon exchanges, which break soft-collinear factorization and invalidate the simple relations in \eqref{eq:simpleRelation}, see also Section~\ref{sec:soft_collinear_matrix_element}. We now discuss the proper treatment of the low-energy matrix elements in the presence of such higher-order effects.

\subsection[Refactorization of \texorpdfstring{$\mathcal{W}_m$}{Wm}]{Refactorization of \texorpdfstring{$\Wm$}{Wm}}

While the low-energy matrix elements $\Wm$ describe both perturbative and non-per\-tur\-ba\-tive low-energy dynamics, in the following we verify explicitly up to the three-loop order that the perturbative soft-collinear physics associated with energies above the scale $Q_0$ factorizes from the non-perturbative physics as described by standard PDFs below that scale. This investigation builds upon~\cite{Becher:2024kmk}, where we  explicitly demonstrated that the double-logarithmic running turns single logarithmic at the scale $Q_0$. 
This suggests that one can perform an additional matching step onto PDF-collinear fields, with virtualities $k^2\sim\Lambda_\mathrm{QCD}^2$~\cite{Bauer:2002nz}, thereby allowing to factorize the PDFs from the low-energy matrix elements.

This can be achieved as follows: first, we match $\Wm$ onto PDF-collinear gauge-invariant building blocks $\bm{\Phi}_k^{\mu A}$ as
\begin{align}
    &[\Wij]\substack{\alpha\beta|\bar\alpha\bar\beta\\ab|\bar{a}\bar{b}}(\{\underline{n}\},Q_0,x_1,x_2,\mu)\nonumber\\&=
    \int d\tilde{t}_1 d\tilde{t}_2
    [\bm{\mathcal{I}}_{ij\to m}^{kl}]\substack{\alpha\beta,\mu\nu|\bar\alpha\bar\beta,\bar\mu\bar\nu\\ab,AB|\bar{a}\bar{b},\bar{A}\bar{B}}(\{\underline{n}\},Q_0,x_1,x_2,\tilde{t}_1,\tilde{t}_2,\mu)
    \,\langle h_1(P_1)|\bar{\bm{\Phi}}^{\bar{\mu}\bar{A}}_k(\tilde{t}_1\nb_1)
    \bm{\Phi}^{\mu A}_k(0)|h_1(P_1)\rangle\nonumber\\
    &\quad\times
    \langle h_2(P_2)|\bar{\bm{\Phi}}^{\bar{\nu}\bar{B}}_l(\tilde{t}_2\nb_2)
    \bm{\Phi}^{\nu B}_l(0)|h_2(P_2)\rangle\,,
\end{align}
where $\bm{\mathcal{I}}$ is the matching coefficient
and contains purely-perturbative partonic physics at scale $Q_0$ for partons $k,l \to i,j$, including perturbative Glauber interactions. Additionally, a sum over initial states $k,l\in \{q,\bar{q},g\}$ is to be understood. The PDF-collinear matrix elements can be evaluated as in~\eqref{eq:PDFSpinSumGeneric} and yield the respective spin sums and PDFs for partons $k,l$
\begin{align}
    &[\Wij]\substack{\alpha\beta|\bar\alpha\bar\beta\\ab|\bar{a}\bar{b}}(\{\underline{n}\},Q_0,x_1,x_2,\mu) \nonumber\\&=
    \int d\tilde{t}_1 d\tilde{t}_2
    \int\frac{d\xi_1}{\xi_1}\frac{d\xi_2}{\xi_2}
    [\widetilde{\bm{\mathcal{I}}}_{ij\to m}^{kl}]\substack{\alpha\beta,\mu\nu|\bar\alpha\bar\beta,\bar\mu\bar\nu\\ab,AB|\bar{a}\bar{b},\bar{A}\bar{B}}(\{\underline{n}\},Q_0,x_1,x_2,\tilde{t}_1,\tilde{t}_2,\mu)e^{i\xi_1 \tilde{t}_1\spac \bar{n}_1\cdot P_1}e^{i\xi_2 \tilde{t}_2\spac \bar{n}_2\cdot P_2}\nonumber\\
    &\quad\times
    \frac{\delta^{A\bar{A}}}{\mathcal{N}_k}\Bigl[\sum_{s}\bm{u}^{(k)}_{s,\mu} \bar{\bm{u}}^{(k)}_{s,\bar{\mu}}\Bigr](\xi_1\spac P_1)\, 
    \frac{\delta^{B\bar{B}}}{\mathcal{N}_l}\Bigl[ \sum_{s}\bm{u}^{(l)}_{s,\nu} \bar{\bm{u}}^{(l)}_{s,\bar{\nu}}\Bigr](\xi_2\spac P_2)\,f_{k/h_1}(\xi_1,\mu) f_{l/h_2}(\xi_2,\mu)\,.
\end{align}
Next, one notices that the spin sums are appropriately contracted with the initial-state particles $k,l$ of $\widetilde{\bm{\mathcal{I}}}(x_1,x_2,\tilde{t}_1,\tilde{t}_2)$, thereby generating the correct spin and color averages.
In addition, the $\tilde{t}_1$ and $\tilde{t}_2$ integrals can be performed to obtain the (purely) momentum-space matching coefficient ${\bm{\mathcal{I}}}(x_1,x_2,\xi_1,\xi_2)$, resulting in
\begin{align}
    &[\Wij]\substack{\alpha\beta|\bar\alpha\bar\beta\\ab|\bar{a}\bar{b}}(\{\underline{n}\},Q_0,x_1,x_2,\mu)\nonumber\\&=
    \int\frac{d\xi_1}{\xi_1}\frac{d\xi_2}{\xi_2}\:
    [{\bm{\mathcal{I}}}_{ij\to m}^{kl}]\substack{\alpha\beta|\bar\alpha\bar\beta\\ab|\bar{a}\bar{b}}(\{\underline{n}\},Q_0,x_1,x_2,\xi_1P_1,\xi_2P_2,\mu)\, f_{k/h_1}(\xi_1,\mu) f_{l/h_2}(\xi_2,\mu)\,.\label{eq:Wij_I_Definition}
    \end{align}
Here, it is understood that $\bm{\mathcal{I}}$ is averaged over the initial-state $k,l$ spins and colors due to the PDFs. Comparing with $\Wm$ in~\eqref{eq:W_function_definition_FT} yields 
\begin{align} \label{eq:I_function_definition}
   [\bm{\mathcal{I}}_{ij\to m}^{kl}]{\substack{ab|\bar{a}\bar{b}\\[-0.68mm]\alpha\beta|\bar{\alpha}\bar{\beta}}}&(\{\underline{n}\},Q_0,x_1,x_2,q_1,q_2,\mu)\nonumber\\
   &=
   \int_{-\infty}^{\infty} \frac{dt_1}{2\pi}\,e^{- i x_1 t_1 \bar{n}_1\cdot P_1}  \int_{-\infty}^{\infty} \frac{dt_2}{2\pi}\,e^{-i x_2 t_2 \bar{n}_2\cdot P_2} 
   {} \hspace{0.15cm}\int\limits_{X}\hspace{-0.54cm}\sum\,\, \theta( Q_0 - E^\perp_{\mathrm{\, out}}) \nonumber \\
   &\quad\times
   \langle  k(q_1) l(q_2)  |\, \bar{\Phi}_{i}^{\bar{\alpha}\bar{c}}(t_1 \bar{n}_1) \, \bar{\Phi}_{j}^{\bar{\beta}\bar{d}}(t_2 \bar{n}_2) \, [\bm{S}_1^\dagger]^{\bar{c}\spac\bar{a}} \,[\bm{S}_2^\dagger]^{\bar{d}\spac\bar{b}}\, \bm{S}_3^\dagger\, \dots\,  \bm{S}_{m+2}^\dagger\,  |X \rangle \nonumber \\
   &\quad\times \langle X | \,[\bm{S}_1]^{ac}\,[\bm{S}_2]^{bd} \,\bm{S}_3\,  \dots\,  \bm{S}_{m+2}\, \Phi_{i}^{\alpha c}(0) \, \Phi_{j}^{\beta d}(0) \, |k(q_1) l(q_2) \rangle \,,
\end{align}
which is spin- and color-averaged over the initial-state partons $k,l$.
In the final step one adds back the hard functions together with the $x_1,x_2$ integrals,
\begin{align}
    \sigma_{2\to M} &= \sum_{m=M}^\infty \int d\Pi_m
    \int\frac{d\xi_1}{\xi_1}\frac{d\xi_2}{\xi_2}
    \int dx_1 dx_2 \:\Bigl\llangle[\Hij]\substack{ab|\bar{a}\bar{b}\\\alpha\beta|\bar{\alpha}\bar{\beta}}(\{\underline{n}\},Q,x_1,x_2,\mu) \nonumber \\
    &\quad\times[\bm{\mathcal{I}}_{ij\to m}^{kl}]\substack{ab|\bar{a}\bar{b}\\\alpha\beta|\bar{\alpha}\bar{\beta}}(\{\underline{n}\},Q_0,x_1,x_2,\xi_1P_1,\xi_2P_2,\mu)\Bigr\rrangle\:
    f_{k/h_1}(\xi_1,\mu) f_{l/h_2}(\xi_2,\mu)
    \,,
\end{align}
where we left the sum over $i,j,k,l \in \{q,\bar{q},g\}$ implicit.
By rescaling $x_i\to z_i\spac \xi_i$, one obtains a more intuitive understanding of the momentum fraction flow.
This results in
\begin{align}
\label{eq:sigma_before_mellin}
    \sigma_{2\to M} &= \sum_{m=M}^\infty \int d\Pi_m
    \int d\xi_1 d\xi_2
    \int dz_1 dz_2 \: \Bigl\llangle[\Hij]\substack{ab|\bar{a}\bar{b}\\\alpha\beta|\bar{\alpha}\bar{\beta}}(\{\underline{n}\},Q,z_1\xi_1,z_2\xi_2,\mu) \nonumber \\
    &\quad\times[\bm{\mathcal{I}}_{ij\to m}^{kl}]\substack{ab|\bar{a}\bar{b}\\\alpha\beta|\bar{\alpha}\bar{\beta}}(\{\underline{n}\},Q_0,z_1,z_2,\xi_1P_1,\xi_2P_2,\mu)\Bigr\rrangle\:
    f_{k/h_1}(\xi_1,\mu) f_{l/h_2}(\xi_2,\mu)
    \,.
\end{align}

\begin{figure}
\centering
\includegraphics[]{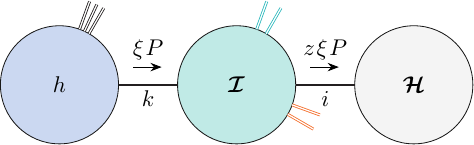}
\caption{The momentum flow through the factorized cross section after PDF factorization. From the hadron $h$ a parton $k$ with momentum $\xi\spac P$ is singled out, where the hadron remnant is depicted by three black lines. The parton $k$ then enters $\bm{\mathcal{I}}$ where generally additional collinear  (depicted in blue) as well as soft emissions (depicted in orange) take place. 
Parton $i$ with momentum fraction $z\spac \xi $ then enters the hard functions $\bm{\mathcal{H}}$.}
\label{fig:momentum:flow_mellin}
\end{figure}

The momentum flow through the different blocks of the factorization theorem is depicted in Figure~\ref{fig:momentum:flow_mellin}. Starting from the hadronic state $h$ with momentum $P$, we single out a parton $k$ with momentum fraction $\xi$. This parton enters the low-energy element $\bm{\mathcal{I}}$, where additional collinear splittings can occur. Consequently, the parton $i$, which is in general different from parton $k$, serves as the initial state of the hard functions with momentum $z\spac \xi \spac P$. While the corresponding momentum flow is related to a Mellin convolution, the low-energy matrix elements $\bm{\mathcal{I}}$ depend on the momentum $P$ through the collinear anomaly~\cite{Chiu:2007yn,Chiu:2007dg,Becher:2010tm,Chiu:2011qc,Chiu:2012ir}.
Keeping track of this additional dependence requires a generalization given by
\begin{equation}
    (r\ast s_n\ast\dots\ast s_1) (P)\equiv\!\! \int d\xi_n d\xi_{n-1}\dots d\xi_1\:
    r (\xi^{(n)}\spac P)\spac s_n(\xi_n,{\xi^{(n-1)}}\spac P)\dots s_1(\xi_1,P)\, ,
    \label{eq:ConvolutionDefinition}
\end{equation}
where $\xi^{(n)}\equiv\xi_1\xi_2\dots\xi_n$. The factorization theorem~\eqref{eq:sigma_before_mellin} can then be conveniently written as
\begin{align}
\label{eq:factorization_final}
    \sigma_{2\to M} &= \sum_{m=M}^\infty \int d\Pi_m \,\Bigl\llangle\Bigl([\Hij]\substack{ab|\bar{a}\bar{b}\\\alpha\beta|\bar{\alpha}\bar{\beta}}
    \ast[\bm{\mathcal{I}}_{ij\to m}^{kl}]\substack{ab|\bar{a}\bar{b}\\\alpha\beta|\bar{\alpha}\bar{\beta}}\ast
    f_{k/h_1} f_{l/h_2}\Bigr)(P_1,P_2)
    \Bigr\rrangle\,,
\end{align}
where the explicit form of the convolution reads
\begin{align}
\label{eq:convolution_new}
   & \Bigl(\Hij
    \ast\bm{\mathcal{I}}_{ij\to m}^{kl}\ast
    f_{k/h_1} f_{l/h_2}\Bigr)(P_1,P_2) \\
    &=\int d\xi_1 d\xi_2
    \int dz_1 dz_2\: \Hij(P_1\spac\xi_1 z_1,P_2\spac \xi_2z_2)\,\bm{\mathcal{I}}_{ij\to m}^{kl}(z_1,z_2,P_1 \spac \xi_1,P_2\spac \xi_2)f_{k/h_1}(\xi_1) f_{l/h_2}(\xi_2)\, \nonumber 
\end{align}
for two incoming momenta $P_1, P_2$. Here, we suppressed all color and Lorentz indices and highlight the momentum flow through the different parts of the factorization theorem. 
As the collinear anomaly which requires the use of the generalized Mellin convolution first appears at three-loop order,
one could still achieve momentum independent soft-collinear matrix elements up to and including two-loop order using the projectors (2.8) in~\cite{Becher:2023mtx} and use the convolution (2.12) therein.

\subsection{Renormalization-group equation and running of the hard functions}

An essential ingredient for the consistency check on PDF factorization is the RG equation of the hard functions,  
which reads~\cite{Becher:2021zkk}
\begin{equation}\label{eq:hardRG}
    \frac{d}{d\ln\mu}\,\bm{\mathcal{H}}_m(\{\underline{p}\},\mu)
    = - \sum_{l=M}^{m} \bigl(\bm{\mathcal{H}}_l(\mu)  \ast \bm{\Gamma}^H_{lm}(\mu)\bigr)(\{\underline{p}\}) \,,
\end{equation}
where the initial state momenta $p_1,p_2$ are convoluted according to~\eqref{eq:ConvolutionDefinition}. Here and below we display only the relevant arguments of the hard functions and anomalous dimensions.
Following the above discussion, there should be two pairs of color and Lorentz indices for the hard functions $\bm{\mathcal{H}}_m$, and correspondingly four pairs of indices for the hard anomalous dimensions $\bm{\Gamma}^H$, which we suppress.
This is sufficient for our analysis, as we are only interested in the leading-logarithmic effect. Therefore, we only consider the one-loop expression for the anomalous dimensions, which can be derived from the soft and collinear limits of the hard functions $\bm{\mathcal{H}}_m$~\cite{Becher:2009qa,Becher:2023mtx}. Indeed, the anomalous dimensions have a diagonal Lorentz structure up to two-loop order. 
Consequently, the open indices carry no relevant information and are not made explicit for ease of notation.
The anomalous dimensions also carry color indices, which are suppressed by working in the color-space formalism for final states and diagonal initial-state splittings, but use explicit color generators for off-diagonal initial-state splittings. The color-space formalism is designed to describe the emission of additional gluons, but not well-suited for the emission of additional (anti-)quarks.

\begin{figure}
\centering
\includegraphics[]{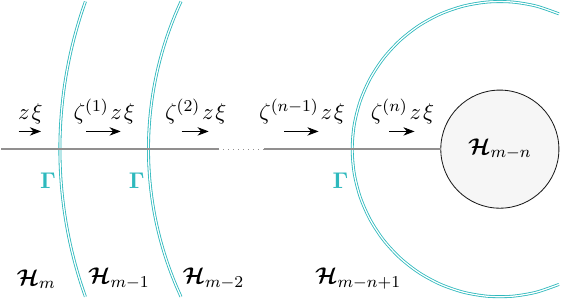}
\caption{Depiction of the momentum flow through the RG evolution of the hard functions from $\bm{\Gamma}^H$. A parton carrying momentum fraction $\xi\spac z $ of the hadron momentum enters the hard functions $\Hm$. Inside the hard functions, each emission driven by $\bm{\Gamma}^H$ changes the momentum fraction entering the subsequent $\bm{\mathcal{H}}_{m-1}$ accordingly. In the end, the parton carries momentum $\xi\spac z \spac\zeta^{(n)}$ with $\zeta^{(n)}\equiv\zeta_1\zeta_2\dots\zeta_n$. }
\label{fig:onion}
\end{figure}

When considering only SLLs, additional collinear splittings can be neglected and the standard form of (2.11) in~\cite{Becher:2023mtx} is recovered.
The solution of the RG equation can then  formally be written in terms of the path-ordered exponential
\begin{equation} \label{eq:U}
   \bm{U}_{lm}(\{\underline{n}\},s,\mu_h,\mu_s) = \mathrm{\textbf{P}} \exp\left[\, \int_{\mu_s}^{\mu_h} \frac{d\mu}{\mu}\, \bm{\Gamma}^H(\{\underline{n}\},s,\mu) \right]_{lm},
\end{equation}
with its action on the hard functions defined as 
\begin{align} \label{eq:Uexp}
   \bm{\mathcal{H}}_l(\mu_h) \ast \bm{U}_{lm}(\mu_h,\mu_s) 
   &= \bm{\mathcal{H}}_m(\mu_h) + \int_{\mu_s}^{\mu_h}\!\frac{d\mu_1}{\mu_1}\,
    \bm{\mathcal{H}}_l(\mu_h) \ast \bm{\Gamma}^H_{lm}(\mu_1) \nonumber\\
   &\quad + \int_{\mu_s}^{\mu_h}\!\frac{d\mu_1}{\mu_1}\,
    \int_{\mu_1}^{\mu_h}\!\frac{d\mu_2}{\mu_2}\,
    \bm{\mathcal{H}}_l(\mu_h) \ast \bm{\Gamma}^H_{ln}(\mu_2) \ast \bm{\Gamma}^H_{nm}(\mu_1)
    + \dots \,.
\end{align}
Here, the anomalous-dimension matrices on the right-hand side are ordered in the direction of decreasing scale values (i.e.\ $\mu_2>\mu_1$ in the second line). A pictorial presentation showcasing the momentum flow through the RG evolution is given in Figure~\ref{fig:onion}.

While~\eqref{eq:hardRG} and~\eqref{eq:Uexp} are valid for arbitrary hard functions, the precise form of the anomalous dimensions depends on whether or not the flux factors, spin-polarization, and energy integration are included in the definition of the hard functions. For the discussion of (super-)leading logarithmic effects, it is convenient to use energy-integrated hard functions $\overline{\bm{\mathcal{H}}}_m$ introduced in~\eqref{eq:hard_function_spin_averaged}, together with the usual form of the anomalous dimensions~\cite{Becher:2009qa}. 
As we show in this article, beyond the two-loop order the hard functions are no longer spin-averaged by the low-energy matrix elements, as these contain additional Lorentz structures sensitive to final state directions. Consequently, the three-loop anomalous dimensions must account for these structures and can no longer be diagonal. Therefore, starting at this order, the spin-averaged hard functions are no longer  sufficient to capture the full complexity of the observable.

\begin{figure}
    \centering
    \includegraphics[scale=1]{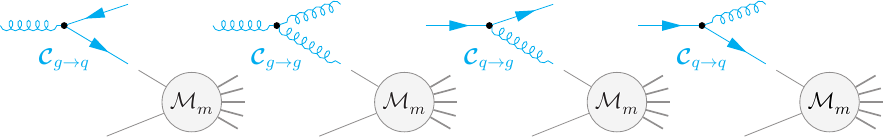}
    \caption{Depiction of the different types of collinear splittings. Additional diagrams obtained from exchanging $q\leftrightarrow\bar{q}$ are not shown. 
    The splittings with emitted gluons are diagonal, while the others are off-diagonal. 
    The SLLs are driven by diagonal splittings.}
    \label{fig:collinear_splittings}
\end{figure}

The anomalous dimensions of the hard functions $\bm{\Gamma}^H$ can be expanded as
\begin{equation}
    \bm{\Gamma}^H = \frac{\alpha_s}{4\pi}\sum_{n=0} \Bigl(\frac{\alpha_s}{4\pi}\Bigr)^n\bm{\Gamma}^H_n\,,
\end{equation}
where the one-loop coefficient reads 
\begin{align}
\label{eq:HardAnomalousDimensionSoftCollinear}
   \GammaHO(z_1,z_2)
    &= \delta(1-z_1)\delta(1-z_2)\bm{\Gamma}^S + \GammaC(z_1,z_2)\,.
\end{align}
The soft and collinear anomalous dimensions are given by 
\begin{equation}
\begin{aligned}
   \bm{\Gamma}^S &= \gamma^{\mathrm{cusp}}_0 \Bigl( \bm{\Gamma}^c
    \ln\frac{\mu^2}{\mu_h^2} +\bm{V}^G \Bigr)
    + \overline{\bm{\Gamma}}\,,\\
    \GammaC(z_1,z_2) &=\bm{\Gamma}^C_1(z_1) \delta(1-z_2) + \delta(1-z_1)\bm{\Gamma}^C_2(z_2)  \,,
    \label{eq:GammaH}
    \end{aligned}
\end{equation}
where we expanded the cusp anomalous dimension $\gamma_{\mathrm{cusp}}(\alpha_s)=\gamma_0^{\mathrm{cusp}}\alpha_s/4\pi+\ldots=\frac{\alpha_s}{\pi}+\dots$,
and the individual pieces are given by
\begin{align} \label{eq:VGGammac}
    \bm{\Gamma}^c
    &= \sum_{i=1,2}\,\big[ C_i\,\bm{1} - \bm{T}_{i,L}\circ\bm{T}_{i,R}\,\delta(n_{l}-n_i) \big] \,,\nonumber\\
    \bm{V}^G
    &= - 2i\pi\,\big( \bm{T}_{1,L}\cdot\bm{T}_{2,L} - \bm{T}_{1,R}\cdot\bm{T}_{2,R} \big)  \,, \nonumber  \\[2mm]  
    \overline{\bm{\Gamma}}
    &= 2 \sum_{i\neq j} \left( \bm{T}_{i,L}\cdot\bm{T}_{j,L} + \bm{T}_{i,R}\cdot\bm{T}_{j,R} \right) 
    \int\frac{d\Omega(n_l)}{4\pi}\,\overline{W}_{ij}^l -4 \sum_{i\neq j} \bm{T}_{i,L}\circ\bm{T}_{j,R}\,
    \overline{W}_{ij}^{l}\left[1-\Theta_{\mathrm{veto}}(n_{l})\right],\nonumber\\[2mm]
    \bm{\Gamma}^C_i &= -2\gamma_0^i\,\delta(1-z_i)\,\bm{1}
    +4
    \spac \overline{\mathcal{P}}_{i \to P}(z_i)\, \bm{\mathcal{C}}^{\phantom{\dagger}}_{i\to P,L}\, \bm{\mathcal{C}}^\dagger_{i\to P,R} \spac \delta(n_k-n_i)\nonumber \\
    &\quad \hspace{0.05cm}+2\gamma_{0}^{\mathrm{cusp}}\delta_{P\spac i}\ln\frac{\mu_h}{2E_i}(C_i\bm{1}\spac\delta(1-z_i)-\bm{T}_{i,L}\circ\bm{T}_{i,R}\spac\delta(n_l-n_i))\,,
\end{align}
where, here and in the following, we suppress the parton-multiplicity indices.
%\comment{delete: Wherever no explicit $z$-dependence is given, an implicit $\delta$-distribution is understood.}
These objects can be understood as follows:
\begin{enumerate}[(i)]
\item $\bm{\Gamma}^c$ is the cusp anomalous dimension. This terms includes a logarithm of the hard scale and generates large double logarithmic contribution under scale evolution. It originates from the soft+collinear limit of the hard functions. To separate it from purely collinear physics, we introduced the arbitrary scale $\mu_h\sim E_i$ with $E_i$ the energy of parton $i$.
Here, $C_i$ denotes the eigenvalue of the quadratic Casimir operator of the respective parton, and the labels $L$ and $R$ of the color generators indicate whether they act from the left or the right on the hard functions.
The notation $\bm{T}_i \circ\bm{T}_{j}\equiv  \bm{T}_i^b\spac \bm{T}_{j}^{b^\prime}$ extends the color-space formalism to accommodate the additional gluon emitted in the $n_l$ direction. The color indices $b$ and $b^\prime$ cannot be contracted immediately as subsequent emissions may attach to them. 
\item Both $\bm{V}^G$ and $\overline{\bm{\Gamma}}$ emerge from the soft limit, where  all imaginary phases from virtual corrections are allocated to the Glauber phase $\bm{V}^G$, while the remaining soft real and virtual contributions are assigned to $\overline{\bm{\Gamma}}$.
Using color conservation, all contributing phase factors can be reshuffled to the initial-state partons, resulting in a simple color structure of the $\bm{V}^G$ term.
The first term of $\overline{\bm{\Gamma}}$ describes the virtual correction and $\overline{W}_{ij}^l$ corresponds to the soft dipole
\begin{align}
\label{eq:def_soft_dipole}
    W_{ij}^l = \frac{n_i\cdot n_j}{n_i\cdot n_l\,n_j\cdot n_l} \,
\end{align}
with collinear singularities subtracted
\begin{equation}
    \overline{W}_{ij}^l = W_{ij}^l - \frac{1}{n_i\cdot n_l}\delta(n_i-n_l) - \frac{1}{n_j\cdot n_l}\delta(n_j-n_l)\,.
\end{equation}
In addition, the second term of $\overline{\bm{\Gamma}}$ in \eqref{eq:VGGammac}, which corresponds to real-emission pieces, contains the veto constraint $1-\Theta_{\mathrm{veto}}=\Theta_{\mathrm{hard}}$.
\item $\GammaC_i$ stems from the purely-collinear limit and contains the splitting functions, which carry information about the DGLAP evolution.
Here, $\gamma^i_0 =\gamma^q=-3C_i$ in the case of quarks and $\gamma_{0}^i=\gamma^g=-\beta_0=-11/3N_c+4/3T_Fn_f$ for gluons.
The quantities $\overline{\mathcal{P}}_{i\to P}$ are defined as the real emission parts of the splitting amplitudes $\mathcal{P}_{i\to P}$~\cite{Catani:1999ss} from which the soft singularities have been subtracted
\begin{align}\label{eq:splitting_functions_final_explicit}
    \overline{\mathcal{P}}_{q \to q}(z) &= \overline{\mathcal{P}}_{\bar{q}\to\bar{q}}(z)  =   (1+z^2)\left[\frac{1}{1-z}\right]_+\ ,\nonumber \\
    \overline{\mathcal{P}}_{q \to g}(z)  &= \overline{\mathcal{P}}_{\bar{q} \to g}(z) = \frac{1+(1-z)^2}{z}\,, \nonumber\\[1mm]
    \overline{\mathcal{P}}_{g\to q}(z) & = \overline{\mathcal{P}}_{g \to\bar{q}}(z)  =  1 - 2 z(1-z) , \nonumber\\
    \overline{\mathcal{P}}_{g \to g}(z)  &=2z\left[\frac{1}{1-z}\right]_++ \frac{2(1-z)}{z} + 2z(1-z).
\end{align}
In these expressions $1-z$ is the momentum fraction carried by the emitted parton. The splitting functions $\overline{\mathcal{P}}_{a \to b}(z)$ are the only terms in the anomalous dimensions with non-trivial dependence on the momentum fraction $z_1$ and $z_2$ of the two incoming partons, all other terms involve only $\delta$-distributions.

The color operators ${\bm{\mathcal{C}}}_{i\to P,L(R)}$ in~\eqref{eq:VGGammac} are connected to the (off-)diagonal splittings, see Figure~\ref{fig:collinear_splittings}.
For diagonal splittings one can employ the color-space formalism, where they are given by\footnote{In~\cite{Becher:2023mtx}, these operators were normalized differently. Namely, the splitting functions included color factors which we do not include in \eqref{eq:splitting_functions_final_explicit}.}
\begin{equation}
     {\bm{\mathcal{C}}}_{i\to P,L}\, {\bm{\mathcal{C}}}^\dagger_{i\to P,R} =\bm{T}_{P,L}\circ\bm{T}_{P,R}\,.
\end{equation}
For off-diagonal splittings, however, the color-space formalism is no longer convenient, and we use explicit definitions.
For $q\to g$ (and analogously for $\bar{q}\to g $), we define
\begin{equation}
    {\bm{\mathcal{C}}}_{q\to g} \,{\bm{\mathcal{C}}}^\dagger_{q\to g} = t_{ij}^a \,t_{\bar{j}\spac \bar{i}}^{\bar{a}}\,,
\end{equation}
where the indices $a,\bar{a}$ are contracted with the lower-multiplicity hard functions, while the indices $i$ and $j$ are the new color indices for the outgoing and incoming quark present in the new hard functions after acting with the operator. We suppress the quark indices in the following and write 
\begin{equation}
    {\bm{\mathcal{C}}}_{q\to g} \,{\bm{\mathcal{C}}}^\dagger_{q\to g} \equiv t_{L}^a \,t_{R}^{\bar{a}}\,.
\end{equation}
The explicit color flow can easily be reconstructed from the diagrams shown in Figure~\ref{fig:collinear_splittings}. For $g\to q$ (and $g\to \bar{q}$ accordingly) we get
\begin{equation}
     {\bm{\mathcal{C}}}_{g\to q}\, {\bm{\mathcal{C}}}^\dagger_{g\to q} \equiv t^a_L t^{\bar{a}}_R\,.
\end{equation}

\end{enumerate} 
Moreover, the operators in~\eqref{eq:VGGammac} satisfy the following useful relations 
\begin{align}\label{eq:usefulRelations}
  [\spac\bm{\Gamma}^c,\overline{\bm{\Gamma}}\spac] = 0 \,, \quad
\big\langle\bm{\mathcal{H}}\spac\bm{\Gamma}^c\otimes\spac\bm{1} \big\rangle = 0 \,, \quad
   \big\langle\bm{\mathcal{H}}\spac\bm{V}^G\otimes\spac\bm{1} \big\rangle = 0\, ,
\end{align}
where $\bm{\mathcal{H}}$ is a generic hard function. The first equality ensures the factorization of soft physics in the absence of Glauber phases, while the second is a consequence of collinear safety. The third equality is a direct consequence of the fact that the Glauber phase is purely imaginary, i.e.\ whenever the Glauber phase is at the end (acting directly on $\bm{1}$), the left- and right-hand side cancel upon taking the trace. 
The purely collinear anomalous dimensions $\bm{\Gamma}^C$ also fulfill useful identities
\begin{align}
\label{eq:usefulRelations2}
    [\spac\bm{\Gamma}^C,\overline{\bm{\Gamma}}\spac] = 0 \,, \quad \langle \bm{\mathcal{H}}\,\bm{\Gamma}^C\otimes \bm{1}\rangle=\langle\bm{\mathcal{H}}\otimes \bm{1}\rangle\, {\Gamma}^C\,.
\end{align}
The first equality is once again the statement of color coherence in the absence of Glauber modes. But while the cusp term vanishes when acting on the unit matrix on the right, the purely collinear term reduces to a non-vanishing, color-diagonal contribution 
\begin{equation}
    \Gamma^C(z_1,z_2) = \Gamma^C_1(z_1) \delta(1-z_2) + \delta(1-z_1)\Gamma^C_2(z_2)\,,
\end{equation}
where we defined the DGLAP anomalous dimensions
\begin{align}
    {\Gamma}^C_i(z_i)= -2 \gamma_0^i\,\delta(1-z_i)\,\bm{1}
+4
\spac C_{i \to P}\,\overline{\mathcal{P}}_{i \to P}(z_i) =4
\spac P_{i \to P}(z_i)\ \,.
\end{align}
Here, $C_{i\to p}$ is the color factor associated with the corresponding splitting and $ P_{i \to P}$ denotes the usual splitting kernel, such that DGLAP evolution is recovered, see Section~\ref{sec:outline} for further details.
A detailed derivation of the relations in~\eqref{eq:usefulRelations} and~\eqref{eq:usefulRelations2} is presented in~\cite{Becher:2023mtx}. As a side remark, note that a similar anomalous dimension lies at the heart of amplitude-level, finite-$N_c$ parton showers, see for example~\cite{Nagy:2007ty,Forshaw:2019ver,Forshaw:2025fif}.

\section{Consistency relation and outline of its verification}
\label{sec:outline}

To validate the operator structure of the low-energy matrix elements $\Wm$ in \eqref{eq:W_function_definition_FT}  and check their consistency with PDF factorization proposed in~\eqref{eq:factorization_final}, we now perform a series of perturbative computations. Before presenting  detailed results in the next two sections, we discuss the form of these cross checks and anticipate our findings.

The consistency checks are rooted in the fact that the full cross section~\eqref{eq:factorizationTheoremNew} is a physical object, and therefore finite and independent of $\mu$.
Hence,~\eqref{eq:factorizationTheoremNew} along with~\eqref{eq:hardRG} implies that
the anomalous dimensions obtained by considering soft and collinear limits of hard amplitudes must match the corresponding counterpart derived directly from the low-energy physics. Our first consistency check is to demonstrate this explicitly at the one-loop order, which is presented in Section~\ref{sec:1_loop_calc}. 
The low-energy matrix elements also contain non-perturbative physics. Evaluating them perturbatively, one encounters infrared singularities. To eliminate these, we analyze the structure using a gluon mass as an IR regulator. With the IR regulator in place, the remaining divergences are of UV origin, so we can directly access the UV physics of the soft-collinear matrix elements. In our computation with the gluon mass, we find that an additional regulator is needed to render the low-energy matrix elements well-defined. The regulator dependence cancels between the soft and collinear sectors, but a logarithm of the large momentum transfer $Q$ is left behind, an effect called the collinear anomaly~\cite{Chiu:2007yn,Chiu:2007dg,Becher:2010tm,Chiu:2011qc,Chiu:2012ir}. Adding up the soft and collinear contributions fully reproduces the anomalous dimensions of the hard functions, including its $Q$ dependence. This constitutes a strong cross check that the UV physics of the low-energy theory matches the scale dependence of hard functions.

However, the one-loop check cannot answer more detailed questions about the structure of the low-energy theory. In particular, one cannot deduce  whether the matrix elements $\bm{\mathcal{W}}$ factorize into perturbative kernels times PDFs, as proposed in~\eqref{eq:factorization_final}, since the low-energy dynamics is modified by the introduction of the IR cutoff. It is also not clear from the one-loop computation if the phase-factors in the matrix elements are purely soft, or if there are genuine Glauber contributions to $\bm{\mathcal{W}}$ at higher orders.
 
As a next step, in Sections~\ref{sec:soft_matrix_element_three_loop} and~\ref{sec:soft_collinear_matrix_element}, we therefore analyze the perturbative part of $\bm{\mathcal{W}}$, which is given by $\bm{\mathcal{I}}$ in~\eqref{eq:factorization_final}, without an IR cutoff and verify order by order in the strong coupling expansion that its divergences are consistent with the factorization theorem. This allows us to identify the intricate structure of the phase factors arising in the low-energy theory, exposing the precise nature of soft-collinear interactions. Since the cusp pieces of the anomalous dimensions first appear in a non-vanishing contribution to $\bm{\mathcal{I}}$ at three loops, the challenge of this approach is the need to perform calculations at least at this order to encounter the full complexity of the structures generated by iterated insertions of the hard anomalous dimensions.

To derive the form of the constraint on the low-energy matrix elements arising from consistency with RG running, we start with the bare hard function 
\begin{align}
    \bm{\mathcal{H}}^\mathrm{bare}=\bm{\mathcal{H}}\ast\bm{Z}\, ,
\end{align}
where all singularities are factorized into the renormalization factor $\bm{Z}$, such that $\bm{\mathcal{H}}$ is finite. With this definition, the soft-collinear matrix elements must be rendered finite through
\begin{align}\label{eq:Wren}
\bm{Z}\ast\bm{\mathcal{W}}^{\mathrm{bare}} = \bm{\mathcal{W}}\, ,
\end{align}
since the full cross-section is finite. 
As we are only interested in the leading pole behavior, we do not need to keep track of the indices on the hard functions, the $\bm{Z}$ factor, as well as on the soft-collinear matrix elements.  Assuming PDF factorization, we have
\begin{align}\label{eq:PDFren}
  \bm{\mathcal{W}}^{\mathrm{bare}}= \bm{\mathcal{I}}^{\mathrm{bare}}\ast f^{\mathrm{bare}}f^{\mathrm{bare}}=\bm{\mathcal{I}}^{\mathrm{bare}} \ast \left(\left[Z^{-1}_{\mathrm{PDF}\spac,\spac 1}\star f_1\right]\left[ Z^{-1}_{\mathrm{PDF}\spac,\spac 2} \star \, f_2\right]\,\right) ,
\end{align}
with $\star$ the standard Mellin convolution.
At one-loop order, the renormalization of each of the PDFs has the form 
\begin{align}
    Z_{\mathrm{PDF}\spac,\spac i}(z_i)=\delta(1-z_i)-\frac{\alpha_s}{4\pi}\frac{1}{2\epsilon}\Gamma^C_i(z_i)+\mathcal{O}(\alpha_s^2)\, .
\end{align}
Together, equations in~\eqref{eq:Wren} and~\eqref{eq:PDFren} imply that the partonic soft-collinear matrix elements renormalize as
\begin{equation}
\bm{\mathcal{I}} = \bm{Z}\ast \bm{\mathcal{I}}^{\mathrm{bare}}\ast \bigl(Z^{-1}_{\mathrm{PDF}\spac,\spac1}\,Z^{-1}_{\mathrm{PDF}\spac,\spac2}\bigr) \stackrel{!}{=} \text{ finite} \,,
\end{equation}
if PDF factorization holds.
Writing out this condition order-by-order in perturbation theory, allows us to predict the structure of the divergences in the bare soft-collinear matrix elements. We expand $\bm{\mathcal{I}}^{\mathrm{bare}}$ in~$\alpha_s$ as follows
\begin{align}
   \bm{\mathcal{I}}^{\mathrm{bare}}  =\bm{1}+\sum_{i=1} \left(\frac{\alpha_s}{4\pi}\right)^i\spac\bm{\mathcal{I}}^{\mathrm{bare},(i)}\, .
\end{align}
The identity matrix is to be understood as identity in both Mellin and color-helicity space.
For the divergences of the one-loop coefficient, we obtain
\begin{align}
    \bm{\mathcal{I}}^{\mathrm{bare},(1)}=\frac{1}{2\epsilon}\left(\GammaHO- \Gamma^C\spac \bm{1} \right) =\frac{1}{2\epsilon}\left(4  \spac\bm{\Gamma}^c
    \ln\frac{\mu^2}{\mu_h^2} +4\spac\bm{V}^G 
    + \,\overline{\bm{\Gamma}} 
    +  \,\bm{\Gamma}^C -  \Gamma^C \spac \bm{1} \right) \,.
\end{align}
Note that from~\eqref{eq:HardAnomalousDimensionSoftCollinear}, the first three terms in the final equality are implicitly accompanied by the respective $\delta(1-z_1)\delta(1-z_2)$. Unless stated otherwise, 
$\delta$-distributions are omitted whenever no explicit 
$z$-dependence is given. 
Inside the color trace with an arbitrary hard function, $\langle \bm{\mathcal{H}} \dots\rangle$, the $\bm{V}^G$, $\bm{\Gamma}^c$ terms vanish, and $\bm{\Gamma}^C$ reduces to $\Gamma^C$, due to relations~\eqref{eq:usefulRelations} and~\eqref{eq:usefulRelations2}. We can therefore simplify the result for $\bm{\mathcal{I}}^{\mathrm{bare},(1)}$ to 
\begin{align}
    \bm{\mathcal{I}}^{\mathrm{bare},(1)}\equalhat\frac{1}{2\epsilon} \overline{\bm{\Gamma}} \,,
\end{align}
where the hat over the equal sign indicates that the equality holds inside the color trace.
The leading poles at the two-loop order are obtained from the product of the one-loop $Z$ factors with the one-loop matrix elements, and the two-loop terms in the $Z$ factors leading to 
\begin{align}
    \bm{\mathcal{I}}^{\mathrm{bare},(2)}&=
    \frac{1}{2\epsilon}\left(\GammaHO- \Gamma^C\right) \ast  \bm{\mathcal{I}}^{\mathrm{bare},(1)}  +\frac{1}{8\epsilon^2}\left(\GammaHO- \Gamma^C\right)\ast \left(\GammaHO- \Gamma^C\right)+\dots\nonumber\\
    &\equalhat\frac{1}{2\epsilon^2}\VG\overline{\bm{\Gamma}}+\dots\, ,
\end{align}
where we focused on the leading pole and single soft emission terms. At this order, the collinear part of the anomalous dimensions does not yet contribute. Collinear contributions first arise at three-loop order, which can be obtained from the expression for the $\bm{Z}$ factor in~\eqref{eq:app:Z_factor}. Similarly, we arrive at the result 
\begin{align}\label{eq:Wm_poles}
   &\bm{\mathcal{I}}^{\mathrm{bare}}
    \equalhat \bm{1} + \frac{\alpha_s}{4\pi}\color{orange}\frac{\overline{\bm{\Gamma}}}{2\varepsilon}\color{black}
    + \left( \frac{\alpha_s}{4\pi} \right)^2 \!\left( \color{orange}\frac{\bm{V}^G\,\overline{\bm{\Gamma}}}{2\varepsilon^2}\color{black} + \dots \right)+ \left( \frac{\alpha_s}{4\pi} \right)^3\! \bigg[ \color{RoyalBlue}
      \frac{\bm{\Gamma}^{c}\spac\bm{V}^G\overline{\bm{\Gamma}}}{3\varepsilon^3} \!
     \left( \frac{11}{6\varepsilon} + \ln\frac{\mu^2}{\mu_h^2} 
     + \frac92\ln\frac{\mu^2}{Q_0^2} \right) \notag\\
   &\hspace{2.07cm} + \color{orange}\frac{\bm{V}^G\spac\bm{V}^G\,\overline{\bm{\Gamma}}}{3\varepsilon^3}\color{black}+\color{Aquamarine}\frac{\left[\bm{\Gamma}^C,\bm{V}^G\spac\overline{\bm{\Gamma}}\right]}{12\varepsilon^3}\color{black}+ \dots \bigg] + \mathcal{O}(\alpha_s^4) \,.
   \end{align}
The orange and darker blue terms feature $\delta(1-z_1)\delta(1-z_2)$ which are suppressed.
Note that $\mu_h\sim Q$ is the reference scale introduced in~\eqref{eq:GammaH} and cancels between the explicit logarithms and the one included in $\GammaC$.
Interestingly, due to the relations in~\eqref{eq:usefulRelations} and~\eqref{eq:usefulRelations2}, the three-loop terms can also be written as double-commutators, e.g.\
\begin{equation} \left[\bm{\Gamma}^C,\bm{V}^G\spac\overline{\bm{\Gamma}}\right] = \left[\left[\bm{\Gamma}^C,\bm{V}^G\right],\overline{\bm{\Gamma}}\right] = \left[\bm{\Gamma}^C, \left[\bm{V}^G,\overline{\bm{\Gamma}} \, \right]\right]\,.
\end{equation}
In Sections~\ref{sec:soft_matrix_element_three_loop} and~\ref{sec:soft_collinear_matrix_element}, we compute all relevant terms appearing in the bare soft-collinear matrix elements $\bm{\mathcal{I}}^{\mathrm{bare}}$ up to the three-loop order and compare their structure to the prediction in~\eqref{eq:Wm_poles}. Our main findings are:
\begin{enumerate}[(i)]
    \item 
    The only perturbative scale within the soft-collinear matrix elements is $Q_0$, resolved by a soft emission entering the gap. All other parts of the one-loop matrix elements are scaleless.
    The divergence of the soft emission precisely matches the contribution of $\overline{\bm{\Gamma}}$ in~\eqref{eq:Wm_poles}. 
    We highlight this and other purely soft terms in orange color.
    \item At two-loop level, the purely collinear as well as the mixed soft and collinear contributions to the low-energy matrix elements are scaleless. Again only the purely soft terms contribute. As we only consider single emissions into the gap, these terms are given by the one-loop soft current. Its imaginary part matches the structure of $\bm{V}^G\,\overline{\bm{\Gamma}}$, and no additional mode beyond soft is present at this order. 
    \item At three-loop order, the orange term proportional to $\VG\VG\Gammabar$, is obtained using the known result for the two-loop soft current~\cite{Duhr:2013msa,Dixon:2019lnw}. However, the terms highlighted in blue in~\eqref{eq:Wm_poles} cannot be purely soft. The darker blue terms contain additional hard logarithms of the form $\sim \ln Q$ as well as additional poles, while the lighter blue terms carry non-trivial dependence on the momentum fraction $z$ through the splitting functions inside $\bm{\Gamma}^C$. Soft radiation cannot resolve the associated collinear dynamics, nor can it change the momentum fraction. Conversely, the collinear modes cannot probe (or enter) the gap region. However, as the presence of the gap provides the only scale for the soft-collinear matrix elements, the collinear physics alone is scaleless, even in the presence of collinear anomalies. A coupling of collinear modes to soft modes, which gives scale  to collinear matrix elements, is therefore required. This means that soft-collinear factorization breaking is necessary for PDF factorization to hold. In Section~\ref{sec:soft_collinear_matrix_element}, we explicitly show the factorization breaking diagrams. They involve genuine Glauber modes and fully reproduce the blue terms in the consistency relation~\eqref{eq:Wm_poles}.
\end{enumerate}
 
While this three-loop check does not constitute a complete factorization proof, it is gratifying that the complicated low-energy dynamics of the Glauber modes has exactly the right structure to match PDF factorization, although all ingredients for the breaking are in place at this order. In our view, this suggests that~\eqref{eq:factorization_final} holds to all orders.

The remainder of this article is structured as follows: in Section~\ref{sec:1_loop_calc} we analyze the one-loop poles in the low-energy matrix elements using parton masses as infrared regulators. This verifies consistency with the hard anomalous dimensions as discussed above.
In Section~\ref{sec:soft_matrix_element_three_loop}, we determine the purely soft, non-factorization-violating contributions, i.e.\ the orange structures appearing in~\eqref{eq:Wm_poles}, up to the three-loop level.
The details of this calculation are not essential to follow the analysis in Section~\ref{sec:soft_collinear_matrix_element}, where we present the complete derivation of the blue terms discussed above and how they arise from genuine Glauber exchange. This section is quite technical, but it contains crucial details relevant for exposing the factorization breaking in the low-energy theory and discovering how this mechanism restores PDF factorization by precisely canceling the collinear factorization-violating terms of the high-energy theory.
Readers who are primarily interested in the physics of the Glauber mode, can focus on the discussion in Section~\ref{sec:discussion}, which summarizes the previous sections, and then highlights the significance of our results for both the understanding of PDF factorization and collinear factorization breaking. We also discuss how our analysis can be extended to a wider class of observables. A summary of notation used throughout the paper is given in Table~\ref{tab:definitions} for convenience.

\begin{table}
    \centering
        \renewcommand{\arraystretch}{1.3}
    \begin{tabular}{c|l}
          $\Hm$ & Hard functions with explicit open initial-state indices as defined in~\eqref{eq:hard_function_definition}.\\
          \hline
           $\bm{\overline{\mathcal{H}}}_m$ & $\Hm$ integrated over energies  and with flux factor~\eqref{eq:hard_function_spin_averaged}.\\
           \hline
          $\Wm$ & Soft-collinear matrix elements as defined in~\eqref{eq:W_function_definition_FT}. \\
          \hline
          $\bm{\mathcal{I}}_m$ & Partonic sofct-collinear matrix elements~\eqref{eq:I_function_definition}.\\
          \hline
          $\llangle\cdots\rrangle$ & Spin and color sum over final-state particles.\\ 
          \hline
          $\langle\cdots\rangle$ & As $\llangle\cdots\rrangle$ but with additional  spin and color average for initial-state particles.\\
          \hline
          $\Theta_{\mathrm{veto}}$ & Function encoding the jet veto prescription. $\Theta_{\mathrm{hard}}=1-\Theta_{\mathrm{veto}}.$ \\
          \hline
           $\star$ & Standard Mellin convolution.\\
           \hline
            $\ast$ & Generalized Mellin convolution as in~\eqref{eq:ConvolutionDefinition}.\\
            \hline
             $\otimes$ & Angular integrals over the
             final-state directions of the hard functions.\\
             \hline
             $\bm{\Gamma}^C$ & Collinear piece of the hard anomalous dimension in color space.\\ 
             \hline
               $\Gamma^C$ & Color diagonal contribution describing DGLAP evolution.
    \end{tabular}
    \caption{Summary of notation used throughout the paper. }
    \label{tab:definitions}
\end{table}

\section{One-loop anomalous dimensions from the low-energy theory}
\label{sec:1_loop_calc}

The anomalous dimensions of the hard functions in~\eqref{eq:GammaH} were extracted from soft and collinear limits of hard amplitudes.  As a first consistency check, we calculate the one-loop anomalous dimensions from the soft-collinear matrix elements and compare them to their hard counterpart. 
We begin with the computation of the purely soft contributions, followed by the purely collinear matrix elements.
While a definition of such purely soft and purely collinear matrix elements to arbitrary order is possible, beyond two loops also soft-collinear interactions occur, such that this separation is no longer useful.

\subsection{Soft function with a gluon mass}
\label{sec:one-loop-soft}
We define the purely soft matrix element as
\begin{align} \label{eq:soft_function_definition}
   {\bm{\mathcal{S}}}_{ij\to m}(Q_0,\mu)=\hspace{0.1cm}
   &\int\limits_{X_s}\hspace{-0.5cm}\sum\,\langle 0\spac |\spac \bm{S}_1^\dagger\spac\dots\spac\bm{S}_{m+2}^\dagger\spac|
    \spac X_s\rangle \spac \langle X_s\spac|\spac\bm{S}_1\spac\dots\spac\bm{S}_{m+2}\spac|\spac 0\rangle \,\theta( Q_0 - E^\perp_{\mathrm{\, out}}) \,,
\end{align}
where $\bm{S}_i$ is the light-like soft Wilson line in direction $n_i$ that appears in~\eqref{eq:W_function_definition}. Its perturbative expansion is given by
\begin{align}
    {\bm{\mathcal{S}}}_{ij\to m}(Q_0,\mu) = \bm{1} + \sum_{k=1}^\infty \spac\Bigl(\frac{\alpha_s}{4\pi}\Bigr)^{k}\bm{\mathcal{S}}^{(k)}(Q_0,\mu)\,,
\end{align}
where $k$ denotes the order in $\alpha_s$.
The virtual correction is given by
\begin{align}
    \frac{\alpha_s}{4\pi}\bm{\mathcal{S}}^{(1)\spac\mathrm{virt}}&=\frac{g_s^2}{2} \sum_{i\neq j}\,\left(\bm{T}_{i,L}\cdot \bm{T}_{j,L}\,I_{ij} +\,\bm{T}_{i,R}\cdot \bm{T}_{j,R}\,I^*_{ij} \right)\,,\label{eq:SoftVirtualIntegralGluonMass}
\end{align}
where $I_{ij}$ is the soft loop integral for the case of two outgoing or incoming Wilson lines. Introducing a gluon mass as the IR regulator, it reads
\begin{align}\label{eq:Iij-def}
   I_{ij} &= i\tilde{\mu}^{2\varepsilon}\int \frac{d^dl_{s}}{\left( 2\pi\right)^d} \frac{n_i\cdot n_j}{(n_i\cdot l_{s} + i0)(n_j \cdot l_{s} - i0)} \frac{1}{l_{s}^2-m^2+i0} \nonumber\\
    &=2i \tilde{\mu}^{2\epsilon}\int \frac{d^dl_{s}}{\left( 2\pi\right)^d}\frac{1}{\left( l_{s,0}-l_{s,z}\right) +i0}\frac{1}{\left( l_{s,0}+l_{s,z}\right) -i0}\frac{1}{l_{s}^2-m^2+i0}\,,
\end{align}
with $\tilde{\mu}^2 =\frac{e^{\gamma_E}}{4\pi}\mu^2$. For the mixed-case, where one of the Wilson lines is outgoing and one incoming, the sign of the $i0$-prescription of one of the eikonal propagators has to be reversed. To arrive at the second line of~\eqref{eq:Iij-def}, we performed the substitution
\begin{equation}
\begin{aligned}
\label{eq:variable_change}
    l_{s,0}&=\frac{1}{2}\sqrt{\frac{2}{n_i\cdot n_j}}\left(n_i\cdot l_{s} + n_j\cdot l_{s}\right)\,,\\
    l_{s,z}&=\frac{1}{2}\sqrt{\frac{2}{n_i\cdot n_j}}\left(n_i\cdot l_{s} - n_j\cdot l_{s}\right)\,.
    \end{aligned}
\end{equation}
We choose the labels $0$, $z$ as these would be the respective components in a back-to-back frame, which is, however, not assumed here.
Since the $l_{s,z}$ integral is ill-defined on its own, we introduce a non-analytic regulator based on~\cite{Rothstein:2016bsq} given by
\begin{equation}\label{eq:RapidityRegulator}
\left(\frac{\nu}{2}\right)^{2\eta}\left(\frac{2}{n_i\cdot n_j}\right)^\eta\lvert l_{s,z}\rvert ^{-2\eta}\,,
\end{equation}
with associated scale $\nu$. Note that for each pair $i,j$ of Wilson lines, a distinct regulator is introduced, which has a double role in the computation. It regulates rapidity integrals, and it isolates the Glauber contribution. At the one-loop order, no distinction between these two regularizations is necessary as the Glauber contribution does not lead to poles in $1/\eta$. However, at higher loop orders two distinct regulators, one for Glaubers and one for rapidity, are needed in general, since the physics associated with the two effects is different~\cite{Moult:2022lfy}. The absolute value in~\eqref{eq:RapidityRegulator} is crucial to extract the Glauber phase, as the $i0$-prescription becomes relevant in this case. The full contribution is
\begin{align}
\bm{\mathcal{S}}^{(1)\spac\mathrm{virt}}
= \sum_{i\neq j}\left(\frac{\mu^2}{m^2}\right)^{\epsilon} 
e^{\varepsilon\gamma_E} 
&\!\biggl[ (\bm{T}_{i,L}\cdot\bm{T}_{j,L} 
+\bm{T}_{i,R}\cdot\bm{T}_{j,R} )
\left(\frac{n_i\cdot n_j}{2}\right)^\eta 
\left(\frac{\nu^2}{m^2}\right)^{\eta}
\frac{\Gamma\left(1-2\eta\right)\Gamma\left(\eta+\epsilon\right)}{\eta\spac\Gamma(1-\eta)}
\nonumber\\ &\;\; 
- i\pi\lambda_{ij}(\bm{T}_{i,L}\cdot\bm{T}_{j,L} 
-\bm{T}_{i,R}\cdot\bm{T}_{j,R} )\Gamma(\epsilon)
\biggr]
\, ,
\end{align}
where the second line corresponds to the phase originating from the exchange of an off-shell potential gluon, which is usually denoted as a ``Glauber'' phase. From this explicit calculation it is clear that this phase arises solely from soft kinematics.
As this term has no pole in $1/\eta$, we already set $\eta\rightarrow 0$.
Moreover, this contribution arises only when the particles $i,j$ are both in- or outgoing, as switching the incoming particle to an outgoing one changes the $i0$ prescription in the corresponding eikonal propagator in~\eqref{eq:SoftVirtualIntegralGluonMass}.
To account for this, we introduced the factor $\lambda_{ij}$, which is $1$ if both particles are in- or outgoing and $0$ otherwise.

In the first term, the dependence on $n_i\cdot n_j$ has the form $\left(n_i\cdot n_j\right/2)^\eta$, which 
will lead to a logarithm of $n_i\cdot n_j$ upon expansion in $\eta$. This logarithmic dependence matches the angular integral over $\overline{W}_{ij}^l$ in the hard functions.
The anomalous dimensions derived from the respective limits of the hard functions are well-defined in dimensional regularization. Therefore, the additional poles in $1/\eta$ must be canceled solely within the soft-collinear matrix elements.
However, due to the color structure, they cannot be compensated by the corresponding soft real-emission contribution.
Indeed, as we show in Section~\ref{sec:one-loop-collinear}, it is instead canceled by the virtual collinear diagrams, as is common for problems featuring a collinear anomaly~\cite{Becher:2010tm}.
In the following, we only present the single-pole terms in both $1/\epsilon$ and  $1/\eta$, as the coefficients for the double poles $1/\epsilon^2$ and  $1/(\epsilon\eta)$ are directly linked to the logarithms in the single-pole terms. We then show that the poles in $1/\eta$ cancel and the pole structure in $1/\epsilon$ reproduces the one of the hard functions. The $1/\epsilon$-pole term of the virtual soft contribution is given by
\begin{equation}
\label{eq:soft-1loop-virt-eps}
\bm{\mathcal{S}}_\epsilon^{(1)\spac\mathrm{virt}}=\frac{1}{\epsilon}\sum_{i \neq j}\left(\bm{T}_{i,L}\cdot \bm{T}_{j,L}+\bm{T}_{i,R}\cdot \bm{T}_{j,R}\right)\left(\int \frac{d\Omega(n_l)}{4\pi}\overline{W}_{ij}^l-\ln{\frac{\mu^2}{\nu^2}}\right)+\frac{2}{\epsilon}\spac \VG\, ,   
\end{equation}
where $\VG$ is the Glauber phase given in~\eqref{eq:VGGammac}.
Here, we summed over all contributing diagrams and used color conservation to express the phase entirely with initial-state generators.
The single pole in $1/\eta$ is  given by
\begin{equation}
\label{eq:soft-1loop-virt-eta}
    \bm{\mathcal{S}}_\eta^{(1)\spac\mathrm{virt}}=-\frac{1}{\eta}\sum_{i \neq j}\left(\bm{T}_{i,L}\cdot\bm{T}_{j,L}+\bm{T}_{i,R}\cdot\bm{T}_{j,R}\right)\ln{\frac{m^2}{\mu^2}}\,.
\end{equation}
The real-emission diagrams can be analyzed along the same lines. In this case, the emission cannot be in the veto region when the energy is above $Q_0$. Extracting the associated UV divergence yields
\begin{equation}
\label{eq:soft_one_loop_real_pole_eps}
    \bm{\mathcal{S}}_\epsilon^{(1)\spac\mathrm{real}}=-\frac{2}{\epsilon}\sum_{i \neq j}\spac\bm{T}_{i,L}\circ\bm{T}_{j,R}\left(\int \frac{d\Omega(n_l)}{4\pi}\spac\overline{W}_{ij}^l\spac\Theta_{\mathrm{hard}}(n_l)-\ln{\frac{\mu^2}{\nu^2}}\right)\,.
\end{equation}
For the collinear divergence, regularized by $\eta$, we obtain
\begin{equation}
   \label{eq:soft-1loop-real-eta}\bm{\mathcal{S}}_\eta^{(1)\spac\mathrm{real}}=\frac{2}{\eta}\sum_{i \neq j}\spac\bm{T}_{i,L}\circ\bm{T}_{j,R}\ln{\frac{m^2}{\mu^2}}\, .
\end{equation}
Note the expected sign difference between the virtual and real emission parts, as well as the different color structure. The real emission always has one color operator acting on the right and one on the left, whereas the virtual contribution has color generators  either both acting on the left or both on the right.

\subsection{Collinear contribution}
\label{sec:one-loop-collinear}
Next, we discuss the contribution from the collinear diagrams. To extract the anomalous dimensions of the operators, it suffices to work  with partonic instead of hadronic states, as long as a gluon mass is used to regulate the IR divergences as in the soft case. As an example, we consider here the diagonal quark channel. For initial state gluons, as well as off-diagonal channels, see Appendix~\ref{sec:app:coll_one_loop}.
At one-loop order, we define the beam function as
\begin{equation}
\label{eq:beam_function}
    \bm{B}_{q/q}(z)=\SumInt_X \delta\left( \bar{n}_i\cdot\left(k-(1-z)p\right)\right)\Big(\frac{\slashed{\bar{n}}_i}{2}\Big)_{\alpha\beta}\langle q(p)\lvert \overline{\chi}_\alpha(0)\rvert X\rangle \langle X\lvert \chi_\beta(0)\rvert q(p)\rangle\, .
\end{equation}
Here, $\chi$ is the gauge-invariant building block of the collinear fermion along the given direction $i=1$ or $i=2$ and $k$ is the total final-state momentum. 
Note that the hard functions carry open color indices, which are contracted with the collinear building blocks via soft Wilson lines~\eqref{eq:W_function_definition}. 
Thus, no color trace is performed in the beam function. 
Instead, it carries open color indices corresponding to the respective partons (here quarks), which we suppress in the following.

The $\alpha_s$ expansion of the beam function is defined as
\begin{equation}
    \bm{B}_{q/q}(z)=\bm{1}+\sum_{k=1}^\infty\left(\frac{\alpha_s}{4\pi}\right)^k\bm{B}_{q/q}^{(k)}(z)\, ,
\end{equation}
where $\bm{1}$ is unity in both color and Mellin space and $k$ denotes the respective order in $\alpha_s$. As before, we split the calculation into real-emission and virtual parts.

\begin{figure}
    \centering
    \includegraphics[scale=1]{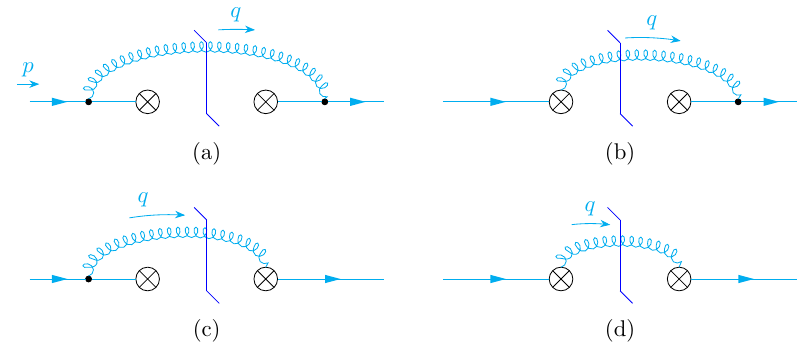}
    \caption{Diagrams depicting the one loop correction to the beam function due to the emission of a collinear gluon.}
    \label{fig:collinear_1loop_emission}
\end{figure}
The relevant diagrams for the emission of collinear gluons are given in Figure~\ref{fig:collinear_1loop_emission}. We use the rapidity regulator given in \eqref{eq:RapidityRegulator} and expand it in the collinear kinematics, e.g.\ $q_z=\bar{n}_i\cdot q+\mathcal{O}(\lambda)$. The phase-space integral in diagram in Figure~\ref{fig:collinear_1loop_emission}a  is of the following form 
\begin{equation}
\tilde{\mu}^{2\epsilon} \nu^{2\eta}\int \frac{d^{d-2}q_\perp}{\left(2\pi\right)^{d-1}}\left(\bar{n}_i\cdot p(1-z)\right)^{1-2\eta}\frac{q_\perp^2}{\bar{n}_i\cdot p\left(q_\perp^2-zm^2\right)^2}\,.
\end{equation}
Here, we already performed the trivial integration over both $q_+$ and $q_-$ components and evaluated the two delta distributions, $\delta{(\bar{n}_i\cdot q-(1-z)\bar{n}_i\cdot p)}$ and $\delta_+{(q^2-m^2)}$, with $\delta_+(q^2-m^2)=\delta(q^2-m^2)\theta(q_0)$.
The gluon mass allows us to extract the UV physics of the diagram, as, without it, the collinear contributions would be scaleless. 
Adding all the diagrams shown in Figure~\ref{fig:collinear_1loop_emission}, we arrive at
\begin{align}
\bm{B}_{q/q}^{(1)\spac\mathrm{real}}(z)&=4t^b_{1,L}t^{b^\prime}_{1,R}\, z (1-z)^{-1-2\eta}\left(\frac{\nu}{\bar{n}_i\cdot p}\right)^{2\eta}\left(\frac{\mu^2}{zm^2}\right)^\epsilon e^{\epsilon\gamma_E}  \Gamma(\epsilon)\nonumber \\
   &\quad+2t^b_{1,L}t^{b^\prime}_{1,R}
   (1-z) \left(\frac{\mu^2}{zm^2}\right)^\epsilon e^{\epsilon\gamma_E}  \left(1-\epsilon\right)^2 \Gamma(\epsilon)\,,
\end{align}
where the first line arises from diagrams in Figure~\ref{fig:collinear_1loop_emission}b and~\ref{fig:collinear_1loop_emission}c and the second line from diagram in Figure~\ref{fig:collinear_1loop_emission}a, while the last diagram in Figure~\ref{fig:collinear_1loop_emission}d gives a vanishing contribution. We set $\eta\to 0$ whenever possible and  explicitly retained the $L$, $R$ labels on the color generators and left the indices $b,b^\prime$ of the emitted gluon open, to allow for additional emissions analogous to the $\circ$ symbol.
Explicitly, the color structure $t^{b\spac(b^\prime)}_{1,L\spac(R)}$ are given by
\begin{equation}
    t^b_{1,L} = t^b_{ki}\,,\quad t^{b^\prime}_{1,R} = t^{b^\prime}_{\bar{i}\bar{k}}\,,
\end{equation}
with $i$ $(\spac\bar{i}\spac)$ the fundamental index of the left (right) external quark and $k$ $(\bar{k})$ the index of the hard (conjugate) amplitude.

\begin{figure}
    \centering
    \includegraphics[scale=1]{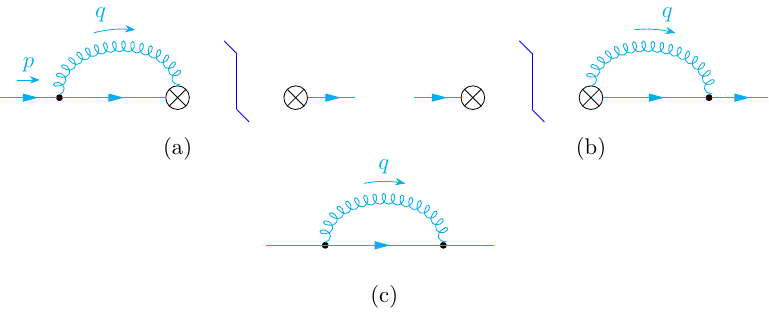}
    \caption{ The one-loop  virtual contributions. The upper two diagrams connect to the Wilson line, while the lower diagram is related to the on-shell renormalization factor.}
    \label{fig:collinear_1loop_virtual}
\end{figure}

The diagrams contributing to the corresponding virtual correction are shown in Figure~\ref{fig:collinear_1loop_virtual}a and~\ref{fig:collinear_1loop_virtual}b. Their calculation proceeds 
along the same lines as the real correction, in the end yielding
\begin{equation}
    \bm{B}_{q/q}^{(1)\spac\mathrm{virt}}(z)=-
    4
    C_F{\delta(1-z)} \left(\frac{\mu^2}{m^2}\right)^\epsilon\left(\frac{\nu}{\bar{n}_i\cdot p}\right)^{2\eta}\frac{\Gamma(-2\eta)\Gamma(2-\epsilon)}{\Gamma(2-2\eta-\epsilon)}e^{\epsilon\gamma_E}\Gamma(\epsilon)\,.
\end{equation}
Note that these contributions carry a factor of $\delta(1-z)$, as there is no intermediate (cut) gluon. Consequently, we set $z=1$ in the above expression outside of the $\delta$-distribution. Additionally, the on-shell renormalization factor of the massive gluon is given by
\begin{equation}
  Z=1 -\frac{\alpha_s}{4\pi}
C_F\left(\frac{\mu^2}{m^2}\right)^{\epsilon}e^{\epsilon\gamma_E}\Gamma(\epsilon)+\mathcal{O}(\alpha_s^2)\, ,
\end{equation}
which has to be multiplied accordingly with the tree-level diagram.
Generalizing the above results to generic representations, including both quarks and gluons, and to both collinear and anti-collinear directions $n_1,n_2$ yields
\begin{align}
\bm{B}_1(z_1)\bm{{B}}_2(z_2)\Big\rvert_{\varepsilon}^{\mathrm{virt}}
&\hspace{-0.05cm}=-\frac{1}{\epsilon}\frac{\alpha_s}{4\pi}
\sum_{i=1,2}\hspace{-0.1cm}C_i{\delta(1-z_{i})}-\frac{4}{\epsilon}\frac{\alpha_s}{4\pi}\sum_{i=1,2\hspace{-0.1cm}}C_i{\delta(1-z_{i})}\hspace{-0.1cm}\left(\ln\frac{2E_i}{\nu}-1\right)\hspace{-0.1cm}+\mathcal{O}(\alpha_s^2) \nonumber \\
&\hspace{-0.05cm}=-\frac{1}{2\epsilon}\frac{\alpha_s}{4\pi}\sum_{i=1,2}\delta(1-z_{i})\biggl(8C_i\ln\frac{2E_i}{\nu}+2
\gamma_{0}^i\biggr)+\mathcal{O}(\alpha_s^2)\, ,
\end{align}
where  $\gamma_{0}^i$ is defined above~\eqref{eq:splitting_functions_final_explicit}. 
Here, we again suppressed the trivial factors $\delta(1-z_{j})$ of the respective other momentum fraction $j\neq i$. The subscript $\epsilon$ indicates that we focus on the $1/\epsilon$ poles of the expression.
Here the distributional relation
\begin{equation}
    \left(1-z\right)^{-1-2\eta}=-\frac{1}{2\eta}\delta(1-z)+\left[\frac{1}{1-z}\right]_++\mathcal{O}\left(\eta\right)\,
\end{equation}
was used. Furthermore, we identified $\bar{n}_i\cdot p\rightarrow 2E_i$.
The contribution to the pole in $1/\eta$ is given by
\begin{equation}
\label{eq:beam_one_loop_virt_eta}
\bm{B}_1(z_1)\bm{B}_2(z_2)\Big\rvert_{\eta}^{\mathrm{virt}}=-\frac{4}{\eta}\frac{\alpha_s}{4\pi}\sum_{i=1,2}C_i\ln\frac{m}{\mu}{\delta(1-z_{i})}+\mathcal{O}(\alpha_s^2)\, .
\end{equation}
Turning our attention to the collinear emission part, we find the following 
\begin{align}
\bm{B}_1(z_1)\bm{B}_2(z_2)\Big\rvert_{\varepsilon}^{\mathrm{real}}&= \frac{1}{\epsilon}\gamma_0^{\mathrm{cusp}}\frac{\alpha_s}{4\pi}\sum_{i=1,2}\bm{T}_{i,L}\circ\bm{T}_{i,R} \ln{\frac{2E_i}{\nu}}\delta(1-z_{i})
\nonumber \\
&\quad+\frac{2}{\epsilon}\frac{\alpha_s}{4\pi}\sum_{i=1,2}\overline{\mathcal{P}}_{i\to P}(z_i)\spac\bm{\mathcal{C}}_{i\to P,L}\,\bm{\mathcal{C}}_{i\to P,R}^\dagger+\mathcal{O}(\alpha_s^2)\,,
\label{eq:coll_one_loop_real_poles_eps}
\end{align}
which holds for all diagonal and off-diagonal splittings. Note that for diagonal splittings the above form is only correct under an integral over the respective momentum fraction.
For the $\eta$-part, we arrive at
\begin{equation} 
\label{eq:beam_one_loop_real_eta}
\bm{B}_1(z_1)\bm{B}_2(z_2)\Big\rvert_{\eta}^{\mathrm{real}}=\frac{4}{\eta}\frac{\alpha_s}{4\pi}\sum_{i=1,2}\bm{T}_{i,L}\circ\bm{T}_{i,R}\ln{\frac{m}{\mu}}\,\delta(1-z_{i})+\mathcal{O}(\alpha_s^2)\, ,
\end{equation}
which arises only in diagonal channels, as the collinear anomaly is only present in these channels.

\subsection{One-loop anomalous dimensions}
The previously obtained UV poles must reproduce the IR poles of the hard functions. The one-loop anomalous dimensions are obtained from these (UV) $1/\epsilon$ poles of the soft and collinear matrix elements and must be independent of $m^2$ and $\nu$.  From the virtual contributions of both the soft and collinear functions, we obtain
\begin{align}
    \bm{\Gamma}_{\mathrm{virt}}^{S+C}=&-\frac{\alpha_s}{4\pi}\sum_{i \neq j}2\left(\bm{T}_{i,L}\cdot\bm{T}_{j,L}+\bm{T}_{i,R}\cdot\bm{T}_{j,R}\right)\int \frac{d\Omega(n_l)}{4\pi}\overline{W}_{ij}^l-\frac{\alpha_s}{4\pi}\gamma_0^{\mathrm{cusp}}\VG\nonumber \\
    &+2\gamma_0^{\mathrm{cusp}}\frac{\alpha_s}{4\pi}\sum_{i=1,2}
    C_i{\delta(1-z_{i})}\ln\frac{2E_i}{\mu}+2\frac{\alpha_s}{4\pi}
    \sum_{i=1,2}\gamma_{0}^i{\delta(1-z_{i})}\,,
\end{align}
while the summed real-emission contribution gives
\begin{align}
    \bm{\Gamma}_{\mathrm{real}}^{S+C}&=4\frac{\alpha_s}{4\pi}\!\sum_{i \neq j}\bm{T}_{i,L}\circ\bm{T}_{j,R}\overline{W}_{ij}\Theta_{\mathrm{hard}}(n_l)-4\frac{\alpha_s}{4\pi}\spac\overline{\mathcal{P}}_{i\to P}\,\bm{\mathcal{C}}_{i\to P,L}\,\bm{\mathcal{C}}_{i\to P,R}^\dagger\delta(n_k-n_i)\nonumber\\
    &\quad
    -2\gamma_0^{\mathrm{cusp}}\frac{\alpha_s}{4\pi}\!\sum_{i=1,2}\bm{T}_{i,L}\circ\bm{T}_{i,R}\spac z_{i}\ln{\frac{2E_i}{\mu}}\delta(1-z_{i})\delta(n_k-n_i)
    \,,
\end{align}
where the $\nu$-dependence completely canceled between the soft and collinear modes. 
In the expressions for the hard anomalous dimensions~\eqref{eq:VGGammac}, the angular integration for real emissions is not yet performed, but left implicit in the $\otimes$ prescription. To directly compare the above expressions with the hard anomalous dimensions, we therefore pulled out the angular integral from the real-emission contributions~\eqref{eq:soft_one_loop_real_pole_eps} and~\eqref{eq:coll_one_loop_real_poles_eps}, inserting angular $\delta$-distributions where necessary. We can now directly confirm the same structure of the anomalous dimensions given in~\cite{Becher:2023mtx} but with a respective minus sign
\begin{equation}
    \bm{\Gamma}^H=-\bm{\Gamma}^{S+C}\, 
\end{equation}
so that the cross section is indeed scale independent.

 It is an interesting question whether one could also derive an evolution equation in the scale $\nu$ to resum rapidity logarithms~\cite{Chiu:2007yn,Chiu:2007dg,Chiu:2011qc,Chiu:2012ir}. Previous applications of this formalism considered cases with a simple multiplicative evolution, which is insufficient for our situation due to color structure.
For the moment, we refrain from identifying the respective $1/\eta$ poles from various sectors with anomalous dimensions. 
However, the $1/\eta$ poles have to cancel within the soft-collinear matrix elements alone, and the explicit check of this cancellation is presented in Appendix~\ref{sec:app:cancellation_nu_dim}.
More discussion on the $\nu$ evolution can be found in Appendix~\ref{sec:app:A1_anomalousdimension}.

This concludes the one-loop consistency check. We have demonstrated that the $\mu$ anomalous dimensions derived from the low-energy theory match the corresponding terms obtained from the high-energy theory and that the $1/\eta$ poles cancel in the low-energy matrix elements alone.

\section{Purely soft matrix elements}
\label{sec:soft_matrix_element_three_loop}

We now turn to the consistency check of the low-energy matrix elements, starting with the calculation of the purely soft part up to the three-loop order. 
Before delving into the details, we remind the reader of the expected form of the soft-collinear matrix elements~\eqref{eq:Wm_poles}. Excluding the collinear contributions, this result reduces to
\begin{equation}\label{eq:Wm_poles_1}
   \bm{\mathcal{I}}_m^{\mathrm{bare}}
    = \bm{1} + \frac{\alpha_s}{4\pi}\,\frac{\overline{\bm{\Gamma}}}{2\varepsilon}
    + \left( \frac{\alpha_s}{4\pi} \right)^2 \left( \frac{\bm{V}^G\,\overline{\bm{\Gamma}}}{2\varepsilon^2} + \dots \right)+\left( \frac{\alpha_s}{4\pi} \right)^3 \left(\frac{\bm{V}^G\spac\bm{V}^G\,\overline{\bm{\Gamma}}}{3\varepsilon^3} +\dots\right) + \mathcal{O}(\alpha_s^4) \,,
\end{equation}
which we now verify order by order.

For the single soft emission, described by $\overline{\bm{\Gamma}}$, the virtual and real contributions combine inside the trace, such that the resulting phase-space integral is indeed only over the veto region. The color structures of the two-loop result, $\VG\overline{\bm{\Gamma}}$, and the three-loop contribution, $\VG\VG\overline{\bm{\Gamma}}$, are given in~\cite{Becher:2023mtx}. Both of these terms involve a single real emission and can therefore be obtained from the one-emission soft current.  
We expand this current describing the emission of a single soft gluon with momentum $l_s$ as
\begin{equation}
\bm{J}^{\mu,a}(l_s) = \sum_{k=0}^\infty  \left(\frac{\alpha_s}{4\pi}\right)^k \bm{J}^{\mu,a(k)}\,,
\end{equation}
and use the known results for one-loop~\cite{Catani:2000pi} and two-loop soft currents. The latter is rather lengthy and can be found in~\cite{Duhr:2013msa,Dixon:2019lnw}.\footnote{Note that the normalizations chosen in~\cite{Dixon:2019lnw} are different from ours. Their color matrices $t^a$ are obtained after multiplying ours with a factor of $\sqrt{2}$ and their structure constants $f^{abc}$ are obtained after multiplying ours with a factor of $i\sqrt{2}$. Additionally, the coupling was rescaled as $\bar{a}=\alpha_s/(4\pi)(e^{-\epsilon\gamma}/(4\pi)^{-\epsilon})$.}

The tree-level soft current is given by the eikonal factor
\begin{equation}\label{eq:TreeLevelSoft}
    \bm{J}^{\mu,a(0)} = g_s\sum_{i=1}^m\, \bm{T}_{i,L}^a\,\frac{n_i^\mu}{n_i\cdot l_s} \,,
\end{equation}
where $l_s$ denotes the momentum of the soft emission. 
The hermitian conjugated current is obtained by changing $L\leftrightarrow R$, such that the conjugate acts to the right of a given hard function.
When computing the cross section with a single soft emission emanating from $\bm{J}^{\mu,a(0)} \bm{J}^{a(0)\dagger}_\mu$, we arrive at
\begin{equation}\label{eq:oneloopsoft}
   \bm{\mathcal{S}}^{(1)}=\frac{2}{\epsilon}\left(\frac{ \mu^2}{4Q_0^2}\right)^{\epsilon} \sum_{i\neq j} \bm{T}_{i,L}\cdot\bm{T}_{j,R} \int \left[d\Omega_l\right] \overline{W}_{ij}^l\Theta_{\mathrm{veto}}(n_l)\, ,
\end{equation}
where the angular integral is defined as
 \begin{equation}
 \label{eq:definition_angular_integral_ddim}
     \int \left[d\Omega_l\right]=\tilde{c}^\epsilon\int \frac{d^{d-2}\Omega_l}{2(2\pi)^{d-3}}\, .
 \end{equation}
Note that the energy is only constrained if the emission is in the veto region. When the soft emission is inside the jet region, the energy is unconstrained, leading to a scaleless contribution. In the veto region, the emission can never become collinear to the emitters $i$ and $j$ so that $\overline{W}_{ij}^l$ and ${W}_{ij}^l$ are identical. We conclude that~\eqref{eq:oneloopsoft} produces exactly the predicted one-loop contribution in~\eqref{eq:Wm_poles_1}, which is given by $\overline{\bm{\Gamma}}$ defined in~\eqref{eq:VGGammac}.

The one-loop soft current is given by~\cite{Catani:2000pi}
\begin{equation}\label{eq:one_loop_soft_current}
   \bm{J}^{\mu,a(1)}= -g_s
\frac{\Gamma^3(1-\varepsilon)\,\Gamma^2(\varepsilon)}{\Gamma(1-2\varepsilon)}
 if^{abc}\sum_{i\neq j}\spac \bm{T}_{i,L}^b\spac\bm{T}_{j,L}^c
\left(\frac{n_i^\mu}{n_i\cdot l_s} - \frac{n_j^\mu}{n_j\cdot l_s}\right) \left[\frac{\overline{W}_{ij}^le^{-i\pi\lambda_{ij}}}{2\spac e^{-i\pi\lambda_{il}}e^{-i\pi\lambda_{jl}}}\right]^\epsilon\, ,
\end{equation}
where $\lambda_{ij}=1$ if both particles $i,j$ are incoming (or both outgoing) and $\lambda_{ij}=0$ otherwise.
For an outgoing soft momentum $l_s$, the analytic continuation of the one-loop soft current reads 
\begin{equation}
\label{eq:analytic_continuation_soft_current}
    \left[\frac{e^{-i\pi\lambda_{ij}}}{\spac e^{-i\pi\lambda_{il}}e^{-i\pi\lambda_{jl}}}\right]^\epsilon\, =1+if_{ij}\pi\epsilon -\frac{\pi^2}{2}\epsilon^2+\mathcal{O}(\epsilon^3)\,,
\end{equation}
with 
\begin{equation}
   f_{ij}= \left\{
\begin{array}{l}
\text{$-1$  if $i,j$ both incoming} \\
\text{$\phantom{-}1$ else.} 
\end{array}
\right.
\end{equation}
While the $\propto\pi^2$ terms are independent of the given kinematics, the $\propto i\pi$ terms are not, and one must be careful when applying color conservation and distinguish incoming and outgoing particles. We find
\begin{align}
\label{eq:result_1loop_soft_current_expanded}
  -\sum_{i \neq j} if^{abc}\bm{T}_{i,L}^b\bm{T}_{j,L}^c\bigg(\frac{n_i^\mu}{n_i\cdot l_s} - \frac{n_j^\mu}{n_j\cdot l_s}\bigg)(f_{ij}i\pi\epsilon)
   %  &-4\pi \epsilon f^{abc}\bm{T}_{1,L}^b\bm{T}_{2,L}^c\left(\frac{n_1^\mu}{n_1\cdot l_s} - \frac{n_2^\mu}{n_2\cdot l_s}\right)-2\pi \epsilon\sum_{i}f^{abc}\bm{T}_{i,L}^b\bm{T}_{i,L}^c\frac{n_i^\mu}{n_i\cdot l_s} =\nonumber \\
     = &-4\pi \epsilon f^{abc}\bm{T}_{1,L}^b\bm{T}_{2,L}^c\left(\frac{n_1^\mu}{n_1\cdot l_s} - \frac{n_2^\mu}{n_2\cdot l_s}\right)
     \nonumber\\
    &-iN_c\pi \epsilon\sum_i\bm{T}_{i,L}^a\frac{n_i^\mu}{n_i\cdot l_s}\,.
\end{align} 
Using this result, keeping only terms containing factors of $i\pi$ from the analytic continuation, and disregarding the factor $[\overline{W}_{ij}^l]^\epsilon$, the one loop soft current in \eqref{eq:one_loop_soft_current} can be written as
\begin{align}
\label{eq:soft_current_pi_squared}
   \bm{J}^{\mu,a(1)}= &-g_s
\frac{\Gamma^3(1-\varepsilon)\,\Gamma^2(\varepsilon)}{\Gamma(1-2\varepsilon)}\bigg[
 if^{abc}\sum_{i \neq j}\spac \bm{T}_{i,L}^b\spac\bm{T}_{j,L}^c
\bigg(\frac{n_i^\mu}{n_i\cdot l_s} - \frac{n_j^\mu}{n_j\cdot l_s}\bigg) \Big(1-\frac{\pi^2}{2}\epsilon^2\Big)\nonumber \\
&\qquad+\pi \epsilon f^{abc}\bm{T}_{1,L}^b\bm{T}_{2,L}^c\bigg(\frac{n_1^\mu}{n_1\cdot l_s} - \frac{n_2^\mu}{n_2\cdot l_s}\bigg)+iN_c\pi \epsilon\sum_i\bm{T}_{i,L}^a\frac{n_i^\mu}{n_i\cdot l_s}\bigg]\, .
\end{align}
The following discussion is correct for general hard functions, but it is natural to use the integrated hard functions $\overline{\bm{\mathcal{H}}}$ in the absence of factorization breaking terms, see also the discussion above~\eqref{eq:hard_function_spin_averaged}.
After reinstating all prefactors, multiplying with the tree-level contribution, adding the complex conjugate, and using cyclicity of the trace we find for the leading pole
\begin{equation}
\label{eq:HtimesS2}
   \left(\frac{\alpha_s}{4\pi}\right)^2 \langle \spac \overline{\bm{\mathcal{H}}}\,\otimes\bm{\mathcal{S}}^{(1)}\rangle =-\left(\frac{\alpha_s}{4\pi} \right)^2\frac{8\pi}{\epsilon^2}\sum_{j>2}f^{abc}\langle \spac\overline{\bm{\mathcal{H}}}\spac\bm{T}_1^a\bm{T}_2^b\bm{T}_j^c\rangle \int \left[d\Omega_l\right]\left(W_{1j}^l-W_{2j}^l\right)\Theta_{\mathrm{veto}}(n_l)\, .
\end{equation}
The angular brackets $\langle\cdots \rangle$ denote the color and spin sum over both initial and final states. In addition we omit the $\otimes \bm{1}$ that multiplies the hard function inside the angular bracket here and in the remainder of the section. The term $\propto N_c$ in~\eqref{eq:soft_current_pi_squared} vanishes after the addition of the complex conjugate.  Moreover, the sum is restricted to $j>2$, as the remaining terms cancel with the hermitian conjugate if $j\in\{1,2\}$. This expression coincides with the $\VG\overline{\bm{\Gamma}}$ term in~\eqref{eq:Wm_poles_1}, the predicted contribution from the consistency relation, demonstrating that at this loop order all required phases arise purely from the soft sector. In the literature, these phases are also sometimes referred to as Glauber phases, but in our language these are soft and do not originate from genuine Glauber modes.

As a next step, we want to reproduce the term $\VG\VG\overline{\bm{\Gamma}}$ arising at the three-loop order. Here, we have to consider two types of terms:
\begin{enumerate}[(i)]
    \item the product of two one-loop soft currents, $\bm{J}^{\mu,a(1)} \bm{J}^{a(1)\dagger}_\mu$,
    \item the product of $\bm{J}^{a(2)}_\mu$ with a tree-level current, including both the tripole and dipole terms, and their hermitian conjugates. 
\end{enumerate}
We start our analysis with the product of two one-loop soft currents. The relevant result can be repurposed from~\eqref{eq:soft_current_pi_squared}, but we now require one current to appear on either side of the hard functions. After straightforward color algebra and using the cyclicity of the trace, we obtain
\begin{align}
\label{eq:J1_J1_dagger_final}
   &\int\frac{d^dl_s}{(2\pi)^d}(-ig_{\mu\nu})(-2\pi i\delta_+(l_s^2))\langle \bm{J}^{\mu,a(1)}\spac\overline{\bm{\mathcal{H}}} \spac \bm{J}^{\nu,a(1)\dagger}\rangle \\
   &=\frac{\alpha_s}{4\pi} \frac{2}{3\spac\epsilon^3}\int \left[d\Omega_l\right]\Theta_{\mathrm{veto}}(n_l)\bigg[
   N_c^2\left(-\pi^2+\zeta_2\right)\sum_{i \neq j}W_{ij}^l
   \langle \spac\overline{\bm{\mathcal{H}}}\spac\bm{T}_i^a\bm{T}_j^a\rangle +N_c^2\pi^2\sum_{i \neq j}W_{ij}^l\langle \spac\overline{\bm{\mathcal{H}}}\spac\bm{T}_i^a\bm{T}_j^a\rangle\nonumber \\
   &\hspace{4.6cm}- 8N_c^2\pi^2
   W_{12}^l
   \langle \spac\overline{\bm{\mathcal{H}}}\spac\bm{T}_1^a\bm{T}_2^a\rangle -32\pi^2f^{abc}f^{ab^\prime c^\prime }W_{12}^l\langle \spac\overline{\bm{\mathcal{H}}}\spac\bm{T}_1^b\bm{T}_2^c\bm{T}_1^{b^\prime}\bm{T}_2^{c^\prime}\rangle\bigg].\nonumber
\end{align}

Here, we performed the energy integral, leading to the restriction $\Theta_{\mathrm{veto}}(n_l)$ in the phase-space integral.

Next, we consider the terms arising from the two-loop soft current. At this loop order, contributions arise due to both the dipole terms, involving two light-like reference vectors, and the tripole terms, involving three. 
The dipole term has the same structure as the one-loop current. Using the results in~\cite{Duhr:2013msa,Dixon:2019lnw}, we perform the $\epsilon$ expansion and keep only the terms $\propto\pi^2$. We do not display $i \pi$ terms since at order $\alpha_s^3$, the two-loop current combines with a tree-level soft current and consequently only its real part is relevant. These considerations lead to
\begin{equation}
\bm{J}^{\mu,a(2)}_{\mathrm{dipole}}=g_s\frac{iN_c}{\epsilon^2}\left(-\pi^2+\zeta_2\right)f^{abc}\sum_{i \neq j}\bm{T}_i^{b}\bm{T}_j^c\left(\frac{n_i^\mu}{n_i\cdot l_s}-\frac{n_j^\mu}{n_j\cdot l_s}\right), 
\end{equation}
where the $\pi^2$ terms arise from analytic continuation and $\zeta_2$ from the expansion of the Gamma functions. Not all $\pi^2$ terms in the cross section are associated with Glauber phases. To isolate the Glauber phases, we later compare the expressions obtained in physical kinematics to the time-like case where these are absent. The $\zeta_2$ term, for example, is also present in time-like kinematics and does not contribute to the Glauber terms we are interested in. 
The tripole terms are more involved, see~\cite{Dixon:2019lnw}. We isolate once again only terms $\propto\pi^2/\epsilon^2$, as all other terms cannot give rise to the structure $\VG\VG\overline{\bm{\Gamma}}$. In this case, only a single contribution is relevant, given by
\begin{equation}
   \bm{J}^{\mu,a(2)}_{\mathrm{tripole}}= g_s\frac{8\pi^2}{\epsilon^2}\sum_{i>2}\bm{T}_i^{c}\bm{T}_1^{d}\bm{T}_2^{e}\left(\frac{n_1^\mu}{n_1\cdot l_s}-\frac{n_2^\mu}{n_2\cdot l_s}\right)f^{acb}f^{bde}\,,
\end{equation}
where  all permutations of the three involved partons $i,j,k$ are taken into account. The only non-vanishing terms involve two incoming and one outgoing Wilson line.
This is fully in line with the fact that Glauber phases can only arise in processes with at least two incoming color-charged particles. In order to render this property more manifest, we apply color conservation to rewrite the expression as 
\begin{equation}
    \bm{J}^{\mu,a(2)}_{\mathrm{tripole}}= -g_s\frac{8\pi^2}{\epsilon^2}f^{acb}f^{bde}\bm{T}_1^{d}\bm{T}_2^{e}\left(\bm{T}_1^{c}+\bm{T}_2^{c}\right)\left(\frac{n_1^\mu}{n_1\cdot l_s}-\frac{n_2^\mu}{n_2\cdot l_s}\right)\,,
\end{equation}
which explicitly demonstrates that this term is only present for hadron colliders, since it only contains contributions from colored incoming legs. Performing the same steps as for the product of two one-loop currents we arrive at
\begin{align}
\label{eq:tripole_final}
    &\int\frac{d^dl_s}{(2\pi)^d}(-ig_{\mu\nu})(-2\pi i\delta_+(l_s^2))\left(\langle \bm{J}^{\mu,a(2)}_{\mathrm{tripole}}\,\overline{\bm{\mathcal{H}}}\spac\bm{J}^{\nu,a(0)\dagger} \rangle +\mathrm{h.c.} \right) \nonumber \\
    &= \frac{\alpha_s}{4\pi}\frac{2\pi^2}{3\epsilon^3}\int \left[d\Omega_l\right]\Theta_{\mathrm{veto}}(n_l)\bigg[8N_c^2W_{12}^l
    \langle \spac\overline{\bm{\mathcal{H}}}\spac\bm{T}_1^a\bm{T}_2^a\rangle+32f^{abc}f^{ab^\prime c^\prime }W_{12}^l
    \langle \spac\overline{\bm{\mathcal{H}}}\spac\bm{T}_1^b\bm{T}_2^c\bm{T}_1^{b^\prime}\bm{T}_2^{c^\prime}\rangle \nonumber\\
    &\quad-8f^{abe}f^{cde}\sum_{j>2} \left(W_{1j}^l-W_{2j}^l\right) \left\langle\spac\overline{\bm{\mathcal{H}}}
    \left(\bm{T}_2^a\left\{
    \bm{T}_1^b,\bm{T}_1^c\right\}-\bm{T}_1^a\left\{\bm{T}_2^b,\bm{T}_2^c\right\}\right) \bm{T}_j^d\right\rangle \bigg]\, .
\end{align}
The second line in this expression gets canceled by the second line in~\eqref{eq:J1_J1_dagger_final}, such that we arrive at
\begin{align}
     &\int\frac{d^dl_s}{(2\pi)^d}(-ig_{\mu\nu})(-2\pi i\delta_+(l_s^2))\left(\langle \bm{J}^{\mu,a(2)}\spac\overline{\bm{\mathcal{H}}}\spac\bm{J}^{\nu,a(0)\dagger} \rangle +\mathrm{h.c.} +\langle \bm{J}^{\mu,a(1)}\spac\overline{\bm{\mathcal{H}}}\spac\bm{J}^{\nu,a(1)\dagger}\rangle\right) \nonumber \\
     &=
   \frac{\alpha_s}{4\pi} \frac{2\pi^2}{3\epsilon^3}\int \left[d\Omega_l\right]\Theta_{\mathrm{veto}}(n_l)\bigg[\Big(3N_c^2\Big[-1+\frac{\zeta_2}{\pi^2}\Big]+N_c^2\Big)\sum_{i \neq j}W_{ij}^l
   \langle \spac\overline{\bm{\mathcal{H}}}\spac \bm{T}_i^a\bm{T}_j^a\rangle\nonumber\\
   &\hspace{1.5cm}-8f^{abe}f^{cde}\sum_{j>2} \left(W_{1j}^l-W_{2j}^l\right) \left\langle \spac\overline{\bm{\mathcal{H}}}
    \left(\bm{T}_2^a\left\{
    \bm{T}_1^b,\bm{T}_1^c\right\}-\bm{T}_1^a\left\{\bm{T}_2^b,\bm{T}_2^c\right\}\right) \bm{T}_j^d\right\rangle 
    \bigg]\, . 
\end{align}
Note that the second line is agnostic about the kinematics, as the sum simply goes over all $i\neq j$. The same contributions also arise in $e^+e^-$ colliders, where Glauber phases are absent. Consequently, these terms do not contribute to quantities proportional to $\sim\VG\VG\overline{\bm{\Gamma}}$ and can be omitted in our analysis. Then, the final result is given by
 \begin{align}
\label{eq:vgvggammabar}
    \left(\frac{\alpha_s}{4\pi}\right)^3\langle \spac\overline{\bm{\mathcal{H}}} \,\otimes  \bm{\mathcal{S}}^{(2)}\rangle &=-\frac{\pi^2}{\epsilon^3}\frac{16}{3}\left(\frac{\alpha_s}{4\pi} \right)^3f^{abe}f^{cde}\int \left[d\Omega_l\right]\Theta_{\mathrm{veto}}(n_l)  \\
   & \quad \times \sum_{j>2}\left(W_{1j}^l-W_{2j}^l\right)\left\langle\spac 
   \overline{\bm{\mathcal{H}}} \left(\bm{T}_2^a\left\{\bm{T}_1^b,\bm{T}_1^c\right\}\bm{T}_j^d-\bm{T}_1^a\left\{\bm{T}_2^b,\bm{T}_2^c\right\}\bm{T}_j^d\right)\right\rangle \nonumber\,,
 \end{align}
which exactly reproduces the predicted result $\propto \VG\VG\overline{\bm{\Gamma}}$ in~\eqref{eq:Wm_poles_1}, see (6.14) in~\cite{Becher:2023mtx}.

In~\cite{Dixon:2019lnw} it was explained that the tripole terms break strict collinear factorization in the space-like limit. For pure QCD processes, this effect vanishes at the cross section level. While these phases are not directly factorization breaking at this loop order, the corresponding terms  are nevertheless crucial, since they exactly match the hard evolution terms $\VG\VG\overline{\bm{\Gamma}}$. 
As a result, all predicted purely-soft terms in~\eqref{eq:Wm_poles_1} are accounted for in the low-energy matrix elements at the perturbative scale $Q_0$, consistent with PDF factorization.

\section{Soft-collinear matrix elements}
\label{sec:soft_collinear_matrix_element}
Having reproduced all soft contributions in the bare matrix elements in the prediction of~\eqref{eq:Wm_poles}, the following two structures are left at three-loop order
\begin{equation}\label{eq:Wm_poles_coll}
   \bm{\mathcal{I}}^{\mathrm{bare}}=
     \left( \frac{\alpha_s}{4\pi} \right)^3\! \bigg[ \frac{\bm{\Gamma}^{c}\spac\bm{V}^G\overline{\bm{\Gamma}}}{3\varepsilon^3} \!
     \left( \frac{11}{6\varepsilon} + \ln\frac{\mu^2}{\mu_h^2} 
     + \frac92\ln\frac{\mu^2}{Q_0^2} \right)+\frac{\bigl[\bm{\Gamma}^C,\bm{V}^G\overline{\bm{\Gamma}}\bigr]}{12\varepsilon^3}\color{black}+ \dots \bigg] + \mathcal{O}(\alpha_s^4) \,.
\end{equation}
The first one is the cusp contribution involving the logarithm of the hard scale, which has to arise from the low-energy matrix elements to reconcile the double-logarithmic running above $Q_0$ with the single-logarithmic DGLAP evolution.
The second commutator turns the color-aware collinear running into the usual color-diagonal DGLAP running below the scale $Q_0$. The first term was analyzed in our previous paper~\cite{Becher:2024kmk}, where we demonstrated that a soft-collinear interaction mediated by a genuine Glauber mode precisely reproduces this structure. Here, we provide more details on this computation and extend it to include the second structure. In contrast to the cusp term, the second contribution also includes off-diagonal splittings, where a quark turns into a gluon, or vice versa.

\begin{figure}
    \centering
    \includegraphics[width=\textwidth]{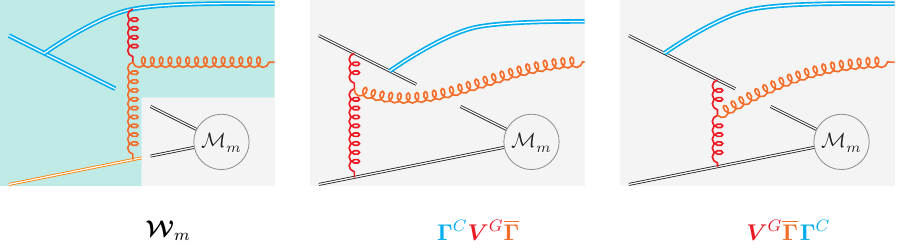}
    \caption{Relation between the color structure of the hard functions (gray shading) and the low-energy matrix elements (blue shading). Blue double lines stand for collinear partons. Glauber contributions are shown in red and soft emissions in orange. The black double lines can stand for hard partons, the orange one for the associated Wilson line.}
    \label{fig:color_commutator}
\end{figure}

As a first step, in Sections~\ref{sec:color_diagonal_splitting} and~\ref{sec:off_diagonal_splittings}, we analyze the color structure of $\left[\bm{\Gamma}^C,\bm{V}^G\overline{\bm{\Gamma}}\right]$ in detail and bring it into a suitable form for comparison with the perturbative calculation of the low-energy matrix elements. While the individual contributions have a different form than the low-energy matrix elements, the commutator as whole arranges itself into the correct structure. This is illustrated in Figure~\ref{fig:color_commutator}, where we display the contributions to the low-energy matrix elements on a blue background, while the hard functions after the action of products of hard anomalous dimensions are indicated by the gray areas. The double lines can be either quarks or gluons, depending on the collinear splitting under consideration.  Evaluating the color structure of the three diagrams, we find that they are inherently connected by color coherence. 

Crucially, we also show that attachments of soft radiation coming from $\overline{\bm{\Gamma}}$ to the additional collinear emissions from $\bm{\Gamma}^C$ vanish. This is important since in the low-energy theory soft emissions cannot directly couple to collinear partons. A complication compared to our earlier results is the generalization to initial-state off-diagonal splittings, where the color-space formalism is not suitable.

Once this is accomplished, we move to the explicit computation of the low-energy matrix elements in Sections~\ref{sec:low-energy-matrix-element} and~\ref{sec:computation-low-energy-matrix-elements}. We briefly review the findings of~\cite{Becher:2024kmk} that the soft-collinear interactions are mediated through a genuine Glauber gluon exchange between soft and collinear partons. Next, we perform the perturbative calculation of
the low-energy matrix elements directly in SCET using the Glauber Lagrangian of~\cite{Rothstein:2016bsq}. In our case, however, all Glauber contributions are genuine and well-defined in dimensional regularization alone. We only introduce a rapidity regulator for the collinear phase-space integral.
We compute all involved three-loop diagrams, and verify the consistency with PDF factorization for both diagonal and off-diagonal splittings. 

This Glauber mode introduces a dependence on soft momenta, leading to non-trivial tensor structures in the low-energy matrix elements connecting to hard functions with gluons in the initial state. A simple spin projection with the polarization sum of the gluons is therefore no longer sufficient.
This justifies the necessity of the double-bracket notation in~\eqref{eq:factorizationTheoremNew} for color \emph{and} spin, which only involves the respective sums over final state partons, but \emph{not} over the initial-state ones. 
However, the leading poles arise from the one-loop anomalous dimensions, which are diagonal in color-helicity space.
For these terms, one can employ the single-bracket, which also includes spin and color average over the initial-state partons, to analyze their color structure in Sections~\ref{sec:color_diagonal_splitting} and~\ref{sec:off_diagonal_splittings}.
It is therefore convenient to again use the integrated hard functions $\overline{\bm{\mathcal{H}}}$ in these two sections.
\subsection{Color structure for diagonal splittings}
\label{sec:color_diagonal_splitting}

In the case of an additional emitted (anti-)collinear gluon, we may directly use the color-space formalism, also for the initial-state partons. Let us consider the two terms in the commutator \eqref{eq:Wm_poles_coll} in turn. 
For diagonal splittings we use~\cite{Becher:2023mtx}
\begin{equation}
\label{eq:diag_splitting_result_old}
  \langle{\spac\overline{\bm{\mathcal{H}}}}\spac\bm{\Gamma}^C_i \VG \overline{\bm{\Gamma}} \rangle=32\pi \left(\overline{\mathcal{P}}_{i\to i}-2\ln\frac{\mu_h}{2E_i}\right)(N_c+2C_i)f^{abc}{\sum_{j>2}}^{\,\prime} J_j\langle \spac{\overline{\bm{\mathcal{H}}}}\spac\bm{T}_1^a\bm{T}_2^b\bm{T}_j^c\rangle  \, ,
\end{equation}
where the prime on the sum indicates 
that the soft emission $j$ does not attach to the collinearly emitted gluon 
produced by $\GammaC_i$, and parton $i$ denotes a (anti-)quark or a gluon. The angular integral in the veto region $J_j$ is given by
\begin{align}
\label{eq:def_Jj}
    J_j\equiv\int\frac{d\Omega(n_l)}{4\pi}\left(W_{1j}^l-W_{2j}^l\right)\Theta_\text{veto}(n_l)\, ,
\end{align}
where the soft dipole $W_{ij}^l$ is found in~\eqref{eq:def_soft_dipole}. Here, and in the following, we omit the $\otimes$-symbol and leave the Mellin convolution implicit. Additionally, we suppress the color-diagonal virtual part of the collinear and cusp anomalous dimensions, as they cancel within the commutator. 
Similarly, if the collinear anomalous dimension stands at the end we arrive at
\begin{equation}
\langle{\spac\overline{\bm{\mathcal{H}}}}\VG\overline{\bm{\Gamma}}\bm{\Gamma}^C_i\rangle =  64\pi\! \left(\overline{\mathcal{P}}_{i\to i}-2\ln\frac{\mu_h}{2E_i}\right)\spac  C_i f^{abc}{\sum_{j>2}}^{\,\prime}J_j \langle\spac{\overline{\bm{\mathcal{H}}}}\spac \bm{T}_1^a\bm{T}_2^b\bm{T}_j^c\rangle  \, .
\end{equation}
The commutator is then given by
\begin{equation}
\label{eq:diagonal_split_c_factor}
\langle\spac\overline{\bm{\mathcal{H}}}\bigl[\bm{\Gamma}^C_i,\VG\overline{\bm{\Gamma}}\bigr]\rangle=32\pi\!\left(\overline{\mathcal{P}}_{i\to i}-2\ln\frac{\mu_h}{2E_i}\right) N_c\spac f^{abc}{\sum_{j>2}}^{\,\prime}J_j\langle \spac\overline{\bm{\mathcal{H}}}\spac\bm{T}_1^a\bm{T}_2^b\bm{T}_j^c\rangle   \, ,
\end{equation} 
such that we recover the structure of the SLL terms. This is not surprising because we can also write the SLL contribution at this order as 
\begin{equation}
\langle{\spac\overline{\bm{\mathcal{H}}}}\spac\bm{\Gamma}^c\VG\overline{\bm{\Gamma}}\rangle = \langle{\spac\overline{\bm{\mathcal{H}}}}\bigl[\bm{\Gamma}^c,\VG\overline{\bm{\Gamma}}\bigr]\rangle\, ,
\end{equation}
as the cusp-anomalous dimension vanishes when left on the right-hand side. In addition, in~\cite{Becher:2024kmk} it was shown  that only a \textit{single} type of Glauber diagram contributes to the diagonal splittings, with a single color structure. Therefore,~\eqref{eq:diagonal_split_c_factor} has indeed to be of the same form as the SLL terms, consistent with PDF factorization.

\subsection{Color structure for off-diagonal splittings}
\label{sec:off_diagonal_splittings}
When emitting a (anti-)quark we can no longer work in the color-space formalism and therefore need to be specific which splitting we consider. In the main text, we treat the $q\to g$ case, the $g\to \bar{q}$ splitting is discussed in Appendix~\ref{app:sec:g_to_q}.

\begin{figure}
    \centering
    \includegraphics[scale=1]{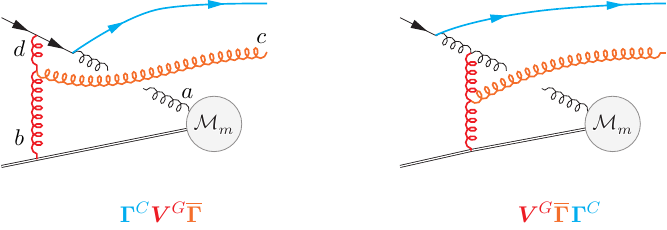}
    \caption{Explicit color structure of the commutator for the $q\to g$ splitting at the three-loop order. When the soft emission attaches to the collinear emission, the respective contribution vanishes. The color indices as used below are explicitly shown in the left-hand diagram with $\bar{a}$, not shown here, connecting to the conjugate hard amplitude.}
    \label{fig:color_commutator_q_to_g}
\end{figure}

As a first step, we show that the color structure does not allow for a soft emission off the collinear emission. In~\cite{Becher:2023mtx} this was shown explicitly for the diagonal case, and it allowed restricting the sum in~\eqref{eq:diag_splitting_result_old}. As we show in the next section, this condition is necessary, as no direct coupling between soft and collinear partons is possible within the low-energy theory.  The relevant color structure for the $q\to g$ splitting is given by
\begin{equation}
    \langle {\spac\overline{\bm{\mathcal{H}}}\spac} {\bm{\Gamma}^C_i}\VG\overline{\bm{\Gamma}}\rangle=-16\pi f^{dbc}\sum_{j>2} J_j\langle{\spac\overline{\bm{\mathcal{H}}}} \spac\bm{\Gamma}^C_i\bm{T}_1^d\bm{T}_2^b\bm{T}_j^c\rangle\, ,
\end{equation}
and the case where ${\bm{\Gamma}^C_i}$ and $\VG$ are interchanged is not relevant, as $\left[{\bm{\Gamma}^C},\overline{\bm{\Gamma}}\right]=0$ holds.
When the soft emission $\bm{T}_j^c$ couples to the new quark produced by ${\bm{\Gamma}^C_i}$, we arrive at a structure
\begin{equation}
 \propto t^dt^{\bar{a}}t^ct^af^{dbc}\bm{T}^b_2\, ,
\end{equation}
where $a,\bar{a}$ are contracted with the open color indices of the hard functions.
To arrive at this form, we have to carefully consider where the soft emission takes place (e.g.\ between the two collinear color operators) and which explicit color operator is used for the Glauber phase.
In this case, we used $\bm{T}_1^d=-t^d$, where the minus sign arises from particle $1$ being a quark, as the Glauber operates \textit{before} the splitting, i.e.\ on hard functions with initial state quarks, not gluons. This can be explicitly seen in Figure~\ref{fig:color_commutator_q_to_g}.
After the collinear splitting, the new hard functions, which the above expression has to be multiplied with, only have incoming gluons. Therefore, the trace over these matrices can be directly performed 
\begin{equation} \Tr\!\left[t^dt^{\bar{a}}t^ct^a\right]f^{dbc}\bm{T}^b_2=0\, ,
\end{equation}
which is indeed vanishing, as can be easily  verified. Therefore, no soft emission from the collinearly emitted quark is allowed and the sum can again be restricted to run over the final states. Furthermore, the above arguments also hold for the $\bar{q}\to g$ splitting.
 
As a next step, we explicitly calculate the structures from the commutator. 
The color structure of the first term in the commutator evaluates to
\begin{equation}
\label{eq:qgqcollinear}
    \langle{\spac\overline{\bm{\mathcal{H}}}}\spac\bm{\Gamma}^C_q(z_1) \VG \overline{\bm{\Gamma}}\rangle=\frac{N_c^2-1}{N_c}64 \pi\spac\overline{\mathcal{P}}_{q\to g}(z_1)f^{dbc}\left(\Tr\! \left[t^dt^at^{\bar{a}}\right]+i\spac T_Ff^{d{\bar{a}}a}\right){\sum_{j>2}}^{\,\prime}J_j\langle{\spac\overline{\bm{\mathcal{H}}}}\spac\bm{T}^b_2\bm{T}^c_j\rangle\, ,
\end{equation}
where we used that $\bm{T}_1^d=-t^d$ in this case, as can be read off from Figure~\ref{fig:color_commutator_q_to_g}.
Note that, computed diagrammatically, the trace appears flipped, and we used
\begin{equation}
    \Tr\!\left[t^{\bar{a}}t^a t^d\right] = \Tr\! \left[t^dt^at^{\bar{a}}\right]+i\spac T_Ff^{d{\bar{a}}a}\,.
\end{equation}
The additional color factor $(N_c^2-1)/N_c$ changes the color trace to the correct normalization for a gluonic initial state, as required for the single-bracket notation which includes the color average over the respective initial-state partons. As shown above, we can restrict the sum to final state partons without emissions off the additional emitted collinear (anti)-quark, denoted by the prime on the sum.
The second structure similarly yields
\begin{equation}
     \langle{\spac\overline{\bm{\mathcal{H}}}} \VG \overline{\bm{\Gamma}}\bm{\Gamma}^C_q(z_1)\rangle= - \frac{N_c^2-1}{N_c}  64 \pi \spac T_F\overline{\mathcal{P}}_{q\to g}(z_1)f^{dbc}{\sum_{j>2}}^{\,\prime}J_j\langle \spac\overline{\bm{\mathcal{H}}}\spac\bm{T}_1^d\bm{T}_2^b\bm{T}_j^c \rangle\, ,
\end{equation}
where the factor $T_F$ in the numerator arises from $\bm{\Gamma}^C$. 
Again, we have to consider on which field the operators act, i.e.\ $\bm{T}_1^d=-if^{d{\bar{a}}a}$ in this case, as the Glauber anomalous dimension acts \textit{after} the collinear splitting, so on hard functions with incoming gluons. This leads to
\begin{equation}
\label{eq:qgqcollinear_trivial}
    \langle\spac{\overline{\bm{\mathcal{H}}}} \VG \overline{\bm{\Gamma}}\bm{\Gamma}^C_q(z_1)\rangle= i \frac{N_c^2-1}{N_c}\spac64 \pi  \spac T_F\overline{\mathcal{P}}_{q\to g}(z_1)f^{dbc}f^{d{\bar{a}}a}{\sum_{j>2}}^{\,\prime}J_j\langle \spac\overline{\bm{\mathcal{H}}}\spac\bm{T}_2^b\bm{T}_j^c\rangle  \,.
\end{equation}
Combining both results, the last term in~\eqref{eq:qgqcollinear} exactly cancels against the contribution in~\eqref{eq:qgqcollinear_trivial} such that
\begin{equation}  
\label{eq:q_to_g_expected_color}\frac{1}{12}\langle\spac{\overline{\bm{\mathcal{H}}}}\bigl[\bm{\Gamma}^C_q(z_1),\VG\overline{\bm{\Gamma}}\bigr]\rangle=\frac{16 \pi}{3}\frac{N_c^2-1}{N_c}\overline{\mathcal{P}}_{q\to g}(z_1)f^{dbc}\spac\Tr\! \left[t^dt^at^{\bar{a}}\right]{\sum_{j>2}}^{\,\prime}J_j\langle\spac{\overline{\bm{\mathcal{H}}}}\spac  \bm{T}^b_2\bm{T}^c_j\rangle\, .
\end{equation}
For the analogous case $\bar{q}\to g$ we get an additional minus sign and $a\leftrightarrow {\bar{a}}$.
The compactness of the color structure in~\eqref{eq:q_to_g_expected_color} is quite remarkable and reflects an underlying symmetry. Diagrammatically, it can be seen that the two structures arising from $\left[\bm{\Gamma}^C,\VG\overline{\bm{\Gamma}}\right]$ are directly connected to a third diagram due to color conservation, where the Glauber phase attaches to the collinear emission. This argument can be generalized to all splittings, as shown in Figure~\ref{fig:color_commutator}: on the left we see the expected color structure for the soft-collinear matrix elements, which corresponds to the two diagrams coming from the commutator.

\subsection{Analysis of the low-energy matrix elements}
\label{sec:low-energy-matrix-element}
\begin{figure}
    \centering
    \includegraphics[]{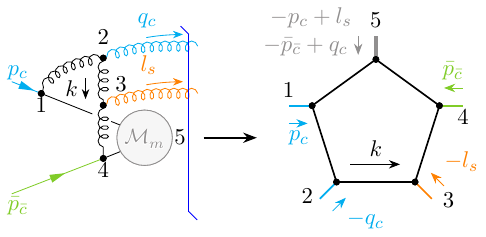}
    \caption{Example graph illustrating the mapping of a diagram contributing to $\bm{\mathcal{I}}_m$ to a pentagon. The collinear lines are depicted in blue, the anti-collinear ones in green and the soft emission into the gap is orange. The scaling of the internal momentum $k$ is left open.}
    \label{fig:pentagon_1}
\end{figure}
As a next step, we analyze the structure of diagrams contributing to the given process for the low-energy matrix elements. At the three-loop order, from the consistency relation~\eqref{eq:Wm_poles_coll} we need to reproduce terms with structure $\sim\left[\bm{\Gamma}^C,\VG\overline{\bm{\Gamma}}\right]$, such that we can focus on a few classes of diagrams:
\begin{enumerate}[(i)]
    \item $\overline{\bm{\Gamma}}$ describes a soft emission into the gap. This provides the necessary scale $Q_0$ for the problem and consequently we need to consider such an emission also from the low-energy matrix elements. 
    \item To reproduce both the hard logarithms $\ln Q$, as well as the non-trivial momentum fraction dependence, we require modes which are sensitive to collinear energies and splittings. Because all final-state directions are integrated out and $\VG$ is inherently connected to space-like splittings, this implies that we look at space-like collinear splittings in the low-energy matrix elements too.
    \item Last but for sure not least we need a virtual contribution producing a phase $\sim \VG$ connecting the two sectors above. This contribution is essential, as collinear matrix elements alone are scaleless, as they cannot resolve the gap region. Thus, when integrated over the perpendicular momentum, they vanish. This implies the existence of an additional mode which breaks factorization between soft and collinear physics---a genuine Glauber mode.
\end{enumerate}
The conditions in (i) and (ii) take up two powers $\alpha_s$ and set the virtualities, i.e.\ one emission with a soft ($l_s=(n\cdot l_s,\nb\cdot l_s,l_{s\perp})\sim (\lambda,\lambda,\lambda)$) and one contribution involving a collinear scaling ($q_c\sim(\lambda^2,1,\lambda)$).\footnote{
We remark that since the only scale within the low-energy matrix elements is $Q_0$, it is not a priori clear what the relative scaling between the soft and collinear modes should be.
This means that it is not obvious if the correct description is in terms of SCET$_{\mathrm{I}}$, featuring hard-collinear and ultrasoft modes with different virtualities, or SCET$_{{\mathrm{II}}}$ where collinear and soft modes have the same virtuality.
However, a detailed method of regions analysis shows that in SCET$_{\mathrm{I}}$, the only appearing virtual modes are ultrasoft and decouple, and the collinear contributions thus remain scaleless.
This verifies that SCET$_{{\mathrm{II}}}$ is indeed the correct description of the underlying physics.} These contributions can be seen in the left-hand side diagram of Figure~\ref{fig:pentagon_1}.
The phase contribution discussed in (iii) should appear as the virtual part of the three-loop contribution, i.e.\ in the loop in Figure~\ref{fig:pentagon_1}, and finding the associated mode presented a challenge. In~\cite{Becher:2024kmk} we showed, using a detailed method-of-regions~\cite{Beneke:1997zp} analysis, that such a hidden, genuine Glauber mode with momentum $k$ indeed appears, connecting the soft and (anti-)collinear sectors, with its components scaling as $k\sim Q(\lambda^2,\lambda,\lambda)$. We achieved this by mapping the loop-integral of the contributions to the soft-collinear matrix elements onto pentagon diagrams, as shown in Figure~\ref{fig:pentagon_1}. By expanding the known results for these pentagons~\cite{Bern:1993kr} in powers of $\lambda$ we directly compared the regions' result to the full expression, ascertaining that no region can be missed. In addition to this Glauber mode, we found a soft-collinear mode, scaling as $k_{sc}\sim Q(\lambda^2,\lambda,\lambda^{3/2})$, which is the only contributing mode in the Euclidean kinematics. In contrast to the Glauber mode, this contribution vanishes when combined with the collinear integrals: The perpendicular momentum of the soft-collinear mode cannot probe the respective collinear momentum and cannot connect the sectors. The collinear integrals therefore remain scaleless in this case.

Furthermore, we showed that, in accordance with the Landau equations~\cite{Ma:2023hrt,Gardi:2024axt}, large cancellations occur within the $\mathcal{F}$ polynomial. These lead to the Glauber region which is hidden inside the Newton polytope of the graph, rather than living on its facets such as e.g.\ soft and collinear momentum regions. The appearance of this genuine Glauber region is a direct consequence of a double pinch, for which the associated contour cannot be deformed in such a way as to avoid it. This also implies that so-called $0$-bin subtractions, which are relevant when a double-counting of regions occurs, are not necessary in this specific case. Further details on the integral under consideration can be found in~\cite{Becher:2024kmk} and an overview of the general theory of asymptotic expansion of loop integrals is given in~\cite{Ma:2025emu}. 
Interestingly,  similar pentagon structures appear in studies of high-energy limit~\cite{Buccioni:2024gzo} and $\pi^0\gamma$ production~\cite{Nabeebaccus:2023rzr,Nabeebaccus:2024mia}.  

\begin{figure}
    \centering
    % Top single figure
    \begin{subfigure}{0.3\textwidth}
        \centering
        \includegraphics[width=\textwidth]{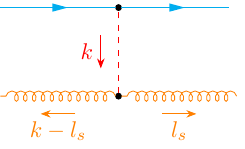}
        \caption{}
    \end{subfigure}

    \vspace{0.cm} % vertical space between top and bottom row

    % Bottom row: three figures side by side
   \centering
    \begin{subfigure}{0.3\textwidth}
    \centering
        \includegraphics[]{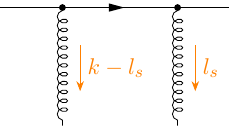}
         \caption{}
    \end{subfigure}
    \begin{subfigure}{0.3\textwidth}
    \centering
        \includegraphics[]{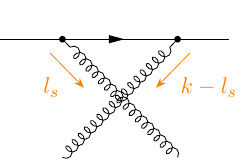}
         \caption{}
    \end{subfigure}
    \begin{subfigure}{0.3\textwidth}
    \centering
        \includegraphics[]{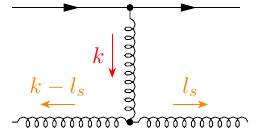}
        \caption{}
    \end{subfigure}

    \caption{Diagram (a) shows the Glauber vertex in the effective theory. All corresponding diagrams in QCD, (b)-(d), exhibit a genuine Glauber region.}
\label{fig:matchingqqggvertex}
\end{figure}

Having found the hidden region in Figure~\ref{fig:pentagon_1}, which already directly matches the color structure of the commutator term analyzed in Section \ref{sec:color_diagonal_splitting}, let us  discuss what other diagrams could contribute. If the virtual gluon line carrying internal momentum $k$ connects directly to the collinear quark instead of the emitted gluon, no Glauber region arises and the respective contributions do not have a  scale in the collinear phase-space integral of the low-energy theory. Contributions from diagrams with a collinear virtual correction, where the collinear line carrying momentum $q_c$ does not cross the cut but rather connects back to the quark line on the amplitude side, are similarly scaleless. In addition to Figure~\ref{fig:pentagon_1}, there are contributions from diagrams where the two soft gluons carrying $l_s$ and $k-l_s$ directly connect to the collinear emission into the final state. These diagrams do not have a propagator which carries a Glauber momentum $k$, but they nevertheless exhibit a Glauber region. In the low-energy theory, all these contributions can be obtained from an effective Glauber vertex connecting the soft gluons to the collinear line \cite{Rothstein:2016bsq}. For the case where the collinear line is a quark, the individual QCD contributions to the effective vertex are shown in Figure~\ref{fig:matchingqqggvertex}. Remarkably, in the effective theory, the entire soft-collinear interaction at three loops therefore arises from the single diagram shown in Figure~\ref{fig:q_to_q_three}.

\subsection{Computation of low-energy diagrams}
\label{sec:computation-low-energy-matrix-elements}

Based on this method-of-regions analysis we can now consider specific diagrams with Glauber scaling, which contribute to the given process and compute them within SCET, using its Glauber extension~\cite{Rothstein:2016bsq}.
Note, however, that we do not need the additional non-analytic regulator, as the relevant integral is well-defined in dimensional regularization alone in our case. If one used this non-analytic regulator, one would also need to consider the additional $0$-bin subtractions.
In the main text we consider the diagonal splitting $q\to q$ to highlight the appearance of SLLs and additional $1/\epsilon$ poles. We also provide a detailed computation of the off-diagonal channel $q\to g$, where additional Lorentz structures appear in sub-leading poles. The $g\to g$ and $g\to \bar{q}$ channels are considered in Appendix~\ref{sec:app:three_loop}.

\begin{figure}
    \centering
    \includegraphics[]{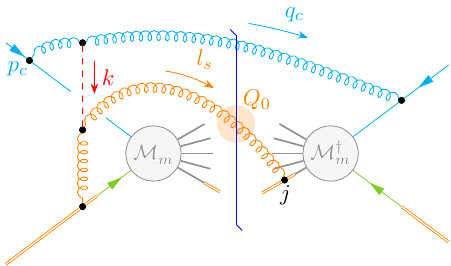}
    \caption{Diagram showing the diagonal splitting of $q\to q$, where the Glauber mode depicted in red mediates between the soft and collinear partons. The soft emission is radiated into the gap, highlighted by the orange circle. This diagrams also gives a contribution to SLLs.}
    \label{fig:q_to_q_three}
\end{figure}

The operator definition of the soft-collinear matrix elements is given in~\eqref{eq:I_function_definition}
\begin{align}
   &[\bm{\mathcal{I}}_{ij\to m}^{kl}]{\substack{ab|\bar{a}\bar{b}\\[-0.68mm]\alpha\beta|\bar{\alpha}\bar{\beta}}}=
   \int_{-\infty}^{\infty} \frac{dt_1}{2\pi}\,e^{- i x_1 t_1 \bar{n}_1\cdot P_1}  \int_{-\infty}^{\infty} \frac{dt_2}{2\pi}\,e^{-i x_2 t_2 \bar{n}_2\cdot P_2} 
   {} \hspace{0.15cm}\int\limits_{X}\hspace{-0.54cm}\sum\,\, \theta( Q_0 - E^\perp_{\mathrm{\, out}}) \\
   &\times
   \langle  k(p_c) l(\bar{p}_{\bar{c}})  |\, \bar{\Phi}_{i}^{\bar{\alpha}\bar{c}}(t_1 \bar{n}_1) \, \bar{\Phi}_{j}^{\bar{\beta}\bar{d}}(t_2 \bar{n}_2) \, [\bm{S}_1^\dagger]^{\bar{c}\spac\bar{a}} \,[\bm{S}_2^\dagger]^{\bar{d}\spac\bar{b}}\, \bm{S}_3^\dagger\, \dots\,  \bm{S}_{m+2}^\dagger\,  |X \rangle \nonumber \\
   &\times \langle X | \,[\bm{S}_1]^{ac}\,[\bm{S}_2]^{bd} \,\bm{S}_3\,  \dots\,  \bm{S}_{m+2}\, \Phi_{i}^{\alpha c}(0) \, \Phi_{j}^{\beta d}(0) \, |k(p_c) l_2(\bar{p}_{\bar{c}}) \rangle \nonumber\,.
\end{align}
Whereas in the previous sections we focused on the purely soft contributions to this expression, arising from the soft Wilson lines alone, we now consider the soft-collinear factorization-breaking contributions. 
The relevant terms for the prediction~\eqref{eq:Wm_poles_coll} at three-loop level arise from matrix elements featuring
\begin{itemize}
    \item one collinear Lagrangian insertion on each side of the cut,
    \item one Glauber operator connecting (anti-)collinear and soft sectors, and
    \item one soft emission connecting to a Wilson line.
\end{itemize}
As the Glauber operator at this loop order can either connect collinear and soft or anti-collinear and soft, the contribution can be written as
\begin{equation}
    \bm{\mathcal{I}}_{i_1 i_2\to m}^{(3)\,k,l\,\mathrm{bare}} \ni \sum_{j} \Bigl(D_{i_1/k}^j(z_1) \Pi_{i_2/l}(z_2) + \Pi_{i_1/k}(z_1) D_{i_2/l}^j(z_2)  + \mathrm{h.c.} \Bigr)\,.
\end{equation}
One example diagram is given in Figure~\ref{fig:q_to_q_three} for the diagonal $q\to q$ splitting.
The objects $D_{i_1/k}^j(z_1)$ carry open color and spin indices connecting to the hard functions, while the initial state partons are averaged in both color and spin.
The functions $D_{i_1/k}^j(z_1)$ can be thought of as the upper half of the diagram containing the non-trivial Glauber interaction on leg $1$.
The color and direction of the second leg can be probed through this Glauber exchange and is located inside $D_{i_1/k}^j(z_1)$ as well.
It therefore depends on the collinear dynamics of leg $1$ and the color charge and direction of leg $2$.
The lower half of the diagram, i.e.\ the trivial second leg, is described by $\Pi_{i_2/l}(z)$, which is  the tree-level part of the partonic matrix elements
\begin{align}
    \Pi_{i_2/l}(z)&=\SumInt_X \delta\!\left( \nb_{2}\cdot\left(k-(1-z)\bar{p}_{\bar{c}}\right)\right)\langle l(\bar{p}_{\bar{c}})\lvert \bar{\Phi}_{i_2\bar{\alpha}}^{\bar{a}}(0)\rvert X\rangle \langle X\lvert \Phi_{i_2\alpha}^a(0)\rvert l(\bar{p}_{\bar{c}})\rangle\,\nonumber\\[2mm]
    &= \frac{1}{\nb_2\cdot \bar{p}_{\bar{c}}}\delta(1-z)
    \begin{dcases}
        \frac{\delta^{a\bar{a}}}{N_c}\biggl[\frac{\nb_2\cdot \bar{p}_{\bar{c}}}{2}\frac{\slashed{n}_2}{2}\biggr]_{\alpha\bar{\alpha}} & \text{for quarks,}\\
        \frac{\delta^{a\bar{a}}}{N_c^2-1}\biggl[-\frac{g_{\perp}^{\alpha\bar{\alpha}}}{d-2} \biggr] & \text{for gluons,}
    \end{dcases}
\end{align}
and vice-versa for the other leg. This produces the correct spin sums for the respective second leg going into the hard functions.
We now calculate the contributions $D_{i_1/k}^j(z_1)$ for each splitting separately.

We start with the diagonal quark channel, as depicted in Figure~\ref{fig:q_to_q_three}.
To keep the notation compact, we now abbreviate $n= n_1$ and $\bar{n}= n_2$ and use the standard normalization $n\cdot\bar{n} =2$. 
Since the contribution would be scaleless without the Glauber mode, choosing a smart order of integrations allows the three-loop computation to factorize into three independent single-scale calculations. First, the $n\cdot k $ and $\nb\cdot k$ integrations can be performed via residues.
Then the collinear phase-space integration is performed.
However, the integral over $q_c^-\equiv\bar{n}\cdot q$ is not well-defined, and following~\cite{Becher:2011dz} we employ a phase-space regulator $(\nu/q_c^-)^{\alpha}$. 
In contrast to~\eqref{eq:RapidityRegulator} this is an analytic regulator. It does not affect the Glauber loop, which is well-defined in dimensional regularization alone. Then the integration over $k_\perp$ is performed to arrive at the result above.
The left-over soft phase-space integral is expressed in terms of the transverse energy $E_l^T = E_l\sin(\theta)$ and the $(d-2)$-dimensional angular integral, yielding
\begin{align}
\label{eq:q_to_q_three_loop_2}
    D_{q/q}^j(z) &= 2i \alpha_s^3 
    f^{eab}f^{ecd} t^a_Lt^b_R \spac\bm{T}_{2,L}^c \spac\bm{T}_{j,R}^d\frac{1}{N_c}\spac \Bigl(\frac{\slashed{n}}{2}\Bigr)_{\!\!\beta\bar{\beta}} \spac\Bigl(\frac{\nu}{(1-z) \nb\cdot p_{c}}\Bigr)^{\alpha}
    e^{2\varepsilon\gamma_E}\frac{\Gamma(2\varepsilon)\Gamma^3(1-\varepsilon)}{\varepsilon\spac\Gamma(1-3\varepsilon)}
    \nonumber\\
    &
    \quad\times
    \int[d\Omega_l]
    (2\spac\nb\cdot n_l\,n\cdot n_l)^{-2\varepsilon}
    \frac{n_{l}^\nu\, g_{\perp,\mu\nu}}{n\cdot n_l\,\bar{n}\cdot n_l}\biggl(-\frac{n_j^\mu}{n_j\cdot n_l} + \frac{n\cdot n_j\,n_{l}^\mu}{n_j\cdot n_l\, n\cdot n_l}\biggr)\Theta_{\mathrm{veto}}(n_l)\nonumber\\
    &\quad\times\mu^{6\varepsilon}
    \int\frac{dE_l}{(2\pi)^2}E_l^{-1-6\varepsilon}\,\theta(Q_0-E_l^T)
    \mathcal{P}_{q\rightarrow q}(z)\,,
\end{align}
where the indices $\beta,\bar{\beta}$ are connected to the hard functions and $\mathcal{P}_{q\rightarrow q}(z)$ is the $d$-dimensional splitting function~\cite{Catani:1999ss} with color factor stripped off
\begin{equation}
    \mathcal{P}_{q\rightarrow q}(z) = \frac{1+z^2}{1-z}-\varepsilon(1-z)\,.
\end{equation}

Note that the integral \eqref{eq:q_to_q_three_loop_2} is proportional to $\slashed{n}$. Since the collinear fields are subject to the constraint $\slashed{n}\spac \chi_c = 0$, the associated Dirac Algebra is effectively four-dimensional and the matrices $\slashed{n}$, $\slashed{n}\gamma_5$, $\slashed{n} \,\gamma_\perp^\mu$ span it in $d=4$. The second structure is parity violating and cannot arise.  The third structure does not contribute for inclusive processes, as there are no transverse vectors available which could contract with it without immediately vanishing. In our case, these structures could in principle appear  due to additional dependence on soft momenta at higher loop orders. However,  the term $\slashed{n} \,\gamma_\perp^\mu$ is chirality violating and does not arise in the massless case. We conclude that there is only a single physical Dirac structure present in the quark case. On the level of the bare matrix elements, additional evanescent structures arise at higher orders, see e.g.~\cite{Becher:2004kk}.
In this specific case, we could therefore have contracted with $\slashed{\bar{n}}/2$ from the beginning. 

There are several remarks regarding~\eqref{eq:q_to_q_three_loop_2}: first, when attaching the soft emission to the collinear emission the result is scaleless. Secondly, the expression also vanishes for $j=1,2$, once the hermitian conjugated contribution is added.   This can be immediately seen from the color structure. In case $j=2$ we e.g.\ get
\begin{equation}
    \propto f^{abc}\bm{T}^b_2\bm{T}^c_2=\frac{iN_c}{2}\bm{T}^a_2\, 
\end{equation}
after taking the color trace with the hard functions. This term is imaginary and vanishes when adding the complex-conjugated diagram. Similarly, also the contribution $j=1$ vanishes after adding the hermitian conjugate.

After performing the sum over all soft attachments $j$, one can use color conservation to rewrite the bracket appearing inside the angular integral in terms of the dipoles $W^l_{ij}$ defined in~\eqref{eq:def_soft_dipole}
as
\begin{align}
\label{eq:DipoleTerms}
    \frac{n_{l}^\nu\, g_{\perp,\mu\nu}}{n\cdot n_l\,\bar{n}\cdot n_l}\biggl(-\frac{n_j^\mu}{n_j\cdot n_l} + \frac{n\cdot n_j\,n_{l}^\mu}{n_j\cdot n_l\, n\cdot n_l}\biggr) &= -\frac{1}{2}\biggl(\frac{n\cdot n_j}{n\cdot n_l\, n_j\cdot n_l} - \frac{\nb\cdot n_{j}}{\bar{n}\cdot n_l\, n_j\cdot n_l}\biggr)\nonumber\\
    &= -\frac{1}{2}(W^l_{1j}-W^l_{2j})\,.
\end{align}
As the integration becomes scales inside the jet region, one can replace $W^{l}_{ij}$ by the collinear-subtracted $\overline{W}^{l}_{ij}$. For the leading poles, the angular integral over these dipoles can be expressed in terms of $J_j$ defined in~\eqref{eq:def_Jj}.

The insertion of the Glauber operator connecting anti-collinear and soft sectors is obtained by substituting $n \leftrightarrow \bar{n}$, up to the fact that the rapidity regulator is always defined with respect to~$\bar{n}$.
In this case, the regulator propagates into the $q_\perp$-integral and leads to higher poles in $1/\epsilon$. We give the leading-pole result for this contribution below.

To compare to the prediction given in~\eqref{eq:Wm_poles_coll}, we extract the leading $1/\epsilon$ poles of these matrix elements and multiply them with the hard functions under the color trace $\llangle\cdots\rrangle$.
Since the result~\eqref{eq:q_to_q_three_loop_2} is proportional to the quark polarization sum, we pull out
\begin{equation}
    \frac{1}{N_c}\frac{1}{\bar{n}\cdot p_c}\left[\frac{\bar{n}\cdot p_c}{2}\frac{\slashed{n}}{2}\right]_{\beta\bar{\beta}}
\end{equation}
to arrive at
\begin{align}
   \int\!\! d\Pi_m\sum_{j} \llangle \bm{\mathcal{H}}\spac  &D^j_{q/q}(z_1) \Pi_{i_2/l}(z_2)\rrangle + \mathrm{h.c.}\nonumber\\
   &= \int\!\! d\Pi_m\sum_{j>2}\llangle \bm{\mathcal{H}} \frac{1}{N_c}\frac{1}{\bar{n}\cdot p_c}\left[\frac{\bar{n}\cdot p_c}{2}\frac{\slashed{n}}{2}\right]\Pi_{i_2/l}(z_2) \spac
     \overline{D}^j_{q/q}
    \rrangle+ \mathrm{h.c.}\nonumber\\
    &= \sum_{j>2} \langle\spac \overline{\bm{\mathcal{H}}} \otimes \overline{D}^j_{q/q}\rangle + \mathrm{h.c.}\,,
\end{align}
where in the last line the polarization sums generate the flux factor in $\overline{\bm{\mathcal{H}}}$ and change the double to the single bracket, now including the spin and color average over the initial state partons of the hard functions.
The second equality holds after including the energy integration over the final-state particles to produce $\overline{\bm{\mathcal{H}}}$ while $\otimes$ contains the angular integrations.

We then find
\begin{align}\label{eq:Icomp_q_to_q}
    &\sum_j \overline{D}_{q/q}^j(z_1)+ \mathrm{h.c.}
    = \frac{i\alpha_s^3}{12\pi^2\varepsilon^3}
    \left( \frac{\mu^2}{Q_0^2} \right)^{3\varepsilon}
    \hspace{-0.2cm}f^{abc} f^{ade} {\sum_{j>2}}^{\,\prime} J_j \\
    &\qquad\times\bigg[t_{1,L}^d\spac t_{1,R}^e\spac
    \bm{T}_{2,L}^b\spac\bm{T}_{j,R}^c 
    \bigg( \left[- \frac{1}{\alpha} - \ln\frac{\nu}{\nb\cdot p_c} \right]\delta(1-z_1)+z^2_1\left[\frac{1}{1-z_1}\right]_++\frac{1+z_1}{2}
     +\dots\,\bigg) \bigg]\nonumber\\
     &\qquad- (L\leftrightarrow R) \,.\nonumber
\end{align}
Here, the hermitian conjugate is realized via $L\leftrightarrow R$. Under the convolution over $z_1$ we can read off the soft-subtracted splitting functions~\eqref{eq:splitting_functions_final_explicit} leading to
\begin{align}\label{eq:Icomp_q_to_q_final}
   &\sum_j \overline{D}_{q/q}^j(z_1)+\mathrm{h.c.}
   = \frac{i\alpha_s^3}{12\pi^2\varepsilon^3}
    \left( \frac{\mu^2}{Q_0^2} \right)^{3\varepsilon}
    f^{abc} f^{ade} {\sum_{j>2}}^{\,\prime} J_j\\
    &\qquad\times
    \bigg[
     \bm{T}_{1,L}^d\spac\bm{T}_{1,R}^e\spac
   \bm{T}_{2,L}^b\spac\bm{T}_{j,R}^c \bigg( \left[- \frac{1}{\alpha} - \ln\frac{\nu}{\nb\cdot p_c} \right]\delta(1-z_1)+\frac{\overline{\mathcal{P}}_{i\to i}(z_1)}{2} +\dots \bigg) \bigg] - (L\leftrightarrow R) \,.\nonumber
\end{align}
The explicit result of the other diagonal splitting $g\to g $ can be found in~\eqref{app:eq:g_to_g_three_loop}. 
While the general structure of this contribution is more involved and contains the aforementioned additional Lorentz structures, the leading pole result can be written in the form of~\eqref{eq:Icomp_q_to_q_final}, which therefore covers both cases.
When considering the same splitting on the anti-collinear leg the result is given by
\begin{align}\label{eq:Icomp_q_to_q_final_anticoll}
   &\sum_j \overline{D}_{q/q}^j(z_2)+\mathrm{h.c.}
   = 
   - \frac{i\alpha_s^3}{12\pi^2\varepsilon^3}
    \left( \frac{\mu^2}{Q_0^2} \right)^{3\varepsilon}
    f^{abc} f^{ade} {\sum_{j>2}}^{\,\prime} J_j \\
    &\quad\times\bigg[ \bm{T}_{2,L}^d\spac\bm{T}_{2,R}^e\spac
   \bm{T}_{1,L}^b\spac\bm{T}_{j,R}^c \bigg( \bigg[\frac{1}{\alpha} + \ln\frac{\nu\spac n\cdot\bar p_{\bar c}}{Q_0^2}-\frac{11}{6 \varepsilon}\bigg]\delta(1-z_2) +\frac{\overline{\mathcal{P}}_{i \to i }(z_2)}{2}+ \dots \bigg)\bigg]
- (L\leftrightarrow R) \,.\nonumber
\end{align}

The $1/\alpha$ poles and the $\nu$-dependence cancel between the attachments of the Glauber at the collinear and anti-collinear legs, i.e.\ between the mirrored diagrams, as this is made manifest when expressed in the color space formalism.\footnote{Depending on the chosen rapidity regulator, the $1/\alpha$ may appear in different sectors. When e.g.\ using a phase-space regulator such as the one employed in~\cite{Becher:2015gsa,Becher:2013xia}, the collinear and anti-collinear contributions take a similar form, and the respective poles then cancel with new poles arising from the soft sector.}
This is only possible because these poles and logarithms are accompanied by $\delta(1-z_1)\delta(1-z_2)$.
Under the color trace with the hard functions, we can simplify
\begin{equation}
     f^{abc} f^{ade}\bm{T}_{2,L}^d\spac\bm{T}_{2,R}^e\spac
   \bm{T}_{1,L}^b\spac\bm{T}_{j,R}^c\equalhat-i\frac{N_c}{2}f^{abc}\bm{T}_2^a\bm{T}_{1}^b\spac\bm{T}_{j}^c.
 \end{equation}
The diagonal contribution is then exactly of the expected form, including the DGLAP splitting kernels
\begin{align}
    \int\!\! d\Pi_m\llangle &\bm{\mathcal{H}}_{i_1 i_2\to m}\spac \bm{\mathcal{I}}^{(3)\,i_1 i_2\,\mathrm{bare}}_{i_1i_2\to m}\rrangle \nonumber\\
    &\ni \frac{i N_c\spac\alpha_s^3}{12\pi^2\varepsilon^3}\,
    i f^{abc}
    \left( \ln\frac{Q_0^2}{Q^2} + \frac{11}{6\varepsilon} 
    + \frac{11}{2}\,\ln\frac{\mu^2}{Q_0^2} -\frac{\overline{\mathcal{P}}_{i_1\to i_1}(z_1)}{2}-\frac{\overline{\mathcal{P}}_{i_2\to i_2}(z_2)}{2}+ \dots \right)  \nonumber\\
    &\quad\times{\sum_{j>2}}^{\,\prime}J_j
    \langle \spac\overline{\bm{\mathcal{H}}}\spac\bm{T}_1^a\spac\bm{T}_2^b\spac\bm{T}_j^c\otimes \bm{1}\rangle\,.\label{eq:Icomp_diag_final}
\end{align}

Here, we defined $Q^2=(n\cdot\bar{p}_{\bar{c}})(\nb\cdot p_c)$.
This result \emph{exactly} reproduces the prediction in~\eqref{eq:diagonal_split_c_factor}, derived from
the general expression~\eqref{eq:Wm_poles_coll}, and expands upon~\cite{Becher:2024kmk}, as it now includes the DGLAP splitting kernels.

\begin{figure}
    \centering
    \includegraphics{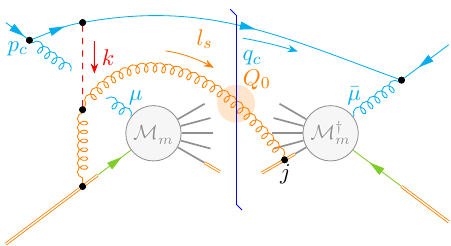}
    \caption{Diagram showing the off-diagonal splitting of $q\to g$. We explicitly show the Lorentz indices $\mu$ and $\bar{\mu}$, which connect to the hard amplitude. The Glauber interaction induces non-trivial Lorentz structures in these indices depending on the vector $n_j$ and the gap constraint.}
    \label{fig:q_to_g_off_diag_three}
\end{figure}

Next, we consider the off-diagonal splitting of a quark into a gluon, $q\to g$, shown in Figure~\ref{fig:q_to_g_off_diag_three}. The computation becomes more involved in this case, as the gluon entering the hard functions carries not only color- but also Lorentz-indices. This allows for a more complex Lorentz structure than just $g_\perp^{\mu\bar{\mu}}$ entering the hard functions, as the indices can propagate through the integration: The collinear sector is no longer decoupled, which means that additional Lorentz-structures involving $l_s^\mu$ appear. Consequently, one can no longer view the hard functions as spin-summed squared amplitudes, which was anticipated in the generalized factorization formula~\eqref{eq:factorizationTheoremNew}. The contribution depicted in Figure~\ref{fig:q_to_g_off_diag_three} is then given by
\begin{align}
\label{eq:three_loop_q_to_g}
D_{g/q}^{j} &= -\alpha_s^3f^{dbc}\bm{T}_{2,L}^b \bm{T}_{j,R}^c \frac{1}{z\spac\nb\cdot p} \frac{1}{N_c}\Tr\!\left[t^{\bar{a}} t^d t^a\right] e^{2\varepsilon\gamma_E}\frac{\Gamma(2\varepsilon)\Gamma^3(1-\varepsilon)}{\varepsilon(1-2\varepsilon)\Gamma(1-3\varepsilon)}\nonumber\\
&\quad\times \int[d\Omega_l](2\nb\cdot n_l\,n\cdot n_l)^{-2\varepsilon}\Theta_{\mathrm{veto}}(n_l) \frac{1}{\nb\cdot n_l\,n\cdot n_l}\Bigl(-\frac{n_{j\alpha}}{n_j\cdot n_l} + n_{l\alpha} \frac{n\cdot n_j}{n\cdot n_l\,n_j\cdot n_l}\Bigr)\nonumber \\
    & \quad \times \mu^{6\varepsilon}\int\frac{dE_l}{(2\pi)^2} E_l^{-1-6\varepsilon}\theta(Q_0-E^T_l)\nonumber\\
    &\quad\times
    \biggl\{\Bigl[4\frac{1-z}{z}
    \Bigl(1-\frac{\varepsilon(1-\varepsilon)}{(1+\varepsilon)(1-3\varepsilon)}
    \Bigr)+2(1-2\varepsilon)z \Bigr]g_{\perp}^{\mu\bar{\mu}}n_{l\perp}^\alpha
    \nonumber\\
    &\qquad
    -\frac{8\varepsilon^2}{(1+\varepsilon)(1-3\varepsilon)}\frac{1-z}{z}
    \Bigl(
    (g_\perp^{\bar{\mu}\alpha}n_{l\perp}^\mu + g_\perp^{\mu\alpha}n_{l\perp}^{\bar{\mu}}) + 2 (1+2\varepsilon)\frac{n_{l\perp}^\mu n_{l\perp}^{\bar{\mu}} n_{l\perp}^\alpha}{\bar{n}\cdot n_l \, n\cdot n_l}
    \Bigr)\biggr\}\,.
\end{align}
The first term in the square bracket can be expressed as
\begin{align}
    \Bigl(2\frac{1-z}{z}
    \Bigl(1-\frac{\varepsilon(1-\varepsilon)}{(1+\varepsilon)(1-3\varepsilon)}
    \Bigr)+(1-2\varepsilon)z \Bigr) &= \frac{2-4\epsilon}{2-2\epsilon}\mathcal{P}_{q\to g}(z)\nonumber\\
    &\quad-\frac{8\varepsilon^2}{(1+\varepsilon)(1-3\varepsilon)}\frac{2\varepsilon}{2-2\varepsilon}\frac{1-z}{z}\,,
    \label{eq:OffDiagCalculationSplitting}
\end{align}
to extract the $d$-dimensional splitting function. The entire curly bracket takes the form
\begin{align}
    \biggl\{\dots\biggr\}&=\frac{g_{\perp}^{\mu\bar{\mu}}n_{l\perp}^\alpha}{1-\varepsilon}(2-4\varepsilon)
    \mathcal{P}_{q\to g}(z)\\
    &\quad
    -\frac{8\varepsilon^2}{(1+\varepsilon)(1-3\varepsilon)}\frac{1-z}{z}\biggl(
    \!2\varepsilon\frac{g_{\perp}^{\mu\bar{\mu}}n_{l\perp}^\alpha}{1-\varepsilon} + (g_\perp^{\bar{\mu}\alpha}n_{l\perp}^\mu + g_\perp^{\mu\alpha}n_{l\perp}^{\bar{\mu}}) + 2 (1+2\varepsilon)\frac{n_{l\perp}^\mu n_{l\perp}^{\bar{\mu}} n_{l\perp}^\alpha}{\bar{n}\cdot n_l \, n\cdot n_l}
    \biggr)\spac.\nonumber
\end{align}
Remarkably, the exact same structure, but with the respective gluon splitting function, appears when calculating the diagonal gluon splitting~\eqref{app:eq:g_to_g_three_loop}. Generically, one can see that the leading and subleading poles are completely described by the splitting function
\begin{equation}
    \mathcal{P}_{q\to g}(z) = \mathcal{P}_{q\to q}(1-z)\,.
\end{equation}

Summing over all soft attachments $j$, using color conservation, and expressing the result in terms of dipoles whenever possible, we find
% A different ordering according to dipole structure
% \begin{align}
% \label{eq:three_loop_q_to_g}
% \sum_j D_{g/q}^{j} &= \alpha_s^3f^{dbc}\sum_j\bm{T}_{2,L}^b \bm{T}_{j,R}^c \frac{1}{z\spac\nb\cdot p} \frac{1}{N_c}\Tr\!\left[t^{\bar{a}} t^d t^a\right] e^{2\varepsilon\gamma_E}\frac{\Gamma(2\varepsilon)\Gamma^3(1-\varepsilon)}{\varepsilon(1-2\varepsilon)\Gamma(1-3\varepsilon)}\\
% &\quad\times \int[d\Omega_l](2\nb\cdot n_l\,n\cdot n_l)^{-2\varepsilon}\Theta_{\mathrm{veto}}(n_l)\,\mu^{6\varepsilon}\int\frac{dE_l}{(2\pi)^2} E_l^{-1-6\varepsilon}\theta(Q_0-E^T_l)\nonumber\\
%     &\quad\times
%     \Bigg\{ \bigg[ \biggl(\frac{2-4\epsilon}{2-2\epsilon}\mathcal{P}_{q\to g}(z)
%     -\frac{8\varepsilon^2}{(1+\varepsilon)(1-3\varepsilon)}\frac{2\varepsilon}{2-2\varepsilon}\frac{1-z}{z}\biggr)
%     g_\perp^{\mu\bar{\mu}} \nonumber\\
%    &\hspace{1.4cm} - \frac{8\epsilon^2(1+2\epsilon)}{(1+\epsilon)(1-3\epsilon)}\,\frac{1-z}{z}\,
%     \frac{n_{l\perp}^\mu n_{l\perp}^{\bar{\mu}}}{\nb\cdot n_l \, n\cdot n_l} \bigg] \left( W_{1j}^l - W_{2j}^l \right)\nonumber \\
%    &\qquad + \frac{8\epsilon^2}{(1+\epsilon)(1-3\epsilon)}\,\frac{1-z}{z}
%     \left[ 2 W_{1j}^l\, \frac{n_{l\perp}^\mu n_{l\perp}^{\bar{\mu}}}{\nb\cdot n_l\,n\cdot n_l}
%     - \frac{1}{n_j\cdot n_l}\,
%     \frac{n_{j\perp}^\mu n_{l\perp}^{\bar{\mu}}+n_{l\perp}^\mu n_{j\perp}^{\bar{\mu}}}{\nb\cdot n_l\, n\cdot n_l} \right] 
%     \Bigg\} \,,\nonumber
% \end{align}
\begin{align}
\label{eq:three_loop_q_to_g_dipole}
\sum_j D_{g/q}^{j} &= \alpha_s^3f^{dbc}\sum_j\bm{T}_{2,L}^b \bm{T}_{j,R}^c \frac{1}{z\spac\nb\cdot p} \frac{1}{N_c}\Tr\!\left[t^{\bar{a}} t^d t^a\right] e^{2\varepsilon\gamma_E}\frac{\Gamma(2\varepsilon)\Gamma^3(1-\varepsilon)}{\varepsilon(1-2\varepsilon)\Gamma(1-3\varepsilon)}\\
&\quad\times \int[d\Omega_l](2\nb\cdot n_l\,n\cdot n_l)^{-2\varepsilon}\Theta_{\mathrm{veto}}(n_l)\,\mu^{6\varepsilon}\int\frac{dE_l}{(2\pi)^2} E_l^{-1-6\varepsilon}\theta(Q_0-E^T_l)\nonumber\\
    &\quad\times
    \Bigg\{ 
    \frac{g_\perp^{\mu\bar{\mu}}}{2-2\epsilon}(2-4\epsilon)\mathcal{P}_{q\to g}(z)\left( W_{1j}^l - W_{2j}^l \right)\nonumber\\
    &\quad-\frac{8\varepsilon^2}{(1+\varepsilon)(1-3\varepsilon)}\frac{1-z}{z}
    \biggl[
    \biggl(2\varepsilon\frac{g_\perp^{\mu\bar{\mu}}}{2-2\varepsilon} + (1+2\varepsilon)
    \frac{n_{l\perp}^\mu n_{l\perp}^{\bar{\mu}}}{\nb\cdot n_l \, n\cdot n_l}\biggr)\left( W_{1j}^l - W_{2j}^l\right)\nonumber \\
   &\qquad -2 W_{1j}^l\, \frac{n_{l\perp}^\mu n_{l\perp}^{\bar{\mu}}}{\nb\cdot n_l\,n\cdot n_l}
    + \frac{1}{n_j\cdot n_l}\,
    \frac{n_{j\perp}^\mu n_{l\perp}^{\bar{\mu}}+n_{l\perp}^\mu n_{j\perp}^{\bar{\mu}}}{\nb\cdot n_l\, n\cdot n_l} \biggr]
    \Bigg\} \,,\nonumber
\end{align}

As for the $q\to q$ splitting case considered earlier, the Glauber contribution induces non-trivial color-correlations between the soft and (anti-)collinear sectors. However, in contrast to this case, we also encounter spin-correlations in $q\to g$. First of all, the term proportional to the polarization sum $g_{\perp}^{\mu\bar{\mu}}$ deviates from the splitting function in the finite terms.
In addition, new non-trivial Lorentz structures arise at the single $1/\varepsilon$ pole level.
 Upon integration, these lead to contributions proportional to $n_{j\perp}^\mu n_{j\perp}^{\bar{\mu}}$, where the $\perp$ refers to the beam direction.
Similar new structures also arise in the case of the diagonal gluon splitting, see~\eqref{app:eq:g_to_g_three_loop}. It is therefore no longer possible to pull out the projector $g_{\perp}^{\mu\bar{\mu}}$, and the adoption of double brackets, which do not average over initial-state spins, polarizations and colors, in~\eqref{eq:factorizationTheoremNew} becomes mandatory.
This calculation of the low-energy theory was performed using a rigorous framework based on SCET, and has implications on both the definition of the hard functions and splitting functions. 
Since this projector would perform the spin-sum resulting in $\overline{\bm{\mathcal{H}}}$, one immediately sees that in full generality, one \emph{cannot} work with spin-averaged hard functions beyond the two-loop level. 
Moreover, as the terms not proportional to $g_{\perp}^{\mu\bar{\mu}}$ need to be matched by the three-loop anomalous dimensions, also the splitting functions have to be of a more general structure, as was anticipated for splitting amplitudes in~\cite{Cieri:2024ytf}. 

However, to compare against the prediction in~\eqref{eq:Wm_poles_coll} only the leading poles are required. As these arise from iteration of one-loop anomalous dimensions, they are proportional to $g_{\perp}^{\mu\bar{\mu}}$.
Similarly to the diagonal $q\to q $ splitting we have to adjust both the spin-average, as well as the flux factor, accordingly by factoring out
\begin{equation}
   \frac{1}{N_c^2-1} \frac{1}{z\spac \nb\cdot p_c}\left[-\frac{g^{\mu\bar{\mu}}_\perp}{ (d-2)}\right]\, 
\end{equation}
where $1/(\nb\cdot p_c \spac z)$ combines into the flux factor to give the hard functions $\overline{\bm{\mathcal{H}}}$. The factor $1/(N_c^2-1)$ adjusts the color average. This is necessary because the anomalous dimensions  depend on the precise definition of the hard functions and~\eqref{eq:Wm_poles_coll} was derived for $\overline{\bm{\mathcal{H}}}$.

Focusing on the leading pole, we take into account all different attachments of the soft gluon by summing over $j$ and add the hermitian conjugate. 
The contributions from the initial-state legs where $j=1,2$ vanish again due to color structure. 
We then find focusing on this particular splitting
\begin{align}
 \int\!\! d\Pi_m\llangle \spac\bm{\mathcal{H}}_{i_1i_2\to m}&\bm{\mathcal{I}}_{g\spac i_2\to m}^{(3)\,q l\,\mathrm{bare}} \rrangle %&\hspace{-5mm}\ni\frac{16\pi}{3\,\epsilon^3}\left(\frac{\alpha_s}{4\pi}\right)^3 \left(2\frac{(1-z)^2}{z}+z+2(1-z)\right)f^{dbc}\spac\frac{N_c^2-1}{N_c}\Tr\!\left[t^{\bar{a}}t^dt^a\right]{\sum_{j>2}}^{\,\prime}J_j\langle \spac\overline{\bm{\mathcal{H}}}\spac\bm{T}_2^b\bm{T}_j^c \rangle\nonumber \\
\ni\frac{\alpha_s^3}{12\,\epsilon^3\pi^2} \frac{N_c^2-1}{N_c}\spac\overline{\mathcal{P}}_{q\to g }(z) f^{dbc}\spac\Tr\!\left[t^{\bar{a}}t^dt^a\right]{\sum_{j>2}}^{\,\prime}J_j\langle \spac\overline{\bm{\mathcal{H}}}\spac \bm{T}_2^b\bm{T}_j^c\otimes\bm{1}\rangle\, ,
\end{align}
which matches the predicted result~\eqref{eq:q_to_g_expected_color}.
When considering the splitting $\bar{q}\to g$ an additional minus sign arises as well as $a\leftrightarrow \bar{a}$. The $g\to \bar{q}$ splitting can be found in Appendix~\ref{sec:app:three_loop}, see in particular~\eqref{app:eq:g_to_q_three_loop}, and similarly reproduces the correct splitting function. We have therefore explicitly verified that both the diagonal and off-diagonal splitting functions, as well as the hard logarithms and additional $1/\epsilon$ poles in~\eqref{eq:Wm_poles_coll} are correctly reproduced within the low-energy theory.

\section{Discussion}
\label{sec:discussion}

It is remarkable that the three-loop consistency check in the previous section has fully worked out. At this loop order, all necessary contributions for factorization breaking are present, as can be seen in \eqref{eq:Wm_poles}. For the low-energy theory to be consistent with PDF factorization, several terms with highly non-trivial structure had to emerge at the perturbative scale $Q_0$: Large logarithms of the form $\ln Q/\mu$, additional higher poles in $1/\epsilon$, as well as color-aware DGLAP splittings. As we have discussed, purely soft or collinear terms cannot reproduce these terms and our detailed method-of-region analysis then showed that exactly two additional modes appear in the relevant QCD diagrams. A Glauber mode scaling as $k_g\sim(\lambda,\lambda^2,\lambda)$ and a soft-collinear mode scaling as  $k_{sc}\sim(\lambda,\lambda^2,\lambda^{3/2})$. While the appearance of a soft-collinear region in sub-diagrams could have further complicated the results, its contribution vanishes after the integration over the collinear emission. It is noteworthy that this contribution only vanishes because the collinear emission is not restricted in $q_T$ for the observables we consider and therefore scaleless on its own. The transverse momentum of the soft-collinear modes scales as $k_{sc,T}\sim  \lambda^{3/2}$, which is parametrically smaller than the transverse momentum of the collinear modes $q_T\sim\lambda$, and hence cannot propagate into the collinear integrals. The collinear integrals therefore remain scaleless. 
This strongly suggests that even at higher loop orders, the soft-collinear mode still drops out due to scalelessness. It is, however, not immediately clear that this mechanism stays intact in the presence of a measurement on $q_T$. 

In contrast, the transverse momentum of the Glauber mode is commensurate with the one of the collinear mode, $k_{g,T}\sim q_T$, and the collinear integrals no longer vanish in the presence of the Glauber mode. This gives rise to a \textit{single} type of Glauber diagram contributing to soft-collinear factorization breaking. Remarkably, the mode does not require additional regulators and is well-defined within dimensional regularization alone. In contrast to computations involving non-analytic regulators, $0$-bin subtractions are not needed. For each type of collinear splitting, the small set of Glauber diagrams matches \textit{all} leading pole structures required by RG consistency, including the correct color and momentum fraction structure. This shows that double-logarithmic running above the scale $Q_0$ is fully consistent with single-logarithmic DGLAP running below. Our computation has explicitly identified the precise mechanism which reconciles collinear factorization breaking with PDF factorization. 

On top of providing the consistency check for the leading poles, our analysis reveals interesting additional Lorentz structures in $i\to g$ splittings at subleading poles In general, hard functions with an incoming gluon have open Lorentz indices~$\mu$ and $\bar{\mu}$ in the amplitude and its conjugate, which are contracted with the low-energy matrix elements. Without soft-collinear factorization breaking, the collinear matrix elements would be proportional to $g_\perp^{\mu\bar{\mu}}$ since they do not depend on any transverse directions. One could then immediately average over the gluon spins in the hard functions. 
However, in the presence of Glauber interactions, the matrix elements now receive contributions depending on the directions of the soft Wilson lines, i.e.\ structures such as $n_{j,\perp}^\mu\spac n_{j,\perp}^{\bar{\mu}}$, associated with final-state parton $j$. In general, also the veto-region can introduce dependence on transverse directions. The new structures arise as single-pole terms at three-loop order and corresponding terms should then also arise in the hard anomalous dimensions, relevant for higher logarithmic resummation. Consequently, also higher order splitting functions must have a more general spin dependence.

\subsection{Towards a proof of factorization}
In this paper, we showed that the perturbative soft-collinear physics associated with $Q_0$ is consistent with PDF factorization, but so far we did not discuss non-perturbative physics below $Q_0$. Crucially, all fields contributing to the soft-collinear matrix elements $\bm{\mathcal{I}}_m$ scale with $k^2\sim\ Q_0^2$ and can therefore not probe physics below $Q_0$. Importantly, the soft-collinear mode scaling as $k^2\sim Q_0^3/Q$ decoupled. It would be disastrous if such a mode contributed to the cross section, as for sufficiently large values of $Q$ this mode can be pushed into the non-perturbative regime for fixed values of $Q_0$, which would constitute a true source of PDF factorization breaking.

Below $Q_0$ we can match onto  collinear fields with $k^2\sim \Lambda_\mathrm{QCD}^2$, along with their respective Wilson lines. 
As we do not impose any constraints on soft or collinear radiation below $Q_0$, the theory below that scale is completely inclusive: After summing over all final states, the soft Wilson lines in \eqref{eq:W_function_definition} cancel and the leading-power collinear operators are exactly the ones relevant for the PDFs~\cite{Bauer:2002nz}. The only obstacle to PDF factorization are non-perturbative Glauber exchanges between the two collinear sectors. The effective theory and the respective operators are identical to the case of Drell-Yan production analyzed in~\cite{Bauer:2002nz}, and the absence of Glauber modes can be inferred from the proof of PDF factorization for the Drell-Yan process~\cite{Collins:1985ue,Collins:1988ig}. Our effective theory framework therefore formalizes the intuitive statement that below the lowest perturbative scale, Glauber gluons cancel and that the standard CSS arguments presented in~\cite{Collins:1985ue,Collins:1988ig} extend to a much wider class of observables. 
To obtain a rigorous proof of PDF factorization, it would be desirable to extend the consistency check performed above to higher orders in perturbation theory. Secondly, one would need to demonstrate to all orders that no effective fields with virtualities below $Q_0^2$, such as the soft-collinear modes, appear within perturbative physics. This is crucial for the CSS proof to work, as only soft, collinear and non-perturbative Glauber modes are considered there. Although it does not seem unreasonable that such additional modes are absent, a rigorous analysis has not been performed yet, even for the inclusive Drell-Yan case.

\subsection{Generalization to other observables}

It is not immediately clear how our analysis generalizes to other observables since it is based on a factorization formula that is specific to gap-between-jet cross section. The proposed factorization theorems for global hadron-collider observables such as $N$-jettiness \cite{Stewart:2010tn} or transverse thrust \cite{Becher:2015gsa} have a much simpler structure. However, the papers~\cite{Banfi:2010xy,Forshaw:2021fxs} argue that coherence violating terms contribute also in these cases, and they predict the order at which they first arise. Global observables are double logarithmic so that these terms are not super-leading. For this reason, the name coherence violating logarithms (CVL) was suggested~\cite{Banfi:2010xy}. It would be very interesting to generalize our consistency check to these observables and to check whether Glauber modes contribute. While simpler than the non-global case, the factorization theorems for the global observables have the same ingredients: collinear fields and associated Wilson lines. Three-loop  diagrams  such as the ones shown in Figures~\ref{fig:q_to_q_three} and~\ref{fig:q_to_g_off_diag_three} are also present in these cases. The only difference are the phase-space constraints, which involve both soft and collinear emissions. It will be important to analyze these cases in detail and to verify whether Glauber contributions are relevant or not. Doing so will answer the long-standing question about Glauber contributions, which were generally assumed to be absent in the derivation of the factorization theorems.

\section{Conclusion}
\label{sec:conclusion}

The factorization of cross sections into perturbative partonic dynamics at high energies and non-perturbative parton distribution functions (PDFs) encoding the hadronic structure is the foundation of high-precision collider physics.
This factorization is formally proven only for the fully-inclusive Drell–Yan process~\cite{Collins:1985ue,Collins:1988ig}, where the cancellation of low-energy Glauber gluon exchanges between the incoming hadrons is essential. At the same time, it is known that higher-loop corrections involve color-coherence violating phases, and in partonic amplitudes involving space-like collinear splittings, these induce collinear factorization violation~\cite{Catani:2011st,Forshaw:2012bi}. This results in physical effects in the form of super-leading logarithms (SLLs)~\cite{Forshaw:2008cq}. The same mechanism also modifies the collinear evolution which is no longer color-diagonal. This raises an important question: how can color-aware DGLAP splitting kernels and double-logarithmic evolution above the veto scale $Q_0$ be reconciled with the single-logarithmic DGLAP evolution implied by PDF factorization?

In this work, we addressed this challenge using a factorization theorem  based on effective field theory, which can be used to derive consistency conditions on the perturbative part of the low-energy matrix elements. We verify these conditions order by order in perturbation theory. The purely collinear contributions vanish to all orders, and up to two-loop order, the soft matrix elements alone produce all relevant terms. 
However, at three-loop level, a genuine double-logarithmic contribution arises that cannot be generated by the soft sector. 
In previous work~\cite{Becher:2024kmk}, we discovered a genuine Glauber mode which connecting the soft and collinear sectors. 
In the present article, we verified that this region not only generates the required hard double logarithm via the collinear anomaly but also the associated color-aware DGLAP evolution. Crucially, this soft-collinear factorization breaking fully compensates the collinear factorization-violating contributions at the perturbative veto scale $Q_0$, ensuring that only standard DGLAP evolution survives below $Q_0$. Additionally, we find new Lorentz structures proportional to final-state direction, which have to be matched by the three-loop anomalous dimensions. This implies that higher-loop splitting functions must have a more general dependence on the whole process.

The explicit verification we performed resolves the tension between the existence of SLLs and DGLAP evolution for non-global jet observables. Having identified the precise mechanism for PDF factorization restoration, it will be very interesting to extend our analysis to other observables, such as global event shapes. The operator formalism employed here should also be instrumental in obtaining a deeper insight into the structure of collinear factorization breaking on the amplitude level. A first step in this direction was the identification of spin correlations in the low-energy matrix elements.

While we verified that PDF factorization holds below $Q_0$, we want to stress that the factorization breaking at higher scales has important physics effects. As for other infrared safe observables, the cancellation of divergences leaves behind footprints in the form of large logarithms. In the present case, these include the SLLs and the logarithms associated with the color-aware collinear evolution. For the SLLs, first steps to assess their phenomenological impact were taken in \cite{Becher:2024nqc} but much remains to be studied, in particular their interplay with non-global logarithms. The numerical effect of the colorful DGLAP evolution above $Q_0$ has not yet been explored, but its presence implies that there is a mismatch between the standard DGLAP evolution of the PDFs and the evolution of the hard functions, which manifests itself through large logarithms in the partonic cross sections. Even forty years after the seminal paper on PDF factorization \cite{Collins:1985ue}, a lot of exciting physics in hadron collisions remains to be explored!

\subsection*{Acknowledgments}

We thank Andrea Banfi, Fabrizio Caola, Jeff Forshaw,  Einan Gardi, Thomas Gehrmann, Jack Holguin, Yao Ma, Davison Soper, and  Michel Stillger for useful discussions.
PH and MN gratefully acknowledge support from the Albert Einstein Center for Fundamental Physics (AEC) at the University of Bern.
TB, SJ and DS would like to thank the Mainz Institute of Theoretical Physics (MITP) for hospitality and support.
This research has received funding from the 
Cluster of Excellence \emph{PRISMA${}^+$} (EXC 2118/1, Project ID 390831469) funded by the German Research Foundation (DFG) within the Germany Excellence Strategy,
from the European Research Council (ERC) under the European Union’s Horizon 2022 Research and Innovation Program (ERC Advanced Grant agreement No.~101097780, EFT4jets), 
from the Swiss National Science Foundation (SNSF) under grant 200021\_219377, and from the SNSF Ambizione grant PZ00P2\_223524.
Views and opinions expressed in this work are those of the authors only and do not necessarily reflect those of the European Union or the European Research Council Executive Agency. Neither the European Union nor the granting authority can be held responsible for them. 
\begin{appendix}
\section{\texorpdfstring{$\bm{Z}$}{Z} factor at three loops and consistency relation}
\label{sec:app:Z_factor}
The renormalization condition for the bare hard functions
\begin{equation}
\bm{\mathcal{H}}^\mathrm{bare}_m=\bm{\mathcal{H}}_l\ast\bm{Z}_{lm}\, 
\end{equation}
leads to the following  RG equation
\begin{equation}
    \frac{d}{d\ln{\mu}}\bm{\mathcal{H}}_m=-\bm{\mathcal{H}}_l\ast\bm{\Gamma}_{lm}\, ,
\end{equation}
where $\ast$ denotes the usual generalized Mellin convolution over the respective momentum fractions.
Then, the renormalization factor is given by
\begin{equation}
    \frac{d}{d\ln{\mu}}\bm{Z}_{km}=\bm{\Gamma}_{kl}\ast\bm{Z}_{lm}\, ,
\end{equation}
with the formal solution given by the path-ordered exponential
\begin{equation}
\bm{Z}=\mathrm{\textbf{P}}\exp{\left[\int_\mu^\infty\frac{d\mu^\prime}{\mu^\prime}\bm{\Gamma}(\{p\},\mu^\prime)\right]}\, .
\end{equation}
In~\cite{Becher:2009qa} an expression for the $\bm{Z}$ factor was given for the case of commuting structures. In our case, this result must be generalized due to non-commuting terms in the anomalous dimensions. This leads to the following general expression
\begin{align}
\label{eq:app:Z_factor}
\bm{Z} \;=\;&
  \bm{1}
  + \frac{\alpha_s}{4\pi}\Bigg[
        -\frac{\bm{\Gamma}_0^{\prime}}{4\epsilon^{2}}
        -\frac{\bm{\Gamma}_0}{2\epsilon}
      \Bigg]                                           \nonumber       \\
&+ \left(\frac{\alpha_s}{4\pi}\right)^{2}\Bigg[
        \frac{\bigl(\bm{\Gamma}_0^{\prime}\bigr)^{2}}{32\epsilon^{4}}
        +\frac{3\,\bm{\Gamma}_0\,\bm{\Gamma}_0^{\prime}
              +\bm{\Gamma}_0^{\prime}\,\bm{\Gamma}_0
              +6\,\beta_{0}\bm{\Gamma}_0^{\prime}}{32\epsilon^{3}}
        +\frac{\bm{\Gamma}_0\bigl(\bm{\Gamma}_0+2\beta_{0}\bigr)}{8\epsilon^{2}}
        -\frac{\bm{\Gamma}_1^{\prime}}{16\epsilon^{2}}
        -\frac{\bm{\Gamma}_1}{4\epsilon}
      \Bigg]                                                \nonumber  \\
&+ \left(\frac{\alpha_s}{4\pi}\right)^{3}\Bigg[
        -\frac{\bigl(\bm{\Gamma}_0^{\prime}\bigr)^{3}}{384\epsilon^{6}}                                    
        -\frac{34\,\bm{\Gamma}_0\,\bigl(\bm{\Gamma}_0^{\prime}\bigr)^{2}
               +13\,\bm{\Gamma}_0^{\prime}\,\bm{\Gamma}_0\,\bm{\Gamma}_0^{\prime}
               + 7\,\bigl(\bm{\Gamma}_0^{\prime}\bigr)^{2}\,\bm{\Gamma}_0}{3456\epsilon^{5}}
        -\frac{3\,\beta_{0}\bigl(\bm{\Gamma}_0^{\prime}\bigr)^{2}}{64\epsilon^{5}}                     \nonumber    \\
&\qquad
        -\frac{11\,\bm{\Gamma}_0^{2}\,\bm{\Gamma}_0^{\prime}
               + 5\,\bm{\Gamma}_0\,\bm{\Gamma}_0^{\prime}\,\bm{\Gamma}_0
               + 2\,\bm{\Gamma}_0^{\prime}\,\bm{\Gamma}_0^{2}}{576\epsilon^{4}}                   
        -\frac{\beta_{0}\,\bigl(
                 33\,\bm{\Gamma}_0\,\bm{\Gamma}_0^{\prime}
               + 12\,\bm{\Gamma}_0^{\prime}\,\bm{\Gamma}_0
               + 44\,\beta_{0}\bm{\Gamma}_0^{\prime}\bigr)}{288\epsilon^{4}}                         \nonumber       \\
&\qquad
        +\frac{7\,\bm{\Gamma}_0^{\prime}\,\bm{\Gamma}_1^{\prime}
               +20\,\bm{\Gamma}_1^{\prime}\,\bm{\Gamma}_0^{\prime}}{1728\epsilon^{4}}                 
        +\frac{5\,\bm{\Gamma}_0\,\bm{\Gamma}_1^{\prime}
               +16\,\bm{\Gamma}_1\,\bm{\Gamma}_0^{\prime}
               + 2\,\bm{\Gamma}_0^{\prime}\,\bm{\Gamma}_1
               + 4\,\bm{\Gamma}_1^{\prime}\,\bm{\Gamma}_0}{288\epsilon^{3}}                           \nonumber  \\
&\qquad
        -\frac{\bm{\Gamma}_0\bigl(\bm{\Gamma}_0+2\beta_{0}\bigr)\bigl(\bm{\Gamma}_0+4\beta_{0}\bigr)}{48\epsilon^{3}}
        +\frac{8\beta_{1}\bm{\Gamma}_0^{\prime}+5 \beta_{0}\bm{\Gamma}_1^{\prime} }{72\epsilon^{3}}                                                      \nonumber \\
&\qquad
        +\frac{(\bm{\Gamma}_0+4\beta_{0})\,\bm{\Gamma}_1
              +2\bigl(\bm{\Gamma}_1+2\beta_{1}\bigr)\,\bm{\Gamma}_0}{24\epsilon^{2}}
        -\frac{\bm{\Gamma}_2^{\prime}}{36\epsilon^{2}}
        -\frac{\bm{\Gamma}_2}{6\epsilon}
      \Bigg]
  + \mathcal{O}\!\bigl(\as^{4}\bigr)\,,
\end{align}
where we have expanded
\begin{align}
\bm{\Gamma}(\alpha_s,\mu) &= \frac{\alpha_s}{4\pi}\bm{\Gamma}_0 + \left(\frac{\alpha_s}{4\pi}\right)^{2} \bm{\Gamma}_1 + \dots\,, \nonumber \\ \beta(\alpha_s) &= -2\alpha_s \left(\frac{\alpha_s}{4\pi}\beta_0 + \left(\frac{\alpha_s}{4\pi}\right)^{2} \beta_1 + \dots\right)\,,
\end{align} 
with  
\begin{equation}
   \beta_0 = \frac{11}{3} C_A - \frac{4}{3} T_F n_f,  
   \qquad \beta_1 = \frac{34}{ 3} C_A^2  - \frac{20}{3}T_FC_A  n_f - 4 T_F C_F  n_f\,,
\end{equation}
and defined
\begin{equation}
\bm{\Gamma}^{\prime} = \frac{\partial}{\partial\ln\mu}\, \bm{\Gamma}\,.
\end{equation}
The result \eqref{eq:app:Z_factor} is valid in a general space of non-commuting anomalous dimension matrices, and we suppress multiplication or convolution signs in the products of the expansion coefficients $\bm{\Gamma}_i$ and $\bm{\Gamma}_i^{\prime}$. Our expression for $\bm{Z}$ reproduces the result given in~\cite{Becher:2009qa} for commuting anomalous dimensions after substituting $\bm{\Gamma}\to - \bm{\Gamma}$. 

The one- and two-loop anomalous dimensions for SLLs are spin diagonal, but, beyond this order, the anomalous dimensions acquire non-trivial Lorentz structure, see the discussion in Section \ref{sec:factorization} and the explicit result in \eqref{eq:three_loop_q_to_g}.

\section{Collinear one-loop calculation}
\label{sec:app:coll_one_loop}
In this appendix, we supplement the discussion presented in Section~\ref{sec:one-loop-collinear} and provide further details regarding the one-loop anomalous dimensions coming from the soft-collinear matrix elements. We use a gluon mass $m$ to regulate the IR singularities and extract the anomalous dimensions. In the case of the $g\to q$ and $g\to \bar{q}$ splittings we furthermore introduce a quark mass, as these diagrams are not regulated via a gluon mass alone. In Section~\ref{sec:one-loop-collinear},  we considered the diagonal quark channel, here we  we calculate the other relevant channels of the one-loop beam function. In general, the beam function is defined as
\begin{equation}
    \bm{B}_{i/j}(z)=\SumInt_X \delta\!\left( \bar{n}_i\cdot\left(k-(1-z)p\right)\right)\mathcal{P}^{(i)}_{\alpha\beta}\langle j(p)\lvert \overline{\Phi}_{i\alpha}(0)\rvert X\rangle \langle X\lvert \Phi_{i\beta}(0)\rvert j(p)\rangle\, ,
\end{equation}
where $j$ denotes the incoming parton and $i$ the field entering the hard functions and the projectors $\mathcal{P}^{(i)}_{\alpha\beta}$ can be found in~\cite{Becher:2023mtx}.
\subsection{Diagonal gluon channel}
\begin{figure}
    \centering
    \includegraphics[scale=1]{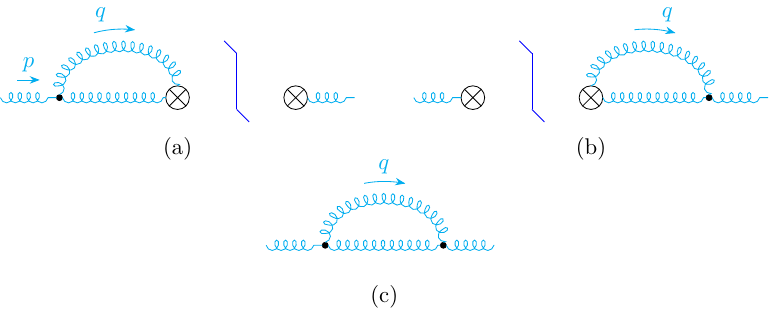}
    \caption{ The virtual contributions to the one-loop beam function for gluons. The upper two diagrams are connected with collinear Wilson lines, the lower represents the renormalization factor.}
    \label{fig:collinear_1loop_virtual_gluons}
\end{figure}
We start with an analysis of the diagonal gluon channels, where $i=j=g$. The virtual contribution (see Figure~\ref{fig:collinear_1loop_virtual_gluons}) is given by 
\begin{equation}
     \bm{B}_{g/g}^{(1)\spac \mathrm{virt}}(z)\big\rvert_{{\mathrm{Figure}}\spac\ref{fig:collinear_1loop_virtual_gluons}a,b}\hspace{-0.1cm}= 2N_c\delta^{ac}\delta^{\bar{a}c^\prime}\delta(1-z)\left(\frac{\mu^{2}}{m^{2}}\right)^{\hspace{-0.1cm}\epsilon}\frac{e^{\epsilon\gamma_E}}{\epsilon}\Gamma(1+\epsilon)\frac{1}{2\eta}\hspace{-0.05cm}\left(1+\frac{1}{1-2\eta}\right)\hspace{-0.1cm}\left(\frac{\nu}{\bar{n}_i\cdot p}\right)^{\hspace{-0.05cm}2\eta} \hspace{-0.1cm},
\end{equation}
where $a$, $\bar{a}$ are the open gluon indices on the hard functions and $c,c^\prime$ are connected to the initial-state trace. As in the main text, we use the rapidity regulator in~\eqref{eq:RapidityRegulator} expanded in the respective kinematics. The renormalization factor (coming from diagram in Figure~\ref{fig:collinear_1loop_virtual_gluons}c) is given by
\begin{equation}
    Z=1-\frac{\alpha_s}{4\pi}\frac{1}{\epsilon}\left(-\frac{5}{3}N_c+\frac{4}{3}T_Fn_f\right) +\mathcal{O}(\epsilon,\alpha_s^2)\, .
\end{equation}

\begin{figure}
    \centering
    \includegraphics[scale=1]{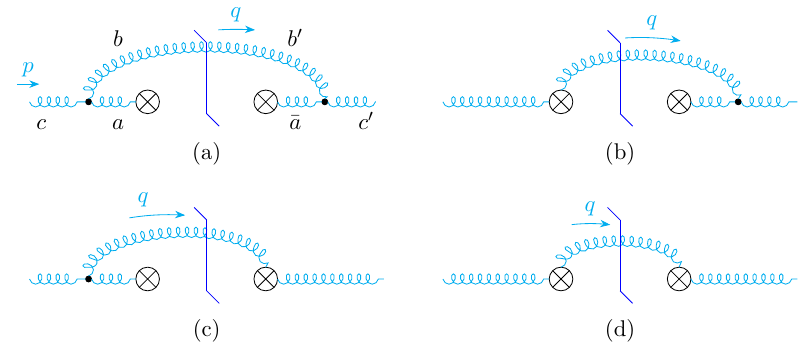}
    \caption{ The real-emission contributions to the one-loop beam function for gluons. The first diagram (a) does not give rise to a pole in $1/\eta$, diagrams (b) and (c) give the contributions to the SLLs. The fourth diagram is vanishing.}
    \label{fig:collinear_1loop_em_gluons}
\end{figure}

As a next step, we compute the contributions from the real-emission diagrams, shown in Figure~\ref{fig:collinear_1loop_em_gluons}. The first diagram is given by
\begin{align}
  \bm{B}_{g/g}^{(1)\spac\mathrm{real}}(z)\big\rvert_{{\mathrm{Figure}}\spac\ref{fig:collinear_1loop_em_gluons}a}&= \big(if^{bca}\big)\big(if^{b^\prime \bar{a} c^\prime}\big)\left(\frac{\mu^2}{m^2}\right)^{\epsilon} e^{\epsilon\gamma_E}\Gamma(\epsilon)(1-z)
  \nonumber \\ &
  \times \hspace{-0.12cm}\bigg(\!\frac{1-\epsilon}{z}(2-z)^2\!+3z+(2-2\epsilon)(2-z)+3(1-2\epsilon)z+\frac{1+z}{1-z}z(1-\epsilon)\!\!\bigg)
  \nonumber \\ &
  +\big(if^{bca}\big)\big(if^{b^\prime \bar{a} c^\prime}\big)\left(\frac{\mu^2}{m^2}\right)^{\epsilon} e^{\epsilon\gamma_E}\Gamma(1+\epsilon)(1+z)z\, .
\end{align}
The color indices $a,\bar{a}$ are the open indices on the hard functions, $b,b^\prime$ represent the open indices to which further soft emissions may couple and $c,c^\prime$ are connected to the initial-state trace as can be seen explicitly in Figure~\ref{fig:collinear_1loop_em_gluons}. On the level of the beam function matrix element, one could contract $b,b^\prime$ and $c,c^\prime$, but it proves advantageous to keep them open for the extraction of the anomalous dimensions. 
Here, we made factors of $i$ manifest such that the correct color-space operator for gluons is recovered. Note that the contribution from the third line is finite.
The last two diagrams then yield

\begin{align}
    \bm{B}_{g/g}^{(1)\spac\mathrm{real}}(z)\big\rvert_{{\mathrm{Figure}}\ref{fig:collinear_1loop_em_gluons}b,c}&= -2\big(if^{bca}\big)\big(if^{b^\prime \bar{a} c^\prime}\big)\left(\frac{\mu^{2}}{m^{2}}\right)^{\epsilon}e^{\epsilon\gamma_E}\Gamma(\epsilon)z(1+z)
    \nonumber \\  & \quad
    \times \left(\frac{\delta(1-z)}{2\eta}-\left[ \frac{1}{1-z}\right]_++\mathcal{O}(\eta)\right)\left(\frac{\nu}{\bar{n}_i\cdot p}\right)^{2\eta}\, ,
\end{align}
where we neglected higher order terms in $\eta$.
\subsection{Off-diagonal channels}
\begin{figure}
    \centering
    \includegraphics[scale=1]{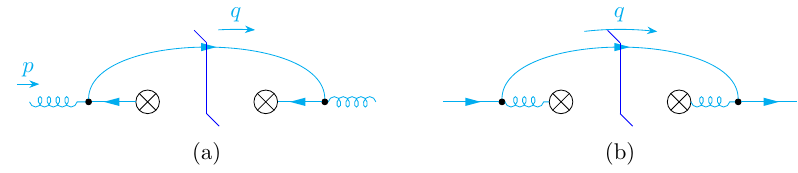}
    \caption{ The off-diagonal diagrams contributing to the splitting functions at one-loop order. The counterparts with $q\Leftrightarrow\bar{q}$ are not shown.}
    \label{fig:collinear_1loop_off_diag}
\end{figure}

In this section, we calculate the one-loop diagrams contributing to the off-diagonal splitting functions shown in Figure~\ref{fig:collinear_1loop_off_diag}.  
The result for the $g\to \bar{q}$ splitting determined from the diagram in Figure~\ref{fig:collinear_1loop_off_diag}a is given by
\begin{align}
    \bm{B}_{\bar{q}/g}^{(1)}(z)&=\left(\frac{\mu^2}{m^2}\right)^{\epsilon}e^{\epsilon\gamma_E}
    (1-\epsilon)\Gamma(\epsilon)
    (1-z)\bigg(\frac{2z^2}{1-z}+2(1-z)-\frac{8\epsilon}{d-2}\bigg)t^c_Lt^{c^\prime}_R\nonumber \\
    &\quad +2\left(\frac{\mu^2}{m^2}\right)^{\epsilon}e^{\epsilon\gamma_E}z^2\Gamma(1+\epsilon)t^c_Lt^{c^\prime}_R\, ,
\end{align}
and the result for the $q \to g$ splitting obtained from the diagram in Figure~\ref{fig:collinear_1loop_off_diag}b is
\begin{equation}
    \bm{B}_{g/q}^{(1)}(z)= \left(\frac{\mu^2}{m^2}\right)^{\epsilon} t^a_Lt^{\bar{a}}_Re^{\epsilon\gamma_E}(1-\epsilon)\Gamma(\epsilon)\bigg(4\frac{(1-z)^2}{z}+4(1-z)+(d-2)z\bigg)(1-z)^{-\epsilon}\, ,
\end{equation}
which exactly matches the structure of the respective splitting functions $\overline{\mathcal{P}}$. Here,  we introduced a quark mass as a regulator for the integral.

%%%%%%%%%%%%%%%%%%%%%%%%%%%%%%
\section{Cancellation of \texorpdfstring{\bm {$1/\eta$}}{1/η} poles}
\label{sec:app:cancellation_nu_dim}
 As the hard functions are independent of $\nu$, the respective dependence on both $\eta$ and $\nu$ has to cancel within the soft-collinear matrix elements alone. The $\nu$-independence was already shown, as the anomalous dimensions match their hard counterpart. Here, we show that also the poles in $1/\eta$ cancel, where it is sufficient to consider the single pole behavior. Using color conservation $\sum_i \bm{T}_i=0$, we may rewrite the  $1/\eta$ poles arising from the soft function~\eqref{eq:soft-1loop-virt-eta} and~\eqref{eq:soft-1loop-real-eta} as 
\begin{align}
\frac{\alpha_s}{4\pi}\bm{\mathcal{S}}_\eta^{(1)\spac\mathrm{virt}} &=\frac{2}{\eta}\frac{\alpha_s}{4\pi}\sum_{i=1,2}C_i\ln{\frac{m^2}{\mu^2}}  \nonumber \, \\
\frac{\alpha_s}{4\pi}\bm{\mathcal{S}}_\eta^{(1)\spac\mathrm{real}}&=-\frac{2}{\eta}\frac{\alpha_s}{4\pi}\sum_{i=1,2}\bm{T}_{i,L}\circ\bm{T}_{i,R}\ln{\frac{m^2}{\mu^2}}\,,
\end{align}
where we restricted the sum to go over initial states only, as the final states cancel directly after taking the trace.
The poles arising from the beam function are already in the correct form, see equations~\eqref{eq:beam_one_loop_virt_eta} and \eqref{eq:beam_one_loop_real_eta}, such that 
\begin{align}
    \bm{B}_1
\bm{B}_2\Big\rvert_{\eta}^{\mathrm{virt}}&=-\frac{2}{\eta}\frac{\alpha_s}{4\pi}\sum_{i=1,2}C_i\ln\frac{m^2}{\mu^2}{\delta(1-z_{i})}+ \mathcal{O}(\alpha_s^2)\, ,\nonumber \\ 
     \bm{B}_1
\bm{B}_2\Big\rvert_{\eta}^{\mathrm{real}}&=\frac{2}{\eta}\frac{\alpha_s}{4\pi}\sum_{i=1,2}\bm{T}_{i,L}\circ\bm{T}_{i,R} \ln{\frac{m^2}{\mu^2}}\delta(1-z_{i})+ \mathcal{O}(\alpha_s^2)\, ,
\end{align}
where we generalized the collinear part to include both gluons and anti-quarks as well. We may then directly see that the poles cancel, even before taking the trace.

\section{Rapidity logarithms and \texorpdfstring{$\bm{\nu}$}{ν} evolution}
\label{sec:app:A1_anomalousdimension}

The cross section must be independent of the unphysical scales $\mu$ and $\nu$ associated with the regularization of UV and rapidity divergences. In the literature, the $\nu$ independence, together with the factorized structure of cross sections is used to resum rapidity logarithms to all orders. This can be done using consistency relations \cite{Chiu:2007yn,Chiu:2007dg} derived from the structure of the collinear anomaly \cite{Becher:2010tm}. Alternatively, one can renormalize away the rapidity divergences and then solve renormalization group equations in the scale $\nu$ \cite{Chiu:2011qc,Chiu:2012ir}.

It would be interesting to perform an all-order analysis of the structure of rapidity logarithms in the low-energy matrix elements $\bm{\mathcal{W}}_m$. However, compared to other cases in the literature, the problem at hand involves several complications. The most obvious one is that the low-energy matrix elements do not factorize into a product of soft and collinear functions due to the appearance of Glauber interactions. One way to side-step this issue is to expand in the number of Glauber exchanges, but this leads to an infinite series of factorized soft and collinear matrix elements. To study the cancellation of  rapidity divergences, the whole tower of operators has to be considered.

A second interesting complication is the non-commuting nature of our anomalous dimensions. Assuming that we manage to define a multiplicative rapidity and UV renormalized cross section, the $\bm{Z}$ factor would fulfill
\begin{equation}\label{eq:munuIndep}
    \left[\frac{\partial}{\partial \ln{\nu}}, \frac{\partial}{\partial \ln{\mu}} \right] \bm{Z} =0\,, 
\end{equation}
if it is a smooth function of the logarithms (as it certainly is at any fixed order in perturbation theory, where the dependence is polynomial). When solving the associated RG equations, this  implies that the resummed result is independent of the chosen path when evolving from initial values $(\mu_1,\nu_1)$ to new scales $(\mu_2,\nu_2)$. Evaluating the above equality explicitly yields
\begin{align}
    \left[\frac{\partial}{\partial \ln{\nu}}, \frac{\partial}{\partial \ln{\mu}} \right]\bm{Z}&=\frac{\partial}{\partial \ln{\nu}}\bm{\Gamma}_\mu \bm{Z}-\frac{\partial}{\partial \ln{\mu}}\bm{\Gamma}_\nu \bm{Z} \nonumber \\
    &=\bm{\Gamma}_\mu \bm{\Gamma}_\nu \bm{Z}+\frac{\partial\bm{\Gamma}_\mu}{\partial \ln{\nu}}\bm{Z}-\bm{\Gamma}_\nu\bm{\Gamma}_\mu \bm{Z}-\frac{\partial\bm{\Gamma}_\nu}{\partial \ln{\mu}}\bm{Z}\, ,   
\end{align}
where we defined the anomalous dimensions
\begin{equation}
\frac{\partial \bm{Z}}{\partial \ln{\mu}} = \bm{\Gamma}_\mu\, \bm{Z}\,,\qquad\frac{\partial \bm{Z}}{\partial \ln{\nu}} = \bm{\Gamma}_\nu\, \bm{Z}\, .
\end{equation}
For $[\bm{\Gamma}_\mu, \bm{\Gamma}_\nu]=0$, this simplifies to to
\begin{equation}
  \frac{\partial\bm{\Gamma}_\nu}{\partial \ln{\mu}}=\frac{\partial\bm{\Gamma}_\mu}{\partial \ln{\nu}}\,.
\end{equation}
As far as we are aware, all previous applications of the rapidity renormalization group have assumed this relation. For the case at hand, Glauber phases $\VG$  spoil this simplification, making the all-order solution significantly more complicated. A more detailed analysis of the structure of the rapidity logarithms in the low-energy matrix elements would be interesting, but is beyond the scope of the present work. 

\section{Color structure for \texorpdfstring{$\bm{g\to \bar{q}}$}{g → q̅}}
\label{app:sec:g_to_q}
In the main text, we analyzed the color structure of the $q\to g$ off-diagonal splitting in Section \ref{sec:off_diagonal_splittings}. Here, we perform the analogous computation for the $g\to \bar{q}$ splitting. As in the former case, our first step is to prove that the soft emission off the additional emitted (anti)-quark vanishes. The relevant structure is given by 
\begin{equation}
\label{eq:gtoqbar-split-color}
    \langle\spac {\overline{\bm{\mathcal{H}}}} {\bm{\Gamma}^C}\VG\overline{\bm{\Gamma}}\rangle=-16\pi f^{abe} \sum_{j>2} J_j\langle\spac{\overline{\bm{\mathcal{H}}}} \spac\bm{\Gamma}^C\bm{T}_1^a\bm{T}_2^b\bm{T}_j^e\rangle\, ,
\end{equation}
and, when the soft gluon couples to the collinear emission, we arrive at
\begin{equation}
   \propto -if^{abe}f^{adc}t^dt^et^{c^\prime} \bm{T}_2^b\, ,
\end{equation}
where $c,c^\prime$ are the open indices on the incoming gluons. Here, the color operator $\bm{T}_1^a$ acts before the splitting, such that it acts on a gluon. When taking the color trace, we take into account a $\delta^{cc^\prime}$, yielding
\begin{equation}
    \propto -if^{dbe}f^{dac}t^at^et^{c} \bm{T}_2^b=0\, .
\end{equation}
This means that no soft emission may take place from the emitted (anti-)collinear particle, be that a (anti-)collinear quark or gluon.
As a next step, we discuss the commutator term in \eqref{eq:Wm_poles_coll}, where, for the first term we get
\begin{align}
\bm{\Gamma}^C_g(z_1)\VG\overline{\bm{\Gamma}}&\equalhat64\pi\overline{\mathcal{P}}_{g\to\bar{q}}(z_1)f^{abc}if^{aed}t^dt^e{\sum_{j>2}}^{\,\prime}J_j\bm{T}_2^b\bm{T}^c_j\nonumber \\
     &=-32\pi\overline{\mathcal{P}}_{g\to\bar{q}}(z_1)N_cf^{abc}t^a{\sum_{j>2}}^{\,\prime}J_j\bm{T}_2^b\bm{T}^c_j\, ,
\end{align}
where we recall that the hat over the equal sign indicates that the equality holds inside the color trace.
To obtain the above, we used that $\bm{T}^a_1=-if^{aed}$, as the Glauber phase acts on a gluon. For the interchanged term we obtain
\begin{equation}
   \VG\overline{\bm{\Gamma}}\bm{\Gamma}^C_g(z_1)\equalhat-64\pi\overline{\mathcal{P}}_{g\to\bar{q}}(z_1)C_Ff^{abc}t^a{\sum_{j>2}}^{\,\prime}J_j\bm{T}_2^b\bm{T}^c_j\, ,
\end{equation}
where we used $\bm{T}^a_1=t^a$ in this case as the incoming particle into the hard functions is  an anti-quark. Together, we then arrive at
\begin{equation}  
\label{eq:g_to_qqbar_Z_factor}\frac{1}{12}\langle\spac{\overline{\bm{\mathcal{H}}}}\left[\bm{\Gamma}^C_g(z_1),\VG\overline{\bm{\Gamma}}\right]\rangle=-\overline{\mathcal{P}}_{g\to\bar{q}}(z_1) \frac{8\pi}{3N_c} \frac{N_c}{N_c^2-1} f^{abc}{\sum_{j>2}}^{\,\prime}J_j\langle\spac{\overline{\bm{\mathcal{H}}}}\spac t^a\bm{T}^b_2\bm{T}^c_j\rangle\, .
\end{equation}
The additional color factor $N_c/(N_c^2-1)$ ascertains the correct normalization of the initial-state color sum.
When considering the splitting $g\to q$ an additional minus sign arises.

\section{Three-loop computation}
\label{sec:app:three_loop}

\begin{figure}
    \centering
    \includegraphics{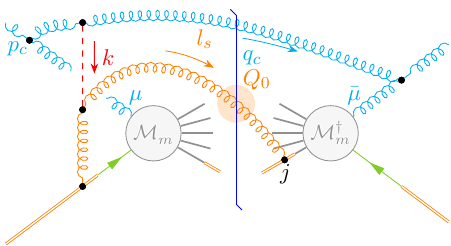}
    \caption{Diagram showing the diagonal $g\to g$ splitting. As in Figure \ref{fig:q_to_g_off_diag_three}, the Glauber exchange induces non-trivial Lorentz structures in the low-energy matrix elements.}
    \label{fig:g_to_g_diag_three}
\end{figure}

In this appendix we give further details for the diagonal $g\to g$ splitting and the off-diagonal $g \to \bar{q}$ splitting not presented in the main text.
For the diagonal $g\to g$, one obtains for the topology shown in Figure~\ref{fig:g_to_g_diag_three} the contribution
\begin{align}
    D_{g/g}^j &= -i\frac{1}{z\spac\nb \cdot p_c}\alpha_s^3 f^{ade}f^{abc}\bm{T}_{1,L}^d \bm{T}_{1,R}^e \bm{T}_{2,L}^b \bm{T}_{j,R}^c e^{2\varepsilon\gamma_E}\frac{\Gamma(2\varepsilon)\Gamma^3(1-\varepsilon)}{\varepsilon(1-2\varepsilon)\,\Gamma(1-3\varepsilon)}\Bigl(\frac{\nu}{(1-z)\nb\cdot p_c}\Bigr)^\alpha
    \nonumber\\
    &\quad\times
    \int[d\Omega_l](2n\cdot n_l\,\bar{n}\cdot n_l)^{-2\varepsilon}
    \frac{1}{n\cdot n_l\,\nb\cdot n_l\,n_j\cdot n_l}\Bigl(-n_{j}^\alpha + \frac{l_\perp^\alpha n\cdot n_j}{n\cdot l}\Bigr)\Theta_{\mathrm{veto}}(n_l)
    \nonumber\\
    &\quad\times
    \mu^{6\varepsilon}\int\frac{dE_l}{(2\pi)^2}E_l^{-1-6\varepsilon} \theta(Q_0-E_l^T)\nonumber\\
    &\quad\times\biggl\{
    \frac{g_{\perp}^{\mu\bar{\mu}} n_{l\perp}^\alpha}{2-2\varepsilon}
    \biggl[
    (2-4\varepsilon)z\Bigl(
    5(1-z)+2(1+z)+\frac{(1+z)^2}{1-z}\Bigr) + (2-2\varepsilon)\frac{4(1-z)}{z}
    \biggr]\nonumber\\
    &\qquad
    -\frac{4(1-z)}{z}
   \frac{\varepsilon}{(1-3\varepsilon)(1+\varepsilon)}\biggl[(1-\varepsilon) g_\perp^{\mu\bar{\mu}}n_l^\alpha\nonumber\\
   &\hspace{4cm}
   +2\varepsilon\Bigl(g_\perp^{\mu\alpha}n_{l\perp}^{\bar{\mu}} + g_\perp^{\bar{\mu}\alpha}n_{l\perp}^\mu + 2(1+2\varepsilon)\frac{n_{l\perp}^\mu n_{l\perp}^{\bar{\mu} }n_{l\perp}^\alpha}{n\cdot n_l\,\bar{n}\cdot n_l}\Bigr)
    \biggr]
    \biggr\}\,.\label{app:eq:g_to_g_three_loop}
\end{align}
The terms proportional to $g^{\mu\bar{\mu}}_\perp/(2-2\varepsilon)$ can be rewritten as
\begin{align}
    &z\biggl[
    (2-4\varepsilon)\Bigl(
    5(1-z)+2(1+z)+\frac{(1+z)^2}{1-z}\Bigr) + (2-2\varepsilon)\frac{4(1-z)}{z^2}
    \biggr]\nonumber\\
    &\quad-\frac{4(1-z)}{z}
   \frac{\varepsilon}{(1-3\varepsilon)(1+\varepsilon)}\biggl[(1-\varepsilon)(2-2\varepsilon)\biggr]\nonumber\\
   &= 2\biggl((2-4\varepsilon)\mathcal{P}_{g\to g}(z) - \frac{8\varepsilon^2}{(1+\varepsilon)(1-3\varepsilon)}\frac{1-z}{z}2\varepsilon\biggr)\,,
\end{align}
which is precisely the same combination as in~\eqref{eq:OffDiagCalculationSplitting} with the respective $d$-dimensional splitting function with color factor stripped
\begin{equation}
    \mathcal{P}_{g\to g} = 2\biggl[\frac{1-z}{z}+\frac{z}{1-z}+z(1-z)\biggr]\,.
\end{equation}

As in the main text, the result vanishes for $j=1,2$ due to color structure when adding the hermitian conjugate.
The topology where the soft emission is attached to the anti-collinear leg can be obtained from switching $n\leftrightarrow\nb$, taking into account the modified loop integral due to the rapidity regulator now depending on $q_\perp$, resulting in higher poles in $\varepsilon$ and $\alpha$.
Summing all soft attachments $j$, including the hermitian-conjugate contribution, and using color conservation $\sum_j T_j=0$, one then finds for the leading pole~\eqref{eq:Icomp_q_to_q_final} for the collinear and~\eqref{eq:Icomp_q_to_q_final_anticoll} for the anti-collinear splitting. Together, they combine precisely into the predicted poles, logarithms, and DGLAP splitting functions~\eqref{eq:Icomp_diag_final}.

\begin{figure}
    \centering
    \includegraphics{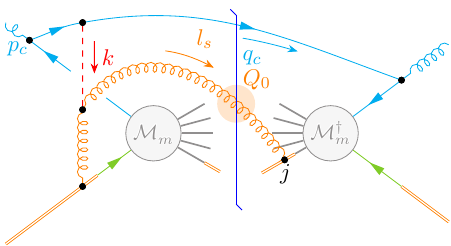}
    \caption{Diagram showing the off-diagonal $g\to \bar{q}$ splitting.}
    \label{fig:g_to_barq_off_diag_three}
\end{figure}

For the off-diagonal $g\to \bar{q}$ splitting, the relevant topology is presented in Figure~\ref{fig:g_to_barq_off_diag_three}, where one finds for the collinear splitting the contribution
\begin{align}
    D_{\bar{q}/g}^j &= 2 \alpha_s^3 f^{ade} \bm{T}_{2,L}^d \bm{T}_{j,R}^e \Bigl(\frac{\slashed{n}}{2}\Bigr)_{\beta\alpha}t^b_R t^a_L t^b_L
    e^{2\varepsilon\gamma_E}\frac{\Gamma(2\varepsilon)\Gamma^3(1-\varepsilon)}{\varepsilon\Gamma(1-3\varepsilon)}
        \nonumber\\
    &\quad\times
    \int[d\Omega_l]
    (2\spac\nb\cdot n_l\,n\cdot n_l)^{-2\varepsilon}
    \frac{n_{l\perp\alpha}}{\nb\cdot n_l\,n\cdot n_l}\Bigl(-\frac{n_j^\alpha}{n_j\cdot n_l} + \frac{n_j\cdot n\, n_{l\perp}^\alpha}{n\cdot n_l\,n_j\cdot n_l}\Bigr)\Theta_{\mathrm{veto}}(n_l)
    \nonumber\\
    &\quad\times
    \mu^{6\varepsilon}
    \int\frac{dE_l}{(2\pi)^2}E_l^{-1-6\varepsilon}\,\theta(Q_0-E_l^T)
    \biggl[
    1-2z(1-z)-\frac{2\varepsilon}{1-\varepsilon}(1-z)
    \biggr]
    \,.\label{app:eq:g_to_q_three_loop}
\end{align}
Similar to above, the splitting function can be made explicit via
\begin{equation}
    \biggl[
    1-2z(1-z)-\frac{2\varepsilon}{1-\varepsilon}(1-z)
    \biggr] = \biggl[
    1-\frac{2z(1-z)}{1-\varepsilon} - \frac{2\varepsilon}{1-\varepsilon}(1-z)^2
    \biggr]
    = \mathcal{P}_{g\to \bar{q}}(z) - \frac{2\varepsilon}{1-\varepsilon}(1-z)^2\,.
\end{equation}

Again, for $j=1,2$ the result vanishes due to color structure after adding the hermitian conjugate.
To obtain the full result, one again adds all attachments $j$, the hermitian conjugates, uses color coherence and includes the anti-collinear splittings. 
The color structure then simplifies to~\eqref{eq:g_to_qqbar_Z_factor}, and one immediately recovers the predicted result, where the correct splitting function~\eqref{eq:splitting_functions_final_explicit} can already be read off from the leading $\varepsilon$-terms in~\eqref{app:eq:g_to_q_three_loop}.

\end{appendix}

\clearpage
\pdfbookmark[1]{References}{Refs}
\bibliography{references.bib}
\end{document}